\documentclass[a4paper,fleqn,usenatbib]{mnras}

\usepackage[T1]{fontenc}
\usepackage{ae,aecompl}

\usepackage{graphicx}	% Including figure files
\usepackage{savesym}
\usepackage{amsmath}
\savesymbol{iint}
\usepackage{txfonts}
\restoresymbol{TXF}{iint}
\usepackage{amssymb}	% Extra maths symbols
\usepackage{longtable,multirow}
\usepackage{lscape}
\usepackage{xtab}

\newcommand{\spitzer}{{\it Spitzer}}
\newcommand{\herschel}{{\it Herschel}}
\newcommand{\iso}{{\it ISO}}
\newcommand{\iras}{{\it IRAS}}
\newcommand{\hipe}{{\texttt{HIPE}}}
\newcommand{\lsim}{\vcenter{\hbox{$<$}\offinterlineskip\hbox{$\sim$}}}
\newcommand{\gsim}{\vcenter{\hbox{$>$}\offinterlineskip\hbox{$\sim$}}}
\newcommand{\hho}{H$_2$O}
\newcommand{\coo}{CO$_2$}
\newcommand{\hii}{H\,{\sc ii}}
\newcommand{\hi}{H\,{\sc i}}
\newcommand{\oi}{[O\,{\sc i}]}

\newcommand{\oiii}{[O\,{\sc iii}]}
\newcommand{\ci}{[C\,{\sc i}]}
\newcommand{\cii}{[C\,{\sc ii}]}

\newcommand{\nii}{[N\,{\sc ii}]}
\newcommand{\sii}{[S\,{\sc ii}]}
\newcommand{\oh}{OH}
\newcommand{\al}{\llap}
\newcommand{\ar}{\rlap}
\newcommand{\lsun}{L$_{\odot}$}

\title[Herschel spectroscopy of Magellanic YSOs]{Herschel spectroscopy of 
Massive Young Stellar Objects in the Magellanic Clouds\thanks{\herschel\ was an ESA space 
observatory with science instruments provided by European-led Principal Investigator 
consortia and with important participation from NASA.}}

\author[Oliveira et al.]{
J.M. Oliveira$^{1}$\thanks{E-mail: j.oliveira@keele.ac.uk},
J.Th. van Loon$^{1}$,
M. Sewi\l{}o$^{2,3}$,
M.-Y. Lee$^{4,5}$, 
V. Lebouteiller$^{6}$, \newauthor
C.-H.R. Chen$^{5}$, 
D. Cormier$^{6}$, 
M.D. Filipovi\'{c}$^{7}$,
L.R. Carlson$^{8}$, 
R. Indebetouw$^{9,10}$, \newauthor
S. Madden$^{6}$, 
M. Meixner$^{11,12}$, 
B. Sargent$^{12}$,
Y. Fukui$^{13}$ 
\\
$^{1}$Lennard Jones Laboratories, School of Chemical \& Physical Sciences,
Keele University, Staffordshire ST5 5BG, UK\\
$^{2}$ CRESST II and Exoplanets and Stellar Astrophysics Laboratory, NASA Goddard 
Space Flight Center, Greenbelt, MD 20771, USA\\
$^{3}$Department of Astronomy, University of Maryland, College Park, MD 20742,
USA\\
$^{4}$ Korea Astronomy and Space Science Institute, 776 Daedeokdae-ro, 34055 Daejeon, 
Republic of Korea\\
$^{5}$Max-Planck-Institut f\"ur Radioastronomie, Auf dem H\"ugel 69, 53121 Bonn,
Germany\\
$^{6}$Laboratoire AIM, CEA/Service d'Astrophysique, B\^at. 709, CEA-Saclay, 91191 
Gif-sur-Yvette Cedex, France\\
$^{7}$ Western Sydney University, Locked Bag 1797, Penrith South DC, NSW 2751, 
Australia\\
$^{8}$ Independent Scholar, Massachusetts 02125, USA\\
$^{9}$Department of Astronomy, University of Virginia, P.O. Box 400325,
Charlottesville, VA 22904, USA\\
$^{10}$National Radio Astronomy Observatory, 520 Edgemont Road, Charlottesville,
VA 22903, USA\\
$^{11}$Department of Physics \& Astronomy, Johns Hopkins University, 3400 N.
Charles Street, Baltimore, MD 21218, USA\\
$^{12}$Space Telescope Science Institute, 3700 San Martin Drive, Baltimore, 
MD 21218, USA\\
$^{13}$Department of Physics, Nagoya University, Chikusa-ku, Nagoya 464-8602,
Japan\\
}

\date{Accepted 2019 October 2. Received 2019 October 1; in original form 2019 July 29
}

\pubyear{2019}

\begin{document}
\label{firstpage}
\pagerange{\pageref{firstpage}--\pageref{lastpage}}
\maketitle

\begin{abstract}
We present \herschel\ Space Observatory Photodetector Array Camera and Spectrometer 
(PACS) and Spectral and Photometric Imaging Receiver Fourier Transform Spectrometer 
(SPIRE FTS) spectroscopy of a sample of twenty massive Young Stellar Objects (YSOs) in the Large and 
Small Magellanic Clouds (LMC and SMC). We analyse the brightest far-infrared (far-IR) 
emission lines, that diagnose the conditions of the heated gas in the YSO envelope and 
pinpoint their physical origin. We compare the properties of massive Magellanic
and Galactic YSOs. We find that \oi\ and \cii\ emission, that originates from the 
photodissociation region associated with the YSOs, is enhanced with respect to the dust 
continuum in the Magellanic sample. Furthermore the photoelectric heating efficiency is 
systematically higher for Magellanic YSOs, consistent with reduced grain charge in low 
metallicity environments. The observed CO emission is likely due to multiple shock 
components. The gas temperatures, derived from the analysis of CO rotational diagrams, are
similar to Galactic estimates. This suggests a common origin to the observed CO 
excitation, from low-luminosity to massive YSOs, both in the Galaxy and the Magellanic 
Clouds. Bright far-IR line emission provides a mechanism to cool the YSO environment. We 
find that, even though \oi, CO and \cii\ are the main line coolants, there is an indication 
that CO becomes less important at low metallicity, especially for the SMC sources. This is
consistent with a reduction in CO abundance in environments where the dust is warmer due
to reduced ultraviolet-shielding. Weak \hho\ and \oh\ emission is detected, consistent 
with a modest role in the energy balance of wider massive YSO environments. 
\end{abstract}


%%%%%%%%%%%%%%%%%%%%%%%%%%%%%%%%%%%%%%%%%%%%%%%%%%

%%%%%%%%%%%%%%%%% BODY OF PAPER %%%%%%%%%%%%%%%%%%

\section{Introduction}

The formation of massive stars has a profound impact on galaxies. Their great 
luminosities, intense ionising radiation, strong stellar winds and often violent
demise help shape the properties of the interstellar medium (ISM) in their host
galaxies. Since the formation of stars in the high-redshift Universe occurred in a 
metal-poor environment, it is important to understand how massive stars form at low 
metallicity.

The Large and Small Magellanic Clouds (LMC and SMC), at distances of 
50.0\,$\pm$\,1.1\,kpc \citep{pietrzynski13} and 62.1\,$\pm$\,2.0\,kpc \citep{graczyk14}
respectively, offer a wide panorama of stellar populations, unencumbered by distance 
ambiguities and foreground dust extinction. Their physical conditions, distinct from 
those prevalent in the Milky Way galaxy, allow us to assess the impact of environmental 
factors like metallicity on ISM properties and on the star formation process. 

The lower metallicities of the LMC and SMC 
\citep[$Z_{\rm LMC}$\,=\,0.3\,$-$\,0.5\,Z$_{\odot}$ and 
 $Z_{\rm SMC}$\,=\,0.2\,Z$_{\odot}$; e.g.,][]{russell92}, imply not only lower gas-phase 
metal abundances but also lower dust abundances \citep[e.g.,][]{duval14}. The reduced
dust shielding in turn results in warmer dust grains 
\citep[e.g.,][]{vanloon10a,vanloon10b}. All these effects have a direct impact on the 
physical and chemical processes that drive and regulate star formation, in particular 
the ability of the contracting cloud core to dissipate its released energy.

The Magellanic Clouds are an interacting system of galaxies. Tidal stripping between the 
LMC and the SMC about 0.2\,Gyr ago \citep{bekki07} gave rise to perturbed \hi\ gas
that is colliding with the pristine \hi\ gas in the LMC disk. These colliding
\hi\ flows are believed to have triggered the formation of the massive O- and B-type
stars in the 30\,Doradus \citep{fukui17} and N\,44 \citep{tsuge19} star forming complexes
in the LMC. Furthermore, by combining dust optical depth maps and \hi\ maps, the gas-to-dust
ratio of the colliding gas is shown to be larger than that of the LMC disk gas 
\citep{fukui17,tsuge19}. In other words, this tidal interaction may have induced
significant metallicity gradients across the disk of the LMC, where the most significant
star formation activity is taking place. Likewise, metallicity differences between
distinct regions of the tidally distorted SMC are also reported \citep{choudhury18}.

The Spitzer Space Telescope \citep[\spitzer,][]{werner04} Legacy Programmes ``Surveying 
the Agents of Galaxy Evolution'' \citep[SAGE,][]{meixner06} and ``Surveying the Agents of
Galaxy Evolution in the Tidally-Disrupted, Low-Metallicity Small Magellanic Cloud'' 
\citep[SAGE-SMC,][]{gordon11} have identified 1000s of previously unknown massive YSO 
candidates, in both Magellanic Clouds \citep[e.g.,][]{whitney08,gruendl09,sewilo13}. The 
Herschel Space Observatory \citep[\herschel,][]{pilbratt10} Key Programme ``HERschel 
Inventory of The Agents of Galaxy Evolution'' \citep[HERITAGE,][]{meixner13} further 
allowed the identification of the most heavily embedded YSOs \citep{sewilo10, seale14}. 
Follow-up programmes with the \spitzer\ Infrared Spectrograph \citep[IRS][]{houck04} in 
the Magellanic Clouds \citep[e.g.,][]{seale09,kemper10,oliveira13} have confirmed the YSO 
nature for 100s of objects 
\citep[see also][]{oliveira09,seale11,woods11,ruffle15,jones17}. 

Heating in the massive YSO near-environment is dominated by the emerging \hii\ regions
and the associated photodissociation regions (PDRs), as well as mechanical (shock)
heating by gas outflows \citep[e.g.,][]{beuther07}. The most efficient cooling of hot and
dense gas occurs at far-infrared (far-IR) wavelengths and therefore rotational bands of
abundant molecules like CO, \hho\ and \oh, and atomic lines of \oi\ and \cii\ are ideal 
tracers to probe the physical mechanism at play. 

The goal of this study is to investigate the properties of massive YSOs in the 
Magellanic Clouds. We present spectroscopic observations, obtained with the \herschel\ 
Photodetector Array Camera and Spectrometer \citep[PACS,][]{poglitsch10} and 
Fourier Transform Spectrometer (FTS) of the Spectral and Photometric Imaging Receiver 
\citep[SPIRE,][]{griffin10}, of twenty massive Magellanic YSOs, targeting the species 
mentioned above. The article is structured as follows. The sample selection, and 
observations and data processing are described in Sections \,2 and 3 respectively. 
Section 4 outlines the method to calculate the total infrared luminosity and dust
temperature for the Magellanic YSOs; Section 5 introduces the sample of massive
Galactic YSOs used for comparison with the Magellanic sample. Sections 6 and 7 detail 
the results for all detected spectral lines, and describe the properties of the 
emitting gas. Section 8 focuses on the contribution of the different gas species to the
cooling budget in the far-IR. We summarise our results in Section 9.

\section{Magellanic massive YSO sample}

%70 micron updated for M113-mid/H7A
\begin{table*}
\begin{center}
\caption{\normalsize Target information for the Magellanic YSO sample analysed 
using \herschel\ spectroscopy. References for the information in the source 
properties column are as follows: 1: \citet{oliveira13}; 2: \citet{breen13}; 3: 
\citet{ward17}; 4: \citet{oliveira09}; 5: \citet{seale09}; 6: \citet{seale11}; 
7: \citet{imai13} and references therein for a compilation of \hho\ maser 
sources in the LMC; 8: \citet{green08} for a compilation of OH and CH$_3$OH 
masers in the LMC; 9: \citet{ward16} and references therein for a recent review 
of the N\,113 region; 10: probable YSO detected with \herschel\ \citep{seale14} 
but not detected with \spitzer; 11: \citet{shimonishi10}. Radio detections for 
SMC YSOs are from \citet[][and references therein]{oliveira13}; archival images 
\citep{hughes07,bozzetto17} were inspected to identify radio counterparts of the
LMC YSOs. Sources \# 7A and 7B are not resolved in the PACS observations 
but a separate PACS spectrum was extracted for \# 7C; these three sources are 
unresolved at SPIRE wavelengths. This results in 20 sources resolved across the 
full \herschel\ (PACS and SPIRE) wavelength range, but PACS spectra were
extracted for 21 sources.  The 
last column indicates which PACS spectral ranges from Table\,\ref{pacsranges} 
are {\it not} observed for each object; all sources are observed with the SPIRE 
FTS except for \#11 (N\,113 YSO-4, see also top panel in Fig.\ref{n113_fov});
a further four sources (\#3, 8, 9 and 16) have been observed with SPIRE FTS but 
the signal-to-noise ratio for the line emission is too low for a reliable 
analysis (Sect.\,\ref{spire_analysis}).}
\label{mcsample}
\begin{tabular}{@{\hspace{0.5mm}}l@{\hspace{0.7mm}}lc@{\hspace{0.6mm}}ccccc}
\hline
\#&Source ID&RA (J2000) &Dec&Source properties&Ref.&Lines not observed \\
&&(h:m:s)&($\degr$:$^{\prime}$:$^{\prime\prime})$&&&or with low SNR\\
 \hline
\multicolumn{7}{c}{SMC YSOs}\\
 \hline
1&IRAS\,00430$-$7326		&00:44:56.3&$-$73:10:11.6&ice; \hho\ maser;
UC\hii; radio &1,2,3& OH 84\,\micron\\                     %H3 
 2&IRAS\,00464$-$7322		&00:46:24.5&$-$73:22:07.3&ice&1& OH 84\,\micron, \oiii, \hho\ 108\,\micron, CO 186\,\micron\\%H4 
 3&S3MC\,00541$-$7319		&00:54:03.6&$-$73:19:38.4&ice&1& low SNR SPIRE FTS\\		                                         %H13
 4&N\,81			&01:09:12.7&$-$73:11:38.4&protocluster; UC\hii; radio &3&OH 79\,\micron\\                       %H2 
 5&SMC\,012407$-$73090 (N\,88A)	&01:24:07.9&$-$73:09:04.1&protocluster; UC\hii&3&\\ 		                                         %H1 
\hline
\multicolumn{7}{c}{LMC YSOs}\\
\hline
 6&IRAS\,04514$-$6931		&04:51:11.4&$-$69:26:46.7&ice; radio&4&OH 79\,\micron\\                             %H11
7A&SAGE\,045400.2$-$691155.4	&04:54:00.1&$-$69:11:55.5&ice; \hho\ maser&5,6,7&\\                  %H6A 
7B&SAGE\,045400.9$-$691151.6	&04:54:00.9&$-$69:11:51.6&ice&5&\\		                         %H6 
7C&SAGE\,045403.0$-$691139.7
	&04:54:03.0&$-$69:11:39.7&ice&5,6& \\	                         %H6B
 8&IRAS\,05011$-$6815		&05:01:01.8&$-$68:10:28.2&\hho, OH, CH$_3$OH masers&7,8& low SNR SPIRE FTS\\ 		 %H17
 9&SAGE\,051024.1$-$701406.5	&05:10:24.1&$-$70:14:06.5&ice&5,6&low SNR SPIRE FTS\\		                         %H16
10&N\,113\,YSO-1  		&05:13:17.7&$-$69:22:25.0&\hho maser; radio &7,8,9&\\	                                                 %H5 
11&N\,113\,YSO-4  	        &05:13:21.4&$-$69:22:41.5&\hho\ maser?; protocluster; UC\hii; radio &7,9& SPIRE FTS\\	                                 %H7A(mid)
12&N\,113\,YSO-3		&05:13:25.1&$-$69:22:45.1&\hho, OH masers; protocluster; UC\hii; radio &7,8,9&\\ 		                                         %H7(maser) 
13&SAGE\,051351.5$-$672721.9	&05:13:51.5&$-$67:27:21.9&ice; radio &5,6&OH 79\,\micron\\  	                                	 %H10
14&SAGE\,052202.7$-$674702.1	&05:22:02.7&$-$67:47:02.1&ice; radio &5,6&\\  		                                         %H19
15&SAGE\,052212.6$-$675832.4	&05:22:12.6&$-$67:58:32.4&ice; radio &5,6&\\  %corrected OH 79                                          %H8 
16&SAGE\,052350.0$-$675719.6	&05:23:50.0&$-$67:57:19.6&ice; radio &5,6& low SNR SPIRE FTS\\		                         %H14
17&SAGE\,053054.2$-$683428.3	&05:30:54.2&$-$68:34:28.3&ice; radio &5,6&OH 79\,\micron\\	%corrected OH 79      	                                         %H9 
18&IRAS\,05328$-$6827		&05:32:38.6&$-$68:25:22.6&ice; radio&4&\\			                                         %H15
19&LMC\,053705$-$694741		&05:37:05.0&$-$69:47:41.0&\herschel\ YSO; radio &10&OH 79\,\micron\\ 	                     	         %H12
20&ST\,01			&05:39:31.2&$-$70:12:16.8&ice; radio&11&\\  		                                         %H18
\hline
\end{tabular}
\end{center}
\end{table*}

In the cold and dense circumstellar envelopes of YSOs, abundant molecules freeze-out to 
form icy mantles on the dust grain surfaces. As a result their IR spectra exhibit 
numerous broad absorption features associated with abundant molecular species like \hho,
CO and CO$_2$ \citep[e.g.,][]{tielens84,gibb04}; such features are thus commonly used to 
identify the most embedded YSOs \citep{woods11,ruffle15,jones17}. In the LMC, 
168 IR sources have been spectroscopically confirmed as bona-fide massive YSOs, 
using a variety of spectral features in the \spitzer\ IRS range 
\citep[see][for a re-evaluation of these classifications]{jones17}, 53 of
which exhibit ice features in their spectrum 
\citep{vanloon05,oliveira09,seale09,shimonishi10,seale11,oliveira13}. In the SMC only 51
massive YSOs have been spectroscopically confirmed,
\citep{oliveira11,oliveira13,ruffle15,ward17}, 14 of which exhibit ice 
absorption features. 

Starting from this sample of 67 YSOs with ice signatures, we inspected 
\spitzer\ and \herschel\ broad-band images to exclude sources located in 
regions with extended complex background, in order to retain only
the YSO sources least likely to be contaminated by ambient ISM emission. Since maser 
emission is another important signpost of the early stages of massive star formation 
\citep[e.g.,][]{fish07}, we included in the sample six maser sources in the LMC and SMC 
\citep[e.g.,][]{oliveira06,green08,ellingsen10,imai13,breen13}. 
The sample further includes an LMC YSO discovered with our \herschel\ 
photometric survey \citep{seale14}, potentially a more embedded, less evolved YSO.
Sources that are fainter than $F_{160\,\micron}$\,$\sim$\,2\,Jy were discarded. 
The final sample comprised 22 regions in the LMC and 6 in the SMC of which
14 regions in the LMC and 5 in the SMC were actually observed before \herschel\
stopped operations. Two LMC pointings include multiple YSOs. In total 20 sources 
with \herschel\ spectroscopy are analysed (see details below and Table\,\ref{mcsample}). 

The properties of the few objects that have also been analysed by other studies at 
{\it higher spatial resolution} are briefly described below. The SMC sample includes
three sources with strong ice detections \citep{oliveira11,oliveira13} and two 
sources listed in Table\,\ref{mcsample} as protoclusters. These five sources were 
recently investigated at high spatial resolution using the adaptive optics assisted 
integral-field unit (IFU) SINFONI at ESO/VLT by \citet{ward17}, with a typical 
field-of-view (FOV) $\sim$\,3\arcsec\,$\times$\,3\arcsec. They found that both 
IRAS\,00430$-$7326 and IRAS\,00464$-$7322 exhibit extended outflow morphologies in 
H$_2$\,1$-$0S(1) at 2.1218\,\micron; IRAS\,00430$-$7326 (a \hho\ maser source, 
\citealt{breen13}) is particularly suggestive of a wide, relatively uncollimated outflow 
that is bound by the presence of a disc detected in CO bandhead emission (the first such
detection in any extragalactic YSO). By contrast S3MC\,00541$-$7319 is a compact emission
line source. \citet{ward17} also analysed two well-known regions, N\,81 and 
SMC\,012407$-$73090 (N\,88A). Both IR sources are in fact resolved into protoclusters. 
N\,81 includes a source that exhibits resolved bipolar H$_2$ emission, while N\,88A is 
dominated by an expanding bubble of ionised gas (seen in Br$\gamma$ emission) surrounded 
by a very conspicuous H$_2$ emission arc. 

\begin{figure}
\begin{center}
\includegraphics[scale=0.46]{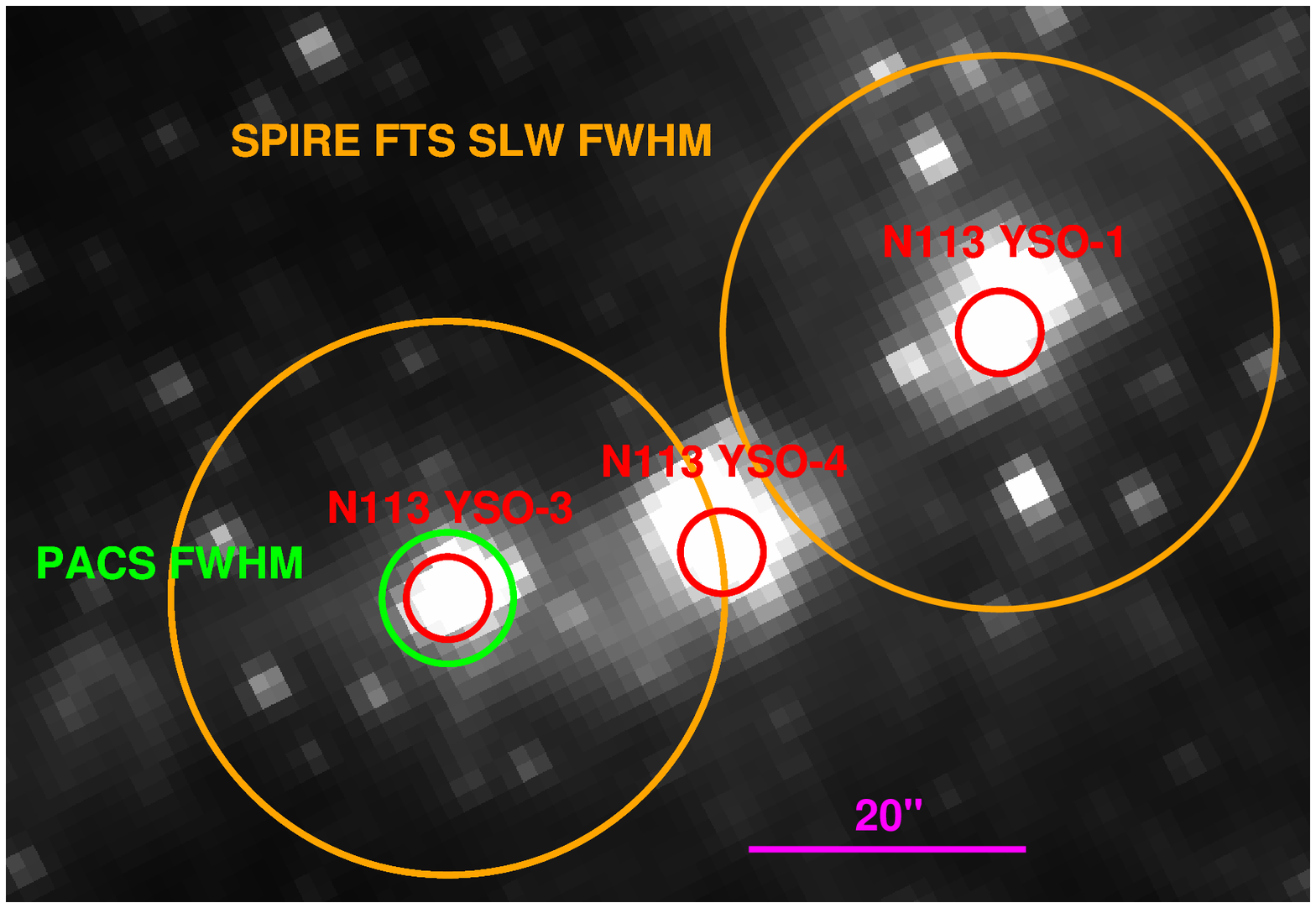}
\includegraphics[scale=0.46]{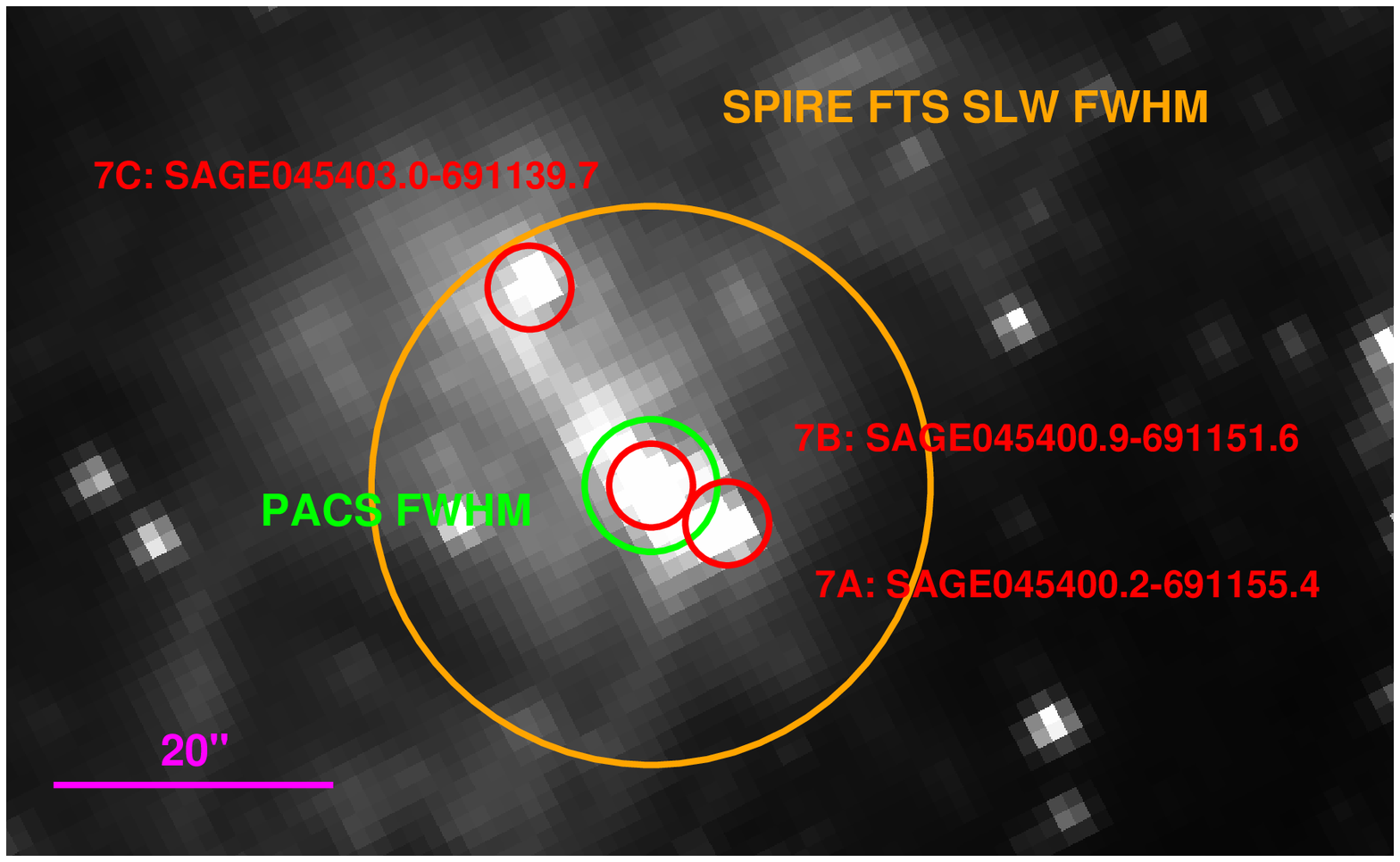}
\end{center}
\caption{\spitzer\ 3.6\,\micron\ image of N\,113 (top) and SAGE\,045400.9$-$691151.6
(bottom), identifying the YSO sources (red circles).  Also indicated is the PACS 
beam full-width-at-half-maximum (green circle, $FWHM=9\rlap{.}\arcsec5$
at the wavelength of the \oi\,63\,\micron\ and \oiii\,88\,\micron\ lines, 
\citealt{altieri13}), and the nominal SPIRE FTS beam (orange circle, $FWHM=40\arcsec$ 
for SLW, \citealt{valtchanov17}). Note that the $FWHM$ for the \cii\ beam is
$\sim$\,20\% larger. North is to the top and East is to the left.}
\label{n113_fov}
\end{figure}

The sample includes YSOs in N\,113, one of the most prominent star forming regions
in the LMC. \citet{sewilo10} provided a compilation of the indicators for 
ongoing star formation in the region (see their Fig.\,2) that we summarise here. A 
dense dust lane seems to be heated and/or compressed by prominent H$\alpha$ emission 
bubbles \citep[e.g.,][]{oliveira06}, seen also in the Magellanic Cloud Emission Line 
Survey (MCELS\footnotemark)  H$\alpha$, \oiii\ and \sii\ images. 
\footnotetext{UM/CTIO MCELS Project/NOAO/AURA/NSF} 
N\,113 also hosts the largest number of \hho\ and OH masers and the brightest 
\hho\ maser in the LMC \citep[e.g.,][]{green08,ellingsen10}.
Figure\,\ref{n113_fov} (top) shows the three conspicuous YSO sources in this region;
N\,113\,YSO-1 and N\,113\,YSO-4 are at distances of $\sim$\,45\arcsec\ and 20\arcsec\ 
respectively from N\,113\,YSO-3\footnotemark
\footnotetext{\citet{sewilo10} labelled N\,113\,YSO-1, while 
N\,113\,YSO-3 and N\,113\,YSO-4 were identified in \citet{ward16}; N\,113\,YSO-2 is
located further away to the North \citep{sewilo10}.}.
These three massive YSOs were analysed by \citet{ward16} using SINFONI/VLT (see 
above). They found that even though the \spitzer\ IRS spectra of the sources are 
similar (dominated by polycyclic aromatic hydrocarbon (PAH) and forbidden line 
emission), the nature of the three sources are in fact quite different. N\,113\,YSO-1 is
a single relatively quiescent and compact source, while N\,113\,YSO-3 is a protocluster 
dominated by an expanding ultra-compact H{\sc ii} region (UCH{\sc ii}) accompanied by 
another more compact YSO and a bright H$_2$ source (suggestive of a YSO with an outflow),
all within a 3\arcsec\ FOV. N\,113\,YSO-4 is in turn resolved into two continuum sources,
one expanding UCH{\sc ii} and a more compact source. The complexity of this region is 
undeniable, and the origin of the observed maser emission remains unclear. 
Using data obtained with the Atacama Large Millimeter/Submillimeter Array (ALMA), 
\citet{sewilo18} reported on the first extragalactic detection of several complex 
organic molecules in two hot cores in the neighbourhood of N\,113\,YSO-1 and 
YSO-3. 

Figure\,\ref{n113_fov} shows the \herschel\ pointings with multiple YSO sources:  
the N\,113 region described above (top) and SAGE\,045400.9$-$691151.6 and its 
neighbours (bottom; \#7 in Table\,\ref{mcsample}). While 
resolved in the \spitzer\ IRAC bands, \#7A and 7B are not resolved beyond 
24\,\micron\ (their separation is approximately 6\arcsec); both sources are strong 
ice sources \citep{seale09,seale11} and are associated with a \hho\ maser 
\citep{imai13}. Another nearby source (\#7C, separation $\sim$\,20\arcsec) is a weak
ice source \citep{seale09,seale11}. These three sources are unresolved at \herschel\
SPIRE wavelengths; \#7A \& 7B are unresolved in the PACS spectra, but a PACS 
spectrum for \#7C could be extracted from the observations (Sect.\,\ref{pacs_data}).
Other sources visible in Fig.\,\ref{n113_fov} (bottom) are relatively blue, for 
instance they are not detected in the PACS 100\,\micron\ images. Henceforth we
will refer to unresolved sources \#7A \& 7B collectively as 
SAGE\,045400.9$-$691151.6. All other pointings are relatively simple point sources at
this resolution.

\section{{\bf \it Herschel} observations and data reduction}

The observations discussed here were obtained as part of two \herschel\ Open Time 
programmes. These two programmes (proposal identifications OT1$\_$joliveir$\_$1 and 
OT2$\_$joliveir$\_$2) amounted to 73.4\,h of \herschel\ observing time. 

\subsection{PACS spectra}
\label{pacs_data}

All the targets in Table\,\ref{mcsample} were observed using the PACS spectrometer: 
an integral field unit made of  $5\times5$ square spaxels -- total FOV 
$\sim$47\arcsec\,$\times$\,47\arcsec) -- operating in the 50\,$-$\,200\,\micron\ 
range. Selected wavelength ranges were targeted to obtain spectra that include 
molecular and atomic lines of interest (Table\,\ref{pacsranges}). All wavelengths
quoted throughout this paper are rest wavelengths in vacuum. The spectral 
resolution varies between $\lambda/\Delta\lambda$\,$\sim$\,3000 
(for \oi\ at 83\,\micron) and $\sim$\,1000 (for \hho\ at 108\,\micron).

\begin{table}
\begin{center}
\caption{\normalsize Transitions observed with the PACS spectrometer (line
spectroscopy mode). As described in the text, 19 pointings translate to
21 YSO sources resolved with PACS;  the last column tallies how many sources are observed 
for each spectral region (see also Table\,\ref{mcsample}).}
\label{pacsranges}
\begin{tabular}{c|c|c|c}
\hline
Species&Transition&Rest $\lambda (\micron)$&N\\
\hline
\oi&$^3P_0$\,--\,$^3P_2$&63.18&21\\
OH&$^2\Pi_{1/2}(J=1/2)$\,--\,$^2\Pi_{3/2}(J=3/2)$&79.11, 79.18&19\\
OH&$^2\Pi_{3/2}(J=7/2$\,--\,$J=5/2)$&84.4, 84.6&16\\
\oiii&$^3P_1$\,--\,$^3P_0$&88.36&20\\
o-\hho&2$_{12}$\,--\,1$_{01}$&179.52&21\\
\cii&$^2P_{3/2}$\,--\,$^2P_{1/2}$&157.74&21\\
o-\hho&2$_{21}$\,--\,1$_{10}$&108.07&20\\
CO&14\,--\,13&185.99&20\\
\hline
\end{tabular}
\end{center}
\end{table}

The targets are unresolved in all \spitzer\ and \herschel\ images, and thus 
single-pointing mode was used. The emission from the ISM surrounding the targets
precluded the use of the chop/nod mode, and instead we observed all targets in
unchopped line scan mode, selecting patches of sky with negligible 160\,\micron\
emission for the offset measurement (dominated by the emission from the 
telescope). For this observing mode the continuum level can be recovered in a 
reliable way only for bright sources \citep{altieri13}; many of our targets are
relatively faint in all or some of the observed spectral ranges, therefore 
continuum uncertainties of at least $\sim$30\% are expected. However, both the 
line profiles and strength are reliably recovered in this observing mode. The number
of objects targeted for each PACS range is detailed in the last column of 
Table\,\ref{pacsranges}; ranges not observed for each target are listed in the last
column of Table\,\ref{mcsample}.

Spectra were reduced as advised within the Herschel Interactive Processing 
Environment (\hipe\footnotemark, v.10.0.2, PACS calibration tree version 48) 
using standard recipes for this type of observations.\footnotetext{\herschel\ 
Interactive Processing Environment \hipe\ is a joint development by the Herschel 
Science Ground Segment Consortium, consisting of ESA, the NASA Herschel Science 
Center, and the HIFI, PACS, and SPIRE consortia.} 
Reduced observations were retrieved from the database but spectral flatfielding
was performed independently step-by-step: this is a crucial task for improving 
the signal-to-noise ratio (SNR) of the final spectra, and it is advised to 
monitor the task's progress closely. For more information, the reader should 
refer to the ``PACS Data Reduction Guide: Spectroscopy''\footnotemark. 
\footnotetext{http://herschel.esac.esa.int/hcss-doc-10.0/} The spectra were 
checked against subsequent database, software and calibration releases; no 
further improvements in the quality of the spectra were forthcoming.

For most pointings the intended target is located in the central spaxel. The 
spectrum for the central spaxel (2,2) was extracted from the rebinned data cubes 
(\texttt{slicedFinalCubes}) and a point source flux loss correction was applied 
(no correction for slight pointing offsets and pointing jitter can be applied to
unchopped observations due to the uncertainties in the continuum level). Line
flux measurements for the lines of interest (Table\,\ref{pacsranges}) were 
performed on these extracted spectra, using standard spectral fitting tools 
within \hipe. All lines are spectrally unresolved. Line fluxes for all lines detected 
with PACS are listed in Table\,\ref{linefluxes_pacs}; example spectra are shown in 
Fig.\,\ref{spec_examples}.

Referring to Table\,\ref{mcsample}, sources SAGE\,045400.2$-$691155.4 and 
SAGE\,045400.9$-$691151.6 (\#7A and 7B) are blended as observed with the PACS 
spectrometer (the observations are actually centred on \#7B); these two sources fall
on the central spaxel (2,2) while SAGE\,045403.0$-$691139.7 (\#7C) falls on spaxel 
(1,3). N\,113 YSO-4 was not directly targeted with a separate PACS observation, but it
falls onto the FOV of the observations for both N\,113 YSO-1 and N\,113 YSO-3; this source 
is however better centred for the observation of N\,113 YSO-3, falling on spaxel (4,3). 
Spectra for those secondary sources are extracted from the spaxels mentioned and the 
point source flux loss correction is applied; centring on spaxels other than the central
one is obviously not optimal, therefore for those sources there are additional flux 
losses that cannot be corrected for. To summarise, the sample comprises 19 PACS 
pointings that result in 21 sources for which spectra were extracted. The analysis of all
lines detected in the PACS range is described in Section\,\ref{pacs_analysis}. 

For some sources, \oi, \cii\ and \oiii\ line emission is detected in spaxels
other than the central spaxel (see Sect.\,\ref{linemorphology}), i.e. a point source
is superposed on an extended environmental contribution. In order to analyse the 
morphology of the emission region and to estimate the environmental contribution to the
on-source line emission, we measured the emission line flux for these lines across 
all spaxels. Since rebinned data cubes do not have a regular sky footprint (they are
organised spatially as a slightly irregular $5\times5$ grid of 25 spaxels), we 
projected the data from the irregular native sky footprint onto a regular 3\arcsec\ 
sky grid, producing line flux maps (we made use of final data product scripts 
available within \hipe). Such maps should not be used for science measurements but 
are adequate to estimate the contribution of the off-source emission. More detail on
how these flux maps are used is given in Sect.\,\ref{linemorphology}. When detected,
\hho, OH and CO emission is present only in the central spaxel; compact OH 
absorption is detected for a small number of sources (see Section\,\ref{otherpacs}).

\subsection{SPIRE FTS spectra}
\label{spire_data}

\begin{table}
\begin{center}
\caption{\normalsize Transitions in the SPIRE FTS range.}
\label{spirelines}
\begin{tabular}{c|c|c|c|c}
\hline
Species&Transition&Rest $\lambda (\micron)$& Rest $\nu$ (GHz)&$E_J$(K)\\
\hline
CO&4\,--\,3&650.25&461.041&55.30\\
\ci&$^3P_1$\,--\,$^3P_0$&609.14&492.161&23.62\\
CO&5\,--\,4&520.23&576.268&83.00\\
CO&6\,--\,5&433.56&691.473&\al{1}16.20\\
CO&7\,--\,6&371.65&806.652&\al{1}54.90\\
\ci&$^3P_2$\,--\,$^3P_1$&369.87&809.342&62.46\\
CO&8\,--\,7&325.23&921.800&\al{1}99.10\\
CO&9\,--\,8&289.12&\al{1}036.912&\al{2}48.90\\
CO&10\,--\,9&260.24&\al{1}151.985&\al{3}04.20\\
CO&11\,--\,10&236.61&\al{1}267.014&\al{3}65.00\\
CO&12\,--\,11&216.93&\al{1}381.995&\al{4}31.30\\
\nii&$^3P_1$\,--\,$^3P_0$&205.18&\al{1}461.130&70.10\\
CO&13$-$12&200.27&\al{1}496.923&\al{5}03.10\\
\hline
\end{tabular}
\end{center}
\end{table}

We observed our YSO targets using the SPIRE FTS in single-pointing mode with sparse 
image sampling to obtain spectra from 447 to 1546\,GHz ($\sim$\,190\,$-$\,650\,\micron) 
with spectral resolution $\Delta \nu$\,=\,1.2\,GHz, covering the CO ladder from 
transitions $^{12}$CO (4$-$3) to (13$-$12) (Table\,\ref{spirelines}). The full 
wavelength coverage was achieved by using two bolometer arrays (SPIRE Short/Long 
Wavelength, respectively SSW and SLW) providing a nominal 2\arcmin\ unvigneted FOV. In 
total 19 SPIRE FTS pointings were performed. No SPIRE FTS spectrum is available for 
N\,113\,YSO-4; even though it falls within the FOV of the observations of both YSO-1 and 
YSO-3 (Fig.\,\ref{n113_fov}), the positions of the individual SSW and SLW bolometers do
not allow for a spectrum to be extracted (see below). Referring to Table\,\ref{mcsample},
a single spectrum was extracted that includes the contributions of sources \#7A, 7B and 
7C (see also bottom panel in Fig.\,\ref{n113_fov}). 

\addtocounter{footnote}{-2}

FTS spectra were reduced as advised within \hipe\footnotemark\,\,(v.10.0.1, SPIRE 
calibration tree spire$\_$cal$\_$11$\_$0) \addtocounter{footnote}{1}
using standard recipes for this type of 
observations (``SPIRE Data Reduction Guide: Spectroscopy''\footnotemark, see also 
\citealt{fulton16}). \footnotetext{http://herschel.esac.esa.int/hcss-doc-11.0/}
Background subtraction was performed using dedicated \hipe\ scripts; we found 
that using carefully selected off-axis detectors for the subtraction provided 
the best results (in terms of spectral shape and agreement between the SSW and 
SLW bands), compared to using dark sky observations from the same operational day
(this is expected for relatively faint sources). The data products were 
checked against subsequent reprocessings with more advanced \hipe\ and SPIRE 
calibration versions for representative sources; no significant improvements 
were found. Finally the FTS spectra of the targets were extracted from the SLWC3
and SSWD4 detectors. Note that given the nature of the observations (single 
pointing sparse map), SSW and SLW detector alignment is only achieved 
for the intended source; thus we cannot extract fully calibrated spectra for 
other sources in the FOV (e.g., N\,113\,YSO-4). 

The majority of SPIRE FTS spectra show a discontinuity between the SSW and SLW 
bands. For most sources the discontinuity is such that the SSW flux drops below 
the SLW flux in the overlap region. This results from the fact that the SLW beam
diameter is approximately a factor of 2 larger than that of SSW and it affects
any sources that are semi-extended at these wavelengths. For such sources the 
final spectra are corrected to an equivalent beam size (a Gaussian beam of 
40\arcsec) using the \texttt{Semi-extended Correction Tool} (\texttt{SECT}, for 
more details see the ``SPIRE Data Reduction Guide: Spectroscopy'' already 
mentioned). It should be noted that there is an implicit assumption that the 
spatial extension of the continuum and line emitting regions is the same. In total 
15 sources out of 19 had a \texttt{SECT} correction applied. 

For remaining four sources the discontinuity is reversed, i.e. the SLW flux drops below 
the SSW flux. This is a calibration effect due to rapidly changing detector temperatures
after the cooler was recycled; it affects observations taken at the beginning of an 
FTS observing block, as is the case for the four sources in question. Specialists at 
the \herschel\ Helpdesk reprocessed and corrected the FTS spectra affected. 

Line emission intensities were obtained by fitting the unapodised spectra with a 
low-order polynomial combined with a $sinc$ profile for each line of interest 
(Table\,\ref{spirelines}), using standard spectral fitting tools within \hipe. Line 
fluxes for all lines and transitions detected with SPIRE FTS are listed in 
Table\,\ref{linefluxes_spire}; example spectra are shown in Fig.\,\ref{spec_examples}. A full characterisation of the SPIRE FTS performance 
can be found in \citet[][and references therein]{hopwood15}, including discussions of 
line flux and velocity measurements and performance (see also 
Section\,\ref{spire_int_vel}). For four of the 19 SPIRE FTS pointings the SNR for 
the CO line fluxes (Table\,\ref{linefluxes_spire}) is deemed too low for a meaningful 
analysis (see also Table\,\ref{mcsample}); for these sources we compute total CO 
luminosities only, and no further analysis is performed. The analysis of emission lines 
in the SPIRE range (i.e. CO, \nii\ and \ci) is described in 
Section\,\ref{spire_analysis}.

\section{Far-IR luminosity and dust temperature}
\label{bb}

Even though the Magellanic sample was selected trying to avoid sources with a 
complex background, this was not always the case. We have also found that the
\spitzer\ IRAC and MIPS Magellanic photometric catalogues \citep{meixner06,gordon11} 
are not always reliable or complete for massive Magellanic YSOs. Such sources 
are often marginally extended to the extent that they are absent from those 
{\it point source} catalogues \citep[see discussion in][]{sewilo13}. Furthermore, 
massive YSO fluxes at 70\,\micron\ often seem anomalously high compared to for
instance PACS fluxes. Since aperture photometry has been shown to perform better 
for samples such as ours \citep{gruendl09,sewilo13}, we instead performed aperture 
photometry tailoring the parameters to each source environment. The measured fluxes 
are presented in Table\,\ref{targetphotometry}. They are used to estimate the 
integrated IR luminosity for each target.

\citet{sewilo10} described the difficulties of using \spitzer-optimised YSO 
spectral energy distribution (SED) fitters \citep[e.g.,][]{robitaille06} to fit 
simultaneous \spitzer\ and \herschel\ photometry: the model components do not fully
account for the envelope of cooler dust and gas further away from the source 
that emits little at shorter wavelengths but contributes significantly in the 
\herschel\ bands. For the line emission analysis presented in this work, it is
crucial to constrain the {\it cold dust emission} that is the spatial counterpart to 
the emission lines measured with PACS and SPIRE. The spatial resolution of the 
observations across the available wavelength range (3.6\,$-$\,500\,\micron) is also 
very disparate. Improved SED models are available \citep{robitaille17} but
limitations remain; namely only Galactic dust emission without the contribution of 
PAH emission is included, and a single source of emission is assumed. Since for our
analysis we simply require an estimate of the dust temperature and total 
integrated far-IR emission, we opt instead for the simpler approach of fitting a 
two temperature component modified blackbody function to the available photometry. 
A cold blackbody component is responsible for the majority of the far-IR emission 
(details below), however a hotter component is needed to fit the \spitzer\ 
photometry at 24\,\micron. 

To fit the two component modified blackbody we use the {\it blackbody} python
code, available as part of the {\texttt{agpy}} collection of astronomical
software\footnotemark. 
\footnotetext{{\texttt{agpy}} is authored by Adam Ginsburg and is available at 
https://github.com/keflavich/agpy.} It uses a Markov-chain Monte Carlo (MCMC) 
Bayesian statistical analysis to constrain the modified blackbodies' 
temperatures and fluxes; we use an emissivity spectral index $\beta$\,=\,1.5 
\citep[see also][]{seale14}. We adopt gas-to-dust ratios of 400 and 1000 
\citep[e.g.,][]{duval14}, and distances of 50\,kpc and 60\,kpc 
\citep*[e.g.,][]{schaefer08,hilditch05}, respectively for the LMC and SMC. For further 
details on the modified blackbody expression and adopted dust opacity $\kappa_\nu$ 
refer to \citet{battersby11}. Examples of the two component modified blackbody fits 
are shown in Fig.\,\ref{sed_examples}.

Integrated total IR luminosity and cold dust temperature are listed in 
Table\,\ref{fir_temp}; $L(>$\,30\,\micron) accounts for the majority of the 
contribution of the cold dust blackbody, while $L(>$\,10\,\micron) includes the 
contribution of the warm blackbody; the ratio $L(>$\,30\micron)/$L(>$\,10\,\micron) ranges
between 67\% and 93\% (median 77\%). Cold dust temperatures vary between 22 and 39\,K 
(median 31.6\,K); the temperature of the warmer blackbody is not well constrained. 
$L(>$\,10\,\micron) varies between $\sim$\,(5\,$-$\,32)\,$\times 10^3$\,\lsun\ 
(median $\sim$\,6.3\,$\times 10^3$\,\lsun). For the same integrated luminosity there is 
a tendency for SMC sources to have higher blackbody temperatures, reflecting higher 
dust temperature \citep[see also][]{vanloon10a,vanloon10b}. 
\citet{seale14} determined temperatures and far-IR luminosities for a subset of our
sources, using \herschel\ photometry only. In general the temperatures agree 
within the uncertainties, but in some cases our derived temperatures are higher
given that we took into account additional photometry (70\,\micron\ MIPS 
fluxes are crucial in constraining the dust temperature). The integrated far-IR 
luminosities from \citet{seale14} are in good agreement with our 
$L(>$\,70\,\micron) fluxes (not tabulated). In the subsequent discussions we 
adopt $L(>$\,10\,\micron) as the object's total IR luminosity $L_{\rm TIR}$. For these 
type of objects $L_{\rm TIR}$ does not differ significantly from the bolometric
luminosity $L_{\rm bol}$.

%data has been changed to match fir_recomputed_jan2016.fits 
%updated for N113-mid/H7A
\begin{table*}
\begin{center}
\caption{\normalsize Parameters for the modified blackbody fits to the available photometry
(see Table\,A1). The emissivity spectral index is 
$\beta$\,=\,1.5. 
Integrated luminosities are tabulated for wavelengths longer than 10\,\micron\ and 30\,\micron,
$L(>$10\,\micron) and $L(>$30\,\micron) respectively. The temperature of the hotter modified 
blackbody that contributes to $L(>$10\,\micron) (see text) is poorly constrained 
(typically $T$\,$\sim$\,100\,K). The comments column indicates missing fluxes and provides an 
empirical assessment of the reliability of the fitted parameters. For
LMC053705$-$694741 the tabulated parameters are particularly uncertain: there
are two bright sources within $\sim 25$\arcsec, and the fit is rather
conservative meaning the luminosities could easily be higher by a factor 2. 
$L(>$10\,\micron) is adopted as $L_{\rm TIR}$ for all subsequent discussions.}
\label{fir_temp}
\begin{tabular}{l|c|c|c|c|l}
\hline
\#&Source ID               &$T$        &$L(>$30\,\micron)&$L(>$10\,\micron)&Comments \\
  &                        &(K)        &($10^3$\lsun)    &($10^3$\lsun)    &\\
\hline
\multicolumn{6}{c}{SMC YSOs}\\
\hline
1 &IRAS\,00430$-$7326	   &35.2$\pm$1.4&\al{4}8.6$\pm$4.1&\al{7}1.4$\pm$4.8&\null\\
2 &IRAS\,00464$-$7322	   &27.9$\pm$0.6&\al{1}0.2$\pm$0.6&\al{1}1.8$\pm$0.6&broad peak, $T$ uncertain\\
3 &S3MC\,00541$-$7319	   &33.5$\pm$2.6&\al{1}6.1$\pm$1.6&\al{2}4.1$\pm$1.9&no fluxes available for $\lambda$\,$\geq$\,250\,\micron\\
4 &N\,81  		   &31.5$\pm$1.9&\al{4}1.6$\pm$9.3&\al{5}3.9$\pm$8.3&poor fit, uncertain parameters\\
5 &SMC\,012407$-$73090 (N\,88A)&38.1$\pm$2.4&\al{1}31$\pm$16&\al{1}95$\pm$25&no fluxes available for $\lambda$\,$\geq$\,250\,\micron\\
\hline
\multicolumn{6}{c}{LMC YSOs}\\
\hline
6 &IRAS\,04514$-$6931	   &32.5$\pm$1.8&\al{5}5.8$\pm$7.7&\al{6}8.2$\pm$7.8&\null\\
7A&SAGE\,045400.2$-$691155.4 &\multirow{2}{*}{31.9$\pm$3.3}&\multirow{2}{*}{\al{1}12$\pm$24}&\multirow{2}{*}{\al{12}8$\pm$24}&\multirow{2}{*}{flux limits for 70\,\micron\ and $\lambda$\,$\geq$\,350\,\micron, extremely uncertain}\\
7B&SAGE\,045400.9$-$691151.6 &\null                        &\null                          &\null                          &\\
7C&SAGE\,045403.0$-$691139.7 &35.6$\pm$2.0&\al{7}6.5$\pm$10&\al{8}3$\pm$10&flux limits for 70\,\micron\ and $\lambda$\,$\geq$\,350\,\micron, extremely uncertain\\
8 &IRAS\,05011$-$6815	   &31.6$\pm$2.1&\al{1}7.7$\pm$2.1&\al{2}0.6$\pm$2.3&\null\\
9 &SAGE\,051024.1$-$701406.5 &23.9$\pm$0.8&\al{1}2.4$\pm$0.8&\al{1}8.4$\pm$1.0&\null\\
10 &N\,113\,YSO-1	   &30.5$\pm$1.6&\al{2}17$\pm$21&\al{2}64$\pm$22&flux limits for $\lambda$\,$\geq$\,350\,\micron\\
11 &N\,113\,YSO-4	   &29.7$\pm$3.6& \al{8}2$\pm$8      &\al{1}14$\pm$8.4      &broad peak, $T$ uncertain, uncertain parameters\\
12 &N\,113\,YSO-3	   &34.1$\pm$1.9&\al{1}88$\pm$20&\al{2}46$\pm$21&flux limits for $\lambda$\,$\geq$\,350\,\micron\\
13 &SAGE\,051351.5$-$672721.9 &31.8$\pm$1.2&\al{8}9.5$\pm$8.2&\al{1}22$\pm$8.9&\null\\
14 &SAGE\,052202.7$-$674702.1 &25.3$\pm$2.4&\al{2}4.0$\pm$6.0&\al{2}8.2$\pm$6.0&flux limits for 70\,\micron, broad peak, $T$ uncertain\\
15 &SAGE\,052212.6$-$675832.4 &38.8$\pm$4.0&\al{2}36$\pm$50&\al{3}10$\pm$54&$T$ possibly too high\\
16 &SAGE\,052350.0$-$675719.6 &30.6$\pm$1.0&\al{4}5.7$\pm$4.4&\al{5}7.2$\pm$4.5&\null\\
17 &SAGE\,053054.2$-$683428.3 &33.5$\pm$2.5&\al{5}5.7$\pm$6.5&\al{7}2.6$\pm$6.8&flux limits for $\lambda$\,$\geq$\,250\,\micron\\
18 &IRAS\,05328$-$6827	   &22.0$\pm$2.6&9.5$\pm$0.9&\al{1}2.8$\pm$1.5&flux limits for 70\,\micron, $T$ uncertain\\
19 &LMC\,053705$-$694741	   &26.1$\pm$4.0&4.6$\pm$1.5&5.0$\pm$1.5&flux limits for
70\,\micron\ and $\lambda \geq 350$\,\micron, extremely uncertain\\
20 &ST\,01 		  
&29.8$\pm$2.1&\al{3}3.1$\pm$3.6&\al{4}0.5$\pm$3.7&broad peak, flux limits for $\lambda$\,$\geq$\,250\,\micron, $T$ uncertain\\
\hline
\end{tabular}
\end{center}
\end{table*}

\section{Galactic massive YSO comparison sample}
\label{isosample}

One of the challenges in observing and interpreting the data of massive YSOs in
the Magellanic Clouds lies with the fact that observations probe different spatial
scales compared to Galactic massive YSOs. A sample of high luminosity Galactic YSOs
was observed with \herschel\ \citep{karska14}, however the spatial scales probed 
(corresponding to the central PACS spaxel only) sample very different spatial 
components in the massive YSO environment compared to the Magellanic \herschel\ 
observations. Furthermore, the \cii\ emission is often saturated for those sources. 
With this in mind, we instead compiled a Galactic comparison sample of massive YSOs 
observed with the Infrared Space Observatory \citep[\iso,][]{kessler96}. In fact, 
the region sampled by PACS at 156\,\micron\ at a distance of the LMC 
($\sim$\,50\,kpc) is equivalent to the region sampled by the \iso\ Long-Wavelength 
Spectrograph \citep[\iso\ LWS,][]{clegg96} at a distance of $\sim$\,8\,kpc 
(\iso-LWS equivalent beam size information from \citealt{gry03}). 

We identified 22 massive YSOs observed with \iso\ LWS with luminosities in the 
range $\sim$\,(5$-$500)\,$\times 10^3$\,\lsun\ and located at distances in the 
range $\sim$\,1$-$10\,kpc (see Table\,\ref{isosources} for individual object 
information); the spectra of these sources were retrieved from the \iso\ Data 
Archive\footnotemark (IDA). 
\footnotetext{https://www.cosmos.esa.int/web/iso/access-the-archive.}
Of these 22 spectra we selected 19 for which both the \oi\ and \cii\ lines are 
in emission. The relatively low SNR of the spectra, especially at shorter 
wavelengths, implies we were only able to measure fluxes for the strongest 
emission lines: \cii\ at 158\,\micron, \oi\ at 63 and 145\,\micron, \oiii\ at 
88\,\micron, \nii\ at 122\,\micron\ and CO at 186\,\micron. To ensure uniformity in
the way fluxes are estimated, we measured all line fluxes from archival spectra 
rather than using published measurements. More details on this massive YSO comparison
sample are provided in Appendix\,\ref{iso_appendix}. The 
Magellanic and Galactic sample properties are further discussed 
in Sect.\,\ref{cooling_top3}.

\section{Results: PACS spectra}
\label{pacs_analysis}

\subsection{\cii, \oi\ and \oiii\ emission}

\subsubsection{Emission line morphology}
\label{linemorphology}

\begin{figure*}
\begin{center}
\includegraphics[scale=0.58]{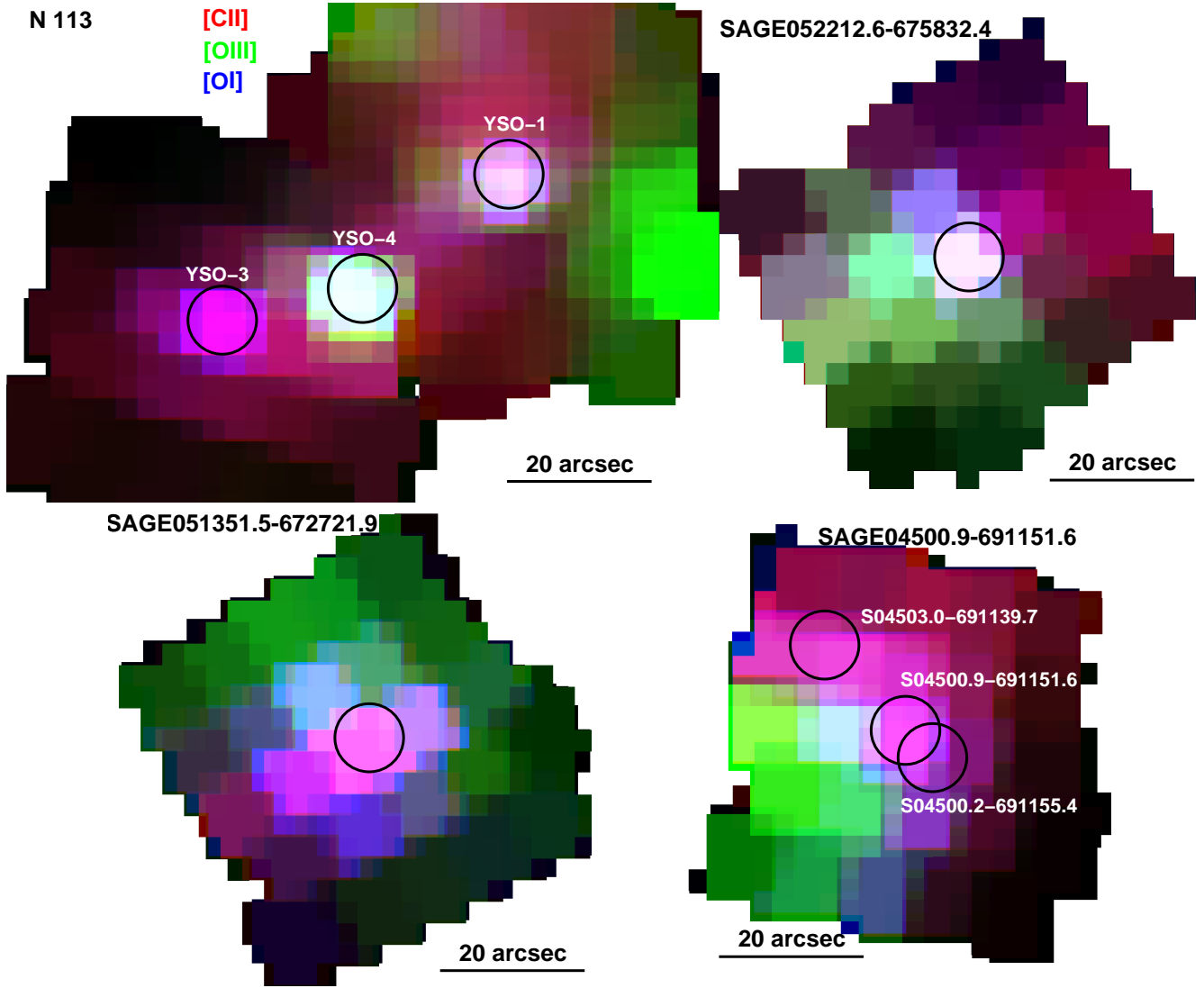}
\end{center}
\caption{Line emission maps for \cii\ (red), \oiii\ (green) and \oi\ (blue). The YSOs 
in N\,113, SAGE\,052212.6$-$675832.4, SAGE\,045400.9$-$691151.6 and 
SAGE\,051351.5$-$672721.9 (clockwise from top left) are shown (black circles with 
diameter 9\rlap{.}\arcsec5, the $FWHM$ of the \oi\ and \oiii\ beams). North is to the
top and East to the left in all maps.}
\label{n113_contours}
\end{figure*}

\cii, \oi\ and \oiii\ emission is often detected beyond the central spaxel. 
\cii\ emission is usually present across the FOV, covering it completely for all
but two sources. \oi\ emission completely covers the FOV in 11 out of 19 pointings, 
and is present beyond the central spaxel in eight others. Crucially there is always 
a flux enhancement in the central spaxel related to the point source targeted, i.e. 
the {\it point source contribution} is superposed on extended diffuse environmental 
emission. 

We investigate the morphology of the extended emission lines observed, making 
use of the line emission maps described in Sect.\,\ref{pacs_data}. Taking into
account the beam size for each line, these maps are used to estimate the environmental 
contribution as a fraction of the peak flux at the source position; that
contribution is subtracted from the measured line flux for the source, and a
point source correction is applied (Sect.\,\ref{pacs_data}). For \oi\ the 
environmental emission accounts for typically 20\% (maximum 70\%) in the LMC and 7\% 
(maximum 11\%) in the SMC; for \cii\ it accounts for typically 42\% (maximum 85\%) 
in the LMC and 30\% (maximum 50\%) in the SMC. Clearly the extended diffuse 
contribution is more important for \cii\ emission than for \oi\ emission 
\citep[see also][]{lebouteiller12}, but it is also more significant for LMC sources 
compared to SMC sources. Furthermore, the morphology of the \cii\ and \oi\ extended
emission follows the dust emission (e.g., 100\,\micron\ PACS emission), even if the 
\oi\ emission is generally less extended (Fig.\,\ref{n113_contours}).

The \oiii\ emission line morphology is as expected somewhat different, given its
distinct physical origin (see discussion in next section). Out of the 18 
observations in this spectral range (IRAS00464$-$7322 was not observed), 
six sources are not detected in \oiii\ line emission and four sources exhibit 
compact emission at the central spaxel only. For the other eight pointings, emission
extends across the FOV: for two of these the intended target (that falls on the 
central spaxel) is the source of the strongest emission; we discuss below the
remaining six pointings in more detail.

As can be seen in Fig.\,\ref{n113_contours} (top left), for the two pointings in 
N\,113, YSO-1 and YSO-4 are strong \oiii\ emission line sources, while YSO-3 is 
actually consistent with environmental emission or contamination from YSO-4 (the 
strongest \oiii\ emitter in this region). Strong emission also originates from 
locations at the FOV's northern and western edges. For four other pointings the peak 
\oiii\ emission is displaced from the central spaxel. Figure\,\ref{n113_contours} 
(bottom right) shows the line emission in the region of SAGE\,045400.9$-$691151.6. 
While the spatial distributions of \oi\ and \cii\ emission are similar (tracing the 
dust emission), the \oiii\ emission is clearly offset. This morphology is very 
suggestive of a large ionised gas bubble (as seen also in MCELS images) with the three
YSOs embedded in the dust at its rim. The observed emission for 
SAGE\,045400.9$-$691151.6 and N\,113 above are reminiscent of the emission line maps of 
LMC-N\,11, in which the spatial distribution of \oiii\ emission seems anti-correlated 
to that of \cii\ \citep{lebouteiller12}.

Two other sources with off-source \oiii\ peak emission are also shown in 
Fig.\,\ref{n113_contours}: SAGE\,051351.5$-$672721.9 (bottom left) and 
SAGE\,052212.6$-$675832.4 (top right) are also detected in the MCELS
\oiii\ map. While there might be a slight problem with source centring on the 
spaxel, it is nevertheless very clear that the \oiii\ emission is not 
point-source like. Instead it originates from the immediate surrounding \hii\ 
regions: SAGE\,051351.5$-$672721.9 is situated $\sim$20\arcsec\ from the ionising B[e]
supergiant Hen\,S22 \citep{chu03}, while SAGE\,052212.6$-$675832.4 is just
$\sim$10\arcsec\ away from an O7V star within N\,44C \citep{chen09}. Therefore we
conclude that the observed emission for SAGE\,051351.5$-$672721.9 and 
SAGE\,052212.6$-$675832.4, as well as N\,113\,YSO-3 and  SAGE\,045400.9$-$691151.6 above,
is likely mostly ambient.

A final source, SAGE\,053054.2$-$683428.3 (not shown in Fig.\,\ref{n113_contours}), 
exhibits compact \oiii\ emission centred on spaxel (2,3) superposed on more extended
environmental emission. Inspection of the emission line centroids for several spaxels 
reveals wavelength shifts that are a tell-tale sign that the source is not well 
centred in the central spaxel and is offset in the dispersion direction 
\citep[i.e. from spaxel (2,2) to spaxel (2,3), for more details refer to][]
{vandenbussche11}. Thus, the observed compact \oiii\ emission is very likely 
associated with the source on the central spaxel but it is affected by poor source 
centring.

In brief, \oiii\ emission associated with the YSO targets is detected for a total of 
nine sources, six in the LMC and three in the SMC. 

\subsubsection{Emission line diagnostics and correlations}
\label{pacs_emission}

In this section we compare line emission for \cii, \oi\ and \oiii\, for the
Magellanic sample and the Galactic \iso\ sample. Emission lines like \oi\ and 
\cii\ are often used to diagnose the environmental conditions of massive YSOs, 
since they are amongst the main contributors to line cooling. The difficulty is 
that such lines can originate from distinct components within the star formation
environment. In particular in their later evolutionary stages, massive YSOs are
copious producers of ultraviolet photons; as a result they often exhibit 
expanding compact \hii\ regions, even while still actively accreting from their 
envelopes \citep[e.g.,][]{beuther07}. These emerging \hii\ regions help shape the 
structure and chemistry of the YSO environment. In schematic terms 
\citep*[see][for an actual diagram and further details]{kaufman06}, two main 
regions can be distinguished: the \hii\ region itself marks the sphere of 
influence of H-ionising photons ($h\nu$\,$\ge$\,13.6\,eV); less energetic far-untraviolet 
(FUV) photons (6\,eV\,$\le$\,$h\nu$\,$\le$\,13.6\,eV) penetrate into the adjacent neutral 
and molecular hydrogen gas and play a significant role in the chemistry, 
heating and ionisation balance of these photodissociation regions 
\citep[e.g.,][]{tielens85}. PDRs include both the neutral dense gas near the 
YSOs but also the neutral diffuse ISM. Energetic outflows are also a ubiquitous
phenomenon in massive star formation, driving shocks through the surrounding gas
\citep[e.g.,][]{beuther07,bally16}. Both PDRs and shocks contribute to the excitation of 
far-IR \oi, \cii\ and CO emission \citep[e.g.,][]{hollenbach89,kaufman99,kaufman06}, while
shocks are particularly important to \hho\ and OH excitation 
\citep[e.g.,][]{vandishoeck11,wampfler13}.

The ionisation potential for neutral carbon C$^{0}$ is just 11.26\,eV, therefore 
\cii\ emission can originate not only in \hii\ regions, but also in PDRs and 
diffuse atomic and ionised gas \citep[e.g.,][]{kaufman99,kaufman06}. On the other 
hand, \oi\ is only found in neutral gas (the ionisation potential for O$^0$ is 
13.62\,eV, just above that for hydrogen), and emission arises from warm, dense 
regions. \oi\ emission can originate from deeper inside the PDR than \cii\ since 
some atomic oxygen remains in regions where all carbon is locked into CO. \oi\ 
emission also originates from shocks in molecular outflows that can contribute in a 
small fraction to \cii\ emission \citep[resulting in \oi/\cii\ flux ratios of 
$\gsim$\,10,][]{hollenbach89}. Given the relatively high ionisation potential for 
O$^{+}$ (35\,eV), \oiii\ emission originates from \hii\ regions, rather than
the diffuse interclump medium \citep[e.g.,][for a thorough description of 
these line properties]{cormier15}. We note that the ionised gas emitting \oiii\ 
emits little \cii\ (the ionisation potential for C$^{+}$ is 24.38\,eV); ionised 
\cii-emitting gas is traced instead by \nii\ emission (the ionisation potential 
for N$^{0}$ is 14.53\,eV).

It is important to quantify the contribution of ionised gas to \cii\ emission, before
comparing \cii\ and \oi\ line fluxes. Considering an integrated PDR and \hii\ region 
model, \citet{kaufman06} find that for solar metallicity \cii\ emission is always 
dominated by the PDR contribution as opposed to the contribution of the ionised gas in
the \hii\ region; furthermore the \hii\ region contribution increases for {\it higher}
metallicity environments. As described in Sect.\,\ref{cii_nii}, we estimated the 
ionised gas contribution to \cii\ for the eleven Magellanic YSOs for which 
\nii\,205\,\micron\ emission is detected with the SPIRE FTS; this contribution is 
typically $\sim$\,20\%. For the Galactic sample, we detect \nii\ emission at 
122\,\micron\ for seven out of 18 sources; the ionised gas contribution is 
$\sim$\,40\% (Appendix\,\ref{iso_appendix}). These contributions are consistent with 
other estimates available in the literature, and with an increased contribution for 
high metallicity environments (further discussion in Sect.\,\ref{cii_nii}).

\begin{figure}
\begin{center}
\includegraphics[scale=0.6]{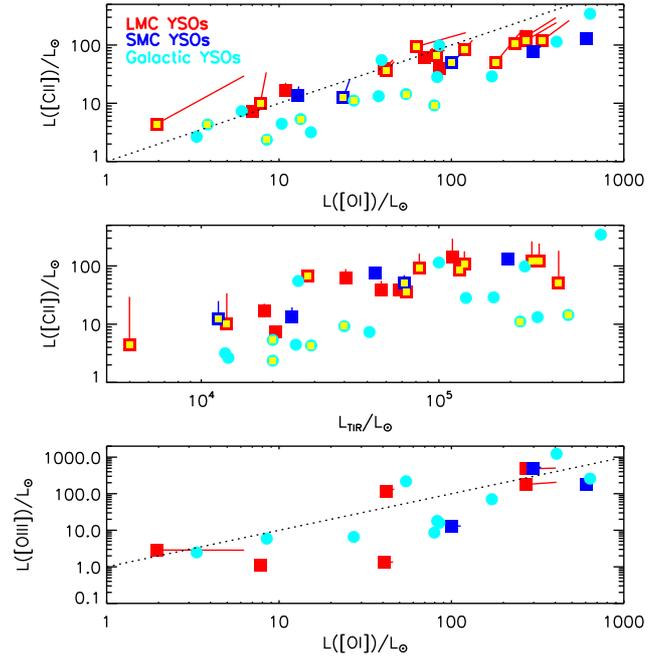}
\end{center}
\caption{FIR line correlations: \cii\ versus \oi\ luminosities (top panel), \cii\ 
luminosity versus $L_{\rm TIR}$ luminosity (middle) and \oiii\ versus \oi\
luminosities (bottom). Squares represent Magellanic YSOs (respectively red and 
blue for the LMC and the SMC) and circles represent Galactic YSOs. Line 
luminosities for both samples have been corrected for the ionised gas contribution 
to \cii\ emission (for sources with \nii\ emission (yellow symbols), see text). Only 
the Magellanic sample has been corrected for contribution of the extended diffuse 
emission (see text for details); the colour-coded lines indicate the size of this 
correction.}
\label{fir_correlations}
\end{figure}

As mentioned in Sect.\ref{linemorphology}, for the Magellanic sample we corrected the 
\oi\ and \cii\ line fluxes for the contribution of more extended diffuse gas; the 
\cii\ ionised gas correction is generally smaller than the extended gas contribution. 
We do not have extended gas estimates for the Galactic sample; however this would 
tend to enhance the observed differences between the Magellanic and Galactic samples 
(see discussion below and Figs.\,\ref{fir_correlations} and \ref{pdr_diagnostic}). 

In Fig.\,\ref{fir_correlations} we compare \cii, \oi\ and TIR luminosities for
the Magellanic and Galactic samples. Firstly, for both samples the \oi\ and 
\cii\ luminosities are strongly correlated (top panel, Spearman's rank 
correlation $\rho \sim 0.89$ and the probability that the two quantities are 
uncorrelated is $p<10^{-6}$). Even though \oi\ emission at 63\,\micron\ can be 
affected by optical depth effects, this strong correlation suggests a common origin 
for the majority of the emission being measured. Secondly, \cii\ luminosity 
correlates with the YSO's $L_{\rm TIR}$ emission (middle panel, $\rho \sim 0.81$, 
$p<10^{-5}$) indicating that photons from the YSO are responsible for its 
excitation. Furthermore, the \oi/\cii\ flux ratios are relatively low ($\sim0.3-5$)
suggesting negligible contribution from shock excitation \citep{hollenbach89}.  Put
together, this points to both \cii\ and \oi\ emission originating predominantly 
from a PDR component, at the spatial scales sampled here (i.e. ``integrated'' over 
the whole complex YSO environment). The ratio \oi/\cii\ correlates with line 
emission, more strongly with \oi\ than \cii\ emission (Spearman's $\rho$ 
respectively 0.86 and 0.56). This is likely related to uncertainties in the large 
corrections for the diffuse emission. Fig.\,\ref{fir_correlations} also shows that 
the line emission to dust continuum ratio is typically higher in the Magellanic 
Clouds compared to the Galaxy (by a factor $\sim 4.5$ for $L($\cii)/$L_{\rm TIR}$ and 
$\sim 3.5$ for $L($\oi)/$L_{\rm TIR}$, comparing line luminosities uncorrected for 
extended emission contribution), as seen also for instance by \citet[]
[and references within]{israel11}.

Figure\,\ref{fir_correlations} (bottom) shows that the Magellanic and Galactic 
samples are not significantly distinct in terms of \oiii\ emission: the mean
$L($\oiii$)/L($\oi$)$ ratio uncorrected for diffuse emission is $\sim$\,1 
with a large scatter for both samples. This is broadly consistent with a typical 
ratio of $\sim$\,0.8 measured for another sample of Magellanic YSOs using \spitzer\ 
MIPS spectroscopy \citep{vanloon10a,vanloon10b}. As described 
in Sect.\,\ref{linemorphology}, the emitting regions are complex and extended, and 
it is not always clear what is the origin of the \oiii\ emission. There is a weak 
correlation between \oiii\ and \oi\ and \cii\ emission, suggesting a mild luminosity 
scaling effect. The $L($\oiii$)/L($\oi) ratio can be a probe of the filling factor of 
ionised gas compared to that of the PDR gas. In dwarf galaxies and resolved 
Magellanic star forming regions (SFRs), this ratio is high \citep[$\sim$\,3,]
[see also \citealt{jameson18}]{cormier15}. However, such 
discussion of relative filling factors of different gas phases is likely only 
meaningful over large scales of whole SFRs or unresolved galaxies, not on smaller 
YSO scales (i.e. on the scales of a single PACS spaxel).

\subsubsection{Photoelectric heating efficiency} 
\label{pdrmodels}

\begin{figure}
\begin{center}
\includegraphics[scale=0.58]{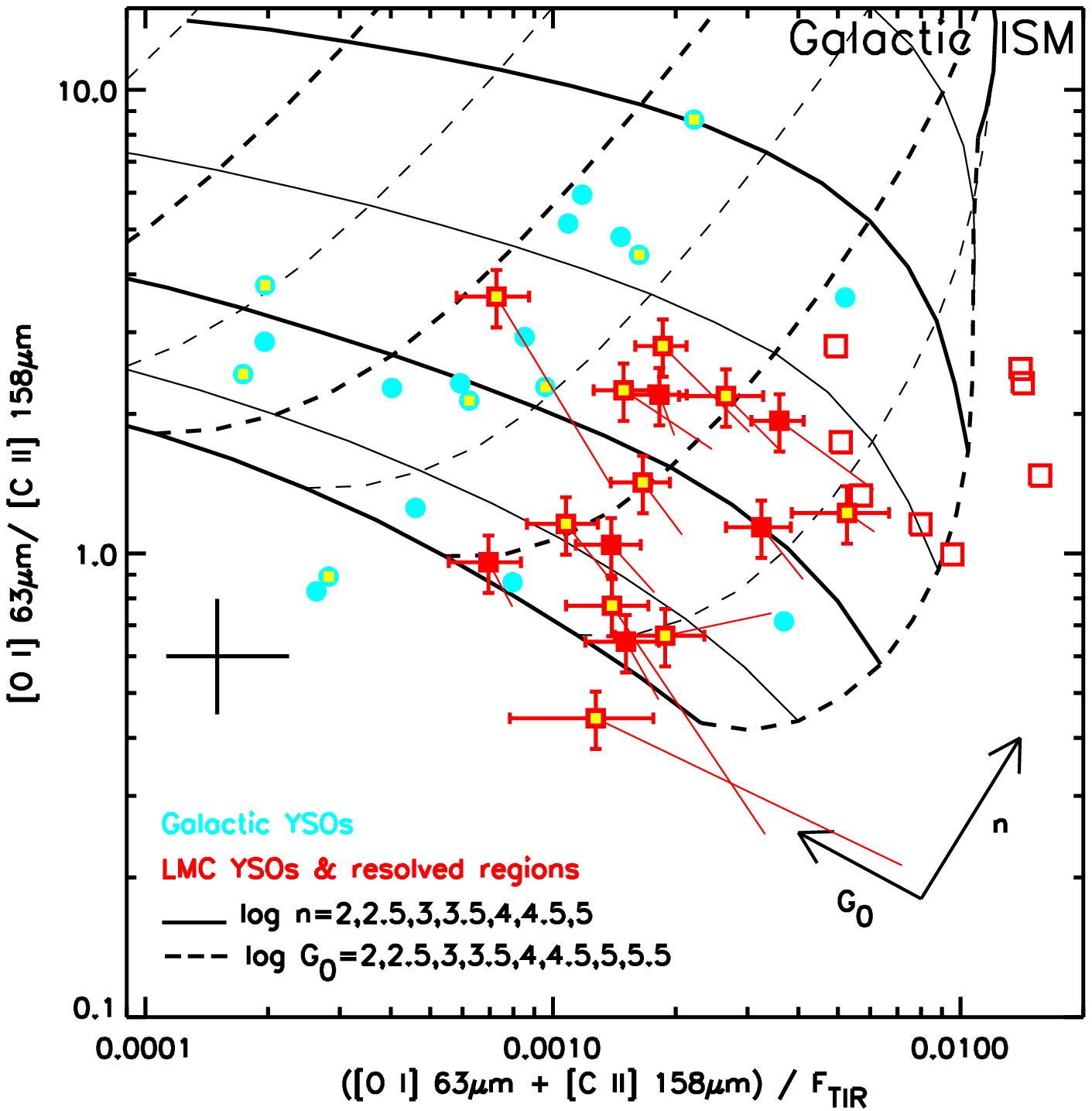}
\includegraphics[scale=0.58]{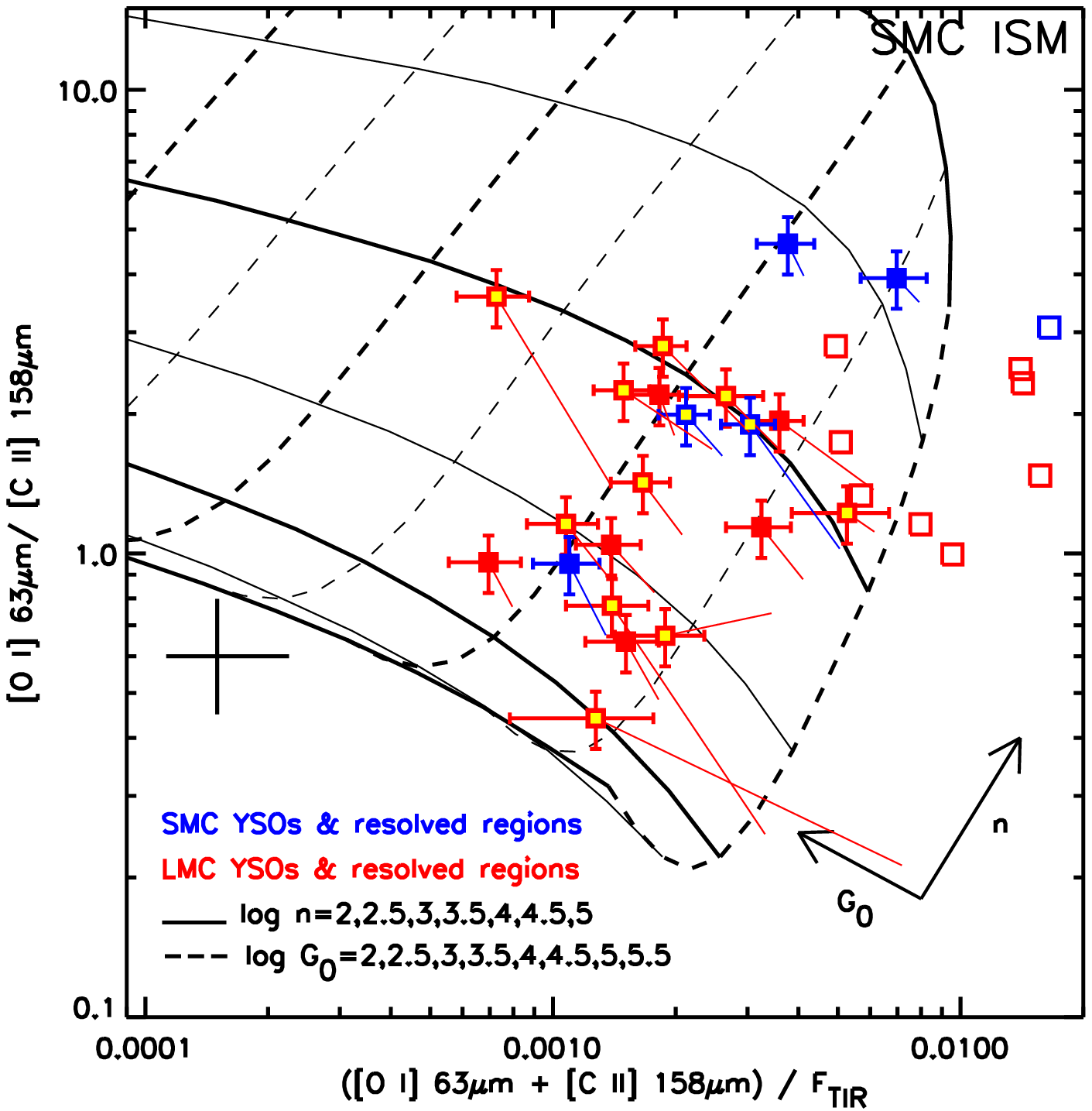}
\end{center}
\caption{PDR diagnostic diagram using \cii, \oi\ and $F_{\rm TIR}$ fluxes, for Galactic
and LMC sources (top) and LMC and SMC sources (bottom). Plotting symbols are as 
in Fig.\,\ref{fir_correlations}; open squares are resolved SFRs in the 
Magellanic Clouds \citep{cormier15}. The black cross gives the size of 25\% 
systematic uncertainties. For the Magellanic sources corrections for the 
ionised \cii\ gas fraction and extended emission are applied; for the Galactic 
sources only an ionised \cii\ fraction correction is applied (see text for further 
details). Observations are compared to models from the PDR Toolbox 
\citep{kaufman99,kaufman06}, for Galactic ISM conditions (top) and 
adapted for SMC ISM conditions \citep{jameson18}.}
\label{pdr_diagnostic}
\end{figure}

Figure\,\ref{pdr_diagnostic} shows the traditional PDR diagram used to diagnose
emitting gas conditions, i.e. line ratio \oi/\cii\ versus the total line
fluxes compared to the total TIR emission (\oi+\cii)/$F_{\rm TIR}$. The line emission is
unresolved, therefore we implicitly assume that the beam filling factor is the same 
for both emission lines. The Galactic and Magellanic YSOs have been corrected for 
the contribution of ionised gas to \cii\ emission, and Magellanic YSOs have further
been corrected for the contribution of more diffuse extended gas (corrections shown 
in Fig.\,\ref{pdr_diagnostic}). It is clear that the YSO samples occupy different 
regions in this diagram: similar line ratios are observed, but the line emission is
more prominent in the Magellanic Clouds for the same dust emission as measured by 
TIR emission, compared to Galactic sources. 

We also include a sample of resolved Magellanic SFRs \citep{cormier15}, for which line 
and dust emission fluxes are summed over the whole regions mapped. Line intensity and 
ratios vary across the regions mapped; furthermore the fraction of line intensity 
compared to dust emission decreases from the diffuse medium to denser regions in SFRs 
\citep[in the LMC,][]{rubin09}. Therefore the massive YSO sample has weaker line 
emission relative to dust emission when compared to integrated SFRs 
\citep[see also][]{jameson18}. As described in \citet{chevance16}, $F_{\rm TIR}$ can 
also include a contribution from ionised gas; such contribution is traced for 
instance by the \oiii\ emission. We find that the (\oi+\cii)/$F_{\rm TIR}$ ratio is not 
anticorrelated with the \oiii/\cii\ ratio, as would be expected if a significant
fraction of $F_{\rm TIR}$ resulted from the ionised gas contribution. Therefore, we 
conclude that $F_{\rm TIR}$ is mostly tracing dust cooling. 

The ratio (\oi+\cii)/$F_{\rm TIR}$ is often used as a proxy for the photoelectric
heating efficiency. Dust grains absorb incident radiation and emit electrons that in
turn heat the gas; given that \oi\ and \cii\ are the main coolants in dense PDRs,
and far-IR continuum emission (as well as PAH emission) cools the dust, this ratio 
provides a measure of the efficiency of the photoelectric heating 
\citep[e.g.,][]{kaufman99}. From Fig.\,\ref{pdr_diagnostic} this efficiency is higher 
for Magellanic YSOs when compared to Galactic YSOs, with medians respectively 0.25\% and 
0.1\%, with a large scatter. These estimates are broadly consistent with other 
estimations in the Galaxy \citep[e.g.,][]{salgado16} and the Magellanic Clouds 
\citep{vanloon10b}. There seems to be no clear difference between the LMC and SMC samples
\citep[as found also by][]{vanloon10b}. The photoelectric heating efficiency can be 
enhanced if the grains are less positively charged, leading to more, and more energetic,
electrons being released. Indeed \citet{sandstrom12} and \citet{oliveira13} suggested 
that observed PAH emission ratios in the SMC are consistent with a predominance of small 
neutral PAHs. 

In Fig.\,\ref{pdr_diagnostic} we also compare the YSO ratios with the PDR model
predictions\footnotemark\, from the PDR Toolbox \citep{kaufman99,kaufman06}. The 
emission is parameterised in terms of the cloud density $n$ and the strength of the
FUV radiation field $G_0$ (in units of the Habing Field, 
1.6\,$\times$\,$10^{-3}$\,ergs\,cm$^{-2}$\,s$^{-1}$). \footnotetext{A factor 2
correction to modelled $F_{\rm TIR}$ is applied, since the observed optically thin dust 
emission arises from the front and back of the cloud, while the model accounts for 
the emission from the FUV exposed face only \citep{kaufman99}.} We show two sets of
PDR models: for standard Galactic conditions (top) and using modified grain 
extinction, grain abundances, and gas-phase abundances appropriate for SMC ISM 
conditions \citep[bottom, full details can be found in][]{jameson18}. 
Figure\,\ref{pdr_diagnostic} suggests that the range of parameterised densities is 
similar ($n$\,$\lsim$\,1\,$-$\,3\,$\times 10^4$\,cm$^{-3}$), however the Magellanic 
YSOs are consistent with lower values of $G_0$ ($G_0$\,$\lsim$\,2\,$\times 10^3$,
bottom), i.e. weaker radiation field at the surface of the PDR, when compared to 
Galactic YSOs ($G_0$\,$\gsim$\,1\,$\times 10^3$, top). Therefore the $G_0/n$ 
ratio is lower for Magellanic YSOs. 

From a study of low metallicity dwarf galaxies, \citet{cormier15} interpreted lower 
$G_0/n$ ratios as an indication of a change in ISM structure and PDR distribution. 
Very schematically, low-metallicity \hii\ regions fill a larger gas volume meaning 
that PDR surfaces are at larger average distances from the FUV source, effectively 
reducing $G_0$. The mean free path of UV photons is longer \citep[UV field dilution, 
see also][]{madden06,israel11} leading to reduced grain charging. There is also 
evidence that the ISM is more porous in lower metallicity galaxies 
\citep{madden06,cormier15}, allowing ionising radiation to more easily leak into 
pristine ISM. This would imply that the region of influence for massive YSOs in the
Magellanic Clouds is generally larger with important consequences for feedback 
processes \citep{ward17}. Our analysis lends support to distinct ISM properties
at lower metallicity also on scales of a few parsecs.

\subsection{Other lines in the PACS range}
\label{otherpacs}

In this section we discuss other lines detected in the PACS spectra. While atomic
line emission seems to originate predominantly from PDRs, \hho\ and OH emission 
originates from shocks impacting on dense protostellar envelopes in complex YSO 
environments \citep[e.g.,][]{vandishoeck11,wampfler13}, with some contribution 
from outer, more quiescent envelopes to ground-state \hho\ emission
\citep{vandertak13}. The origin of CO emission in either PDRs or shocks is 
discussed in the next section.

\subsubsection{\hho\ lines}

For our sample of 21 sources with PACS spectra all sources were observed in the 
range that includes the \hho\ line at 179.5\,\micron, and all but one SMC source 
(IRAS\,00464$-$7322) were observed in the range that includes the \hho\ line at 
108\,\micron\ (Tables\,\ref{mcsample} and \ref{pacsranges}); the \hho\ 
180.5\,\micron\ line falls too close to the edge of the spectrum to be usable. In 
the LMC six sources exhibit \hho\ emission; in the SMC there is only one 
source with a tentative \hho\ emission detection. \hho\ absorption is not detected 
in the spectra of any LMC or SMC source. 

In the LMC sample the following sources exhibit 179.5\,\micron\ \hho\ emission: 
N\,113\,YSO-1, N\,113\,YSO-3, N\,113\,YSO-4, IRAS\,05011$-$6815 (all \hho\ maser 
emitters, e.g., \citealt{imai13}), and IRAS\,04514$-$6931 (YSO with strong 15\,\micron\ 
\coo\ ice absorption in its \spitzer-IRS spectrum, from which strong \hho\ ice absorption
can be inferred, \citealt{oliveira09}). The remaining LMC \hho\ maser source in the 
sample, SAGE\,045400.9$-$691151.6 (\#7A\&B in Table\,\ref{mcsample}), exhibits no 
detectable \hho\ emission lines, but another source in that protocluster, 
SAGE\,045403.0$-$691139.7 (\#7C) does. N\,113\,YSO-3 is the only source with definite 
\hho\ emission both at 179.5 and 108\,\micron. Towards the \hho\ maser source in 
the SMC (IRAS\,00430$-$7326, \citealt{breen13}) emission at 108\,\micron\ is tentatively 
detected. Spectra are shown in Fig.\,\ref{hho_spec}.

\citet{karska14} analysed PACS range spectroscopy (55\,$-$\,190\,\micron) for ten 
Galactic massive YSOs, covering a range of luminosities 
$\sim$\,(1$-$5)$\times10^4$\,\lsun\ (see their Table\,1). All but two sources
in that sample show a combination of \hho\ emission and absorption lines; two 
sources show \hho\ absorption lines only. Only W3\,IRS5 exhibits 179.5\,\micron\
emission, accounting for less than 2\% of the total \hho\ line emission. This 
source is one of the most evolved in the \citet{karska14} sample, and it is also 
one of the sources with the largest contribution of \hho\ luminosity to the total 
molecular cooling in the PACS range ($\sim$\,35\%, corresponding to $\sim$\,30\% of 
the total, atomic and molecular, line cooling). W3\,IRS5 also exhibits \hho\ maser 
emission and ice absorption features \citep[][and references therein]{gibb04}. 

More recently, \citet{karska18} analysed a large sample of low-luminosity Galactic
YSOs. All sources exhibit \hho\ emission; 55\% of the sources show emission at 
either 179.5 or 108\,\micron\ \citep[see also][]{karska13,mottram17}. The 
179.5\,\micron\ line accounts for $\sim$\,8\% of the total \hho\ 
luminosity\footnotemark  \,with a large scatter \citep{karska13,karska18}. 
Typically the ratio of 179.5\,\micron\ to 108\,\micron\ emission is $\sim$\,0.9 
\citep[range 0.6\,$-$\,1.3,][]{karska13}, therefore both lines together account for 
$\sim$\,18\% of the total \hho\ line luminosity. These fractions will be used in
Sect.\,\ref{coolingbudget} to estimate the total \hho\ line luminosity for the 
Magellanic sample.
\footnotetext{By total \hho\ and OH luminosities we mean integrated luminosities 
over {\it all lines} in the PACS range spectroscopy mode: 50\,$-$\,210\,\micron\ 
\citep[see e.g.,][for full details]{karska18}.} 

\subsubsection{OH lines}

The full sample of 21 sources with PACS spectra was observed in spectral ranges 
that cover at least one of the OH doublets listed in Table\,\ref{pacsranges}: for 
twelve LMC and two SMC sources we have spectra for both OH doublets at 84 and 
79\,\micron; for a further four LMC and three SMC sources we have observations for 
only one doublet (see Table\,\ref{mcsample}). For the 84\,\micron\ doublet, we have 
detected emission for the bluest component (84.4\,\micron) for two sources: 
N\,113\,YSO-3 and IRAS\,04514$-$6931; no absorption features are detected. For the 
79\,\micron\ doublet, two sources exhibit weak absorption (SAGE\,052350.0$-$675719.6 
and SAGE\,052212.6$-$675832.4), three sources show emission (N\,113\,YSO-3, N\,113\,YSO-4 
and SAGE\,045400.9$-$691151.6), and N\,113\,YSO-1 shows emission for the bluest 
component (79.11\,\micron) only. Most sources that show OH emission exhibit
\hho\ emission (the exception is SAGE\,045400.9$-$691151.6). No \oh\ emission or 
absorption is detected for SMC targets. Spectra are shown in Fig.\,\ref{oh_spec}.

\addtocounter{footnote}{-1}

Referring to the \citet{karska14} study of Galactic massive YSOs, while most sources 
show some OH emission, the OH doublets at 84\,\micron\ and 79\,\micron\ are seen 
mostly in absorption for most sources. This is in contrast with low- and intermediate-mass 
YSOs, for which these OH doublets are seen mostly in 
emission, e.g., 63\% sources show the 84\,\micron\ doublet in emission \citep{karska18}. 
Based on the samples described in \citet{wampfler13}, the typical flux ratios are 
F(79.11)/F(79.18)\,$\sim$\,1.0 (range 0.6\,$-$\,1.8) and 
F(84.42)/F(84.60)\,$\sim$\,1.34 (range 0.8\,$-$\,2.9) for the doublet components, 
F(79.18)/F(84.42)\,$\sim$\,0.7 (range 0.3\,$-$\,0.86), and 
F(79)/F(84)\,$\sim$\,0.77 (range 0.4\,$-$\,1.23). In terms of fraction of total OH 
luminosity\footnotemark, the 79 and 84\,\micron\ doublets account for $\sim 24$\% and
$\sim 31$\%, respectively. 

The observed ratios for the Magellanic sources are consistent with the values
above, with large uncertainties. Where only one doublet component is detected, the 
upper limits are also consistent with these ratios. We take the estimated luminosity
fractions above to predict total OH luminosities from our measured line fluxes. We 
will discuss the emission line budget for the LMC and SMC sources in 
Section\,\ref{coolingbudget}. 

\subsubsection{CO\,(14$-$13) line emission}
\label{pacs_co}

We only detected CO\,(14$-$13) emission for eight sources (out of 19 sources
observed); since we detected CO emission lines in the SPIRE range for all sources 
(see next section), this is probably just due to the low SNR ratio of the PACS 
spectra. Given the very different beam sizes for PACS and SPIRE and the fact that 
the emission beam filling factor is unconstrained, our analysis of the CO rotational
diagram is based solely on those lines in the SPIRE spectral range.

\section{Results: SPIRE FTS spectra}
\label{spire_analysis}

Table\,\ref{mcsample} provides an overview of the SPIRE FTS observations. One
source in the sample was not observed with this instrument mode. A further
four sources (three in the LMC and one in the SMC) resulted in FTS spectra with 
low continuum and line SNR ($<5$ for all CO ladder transitions); for those 
sources we only compute the total CO luminosity $L_{\rm CO}$ but we are not able to 
reliably identify other emission lines nor analyse the CO rotational diagrams. 
That leaves fifteen sources that are discussed in more detail in this section.

\subsection{Line identifications}
\label{spire_int_vel}

While most CO ladder transitions for these fifteen sources are usually well 
identified and measured, the process is somewhat more complicated for weaker 
emission lines that are detected at generally lower SNR, as is the case for the 
\ci\ and \nii\ lines (Table\,\ref{spirelines}). As described in \citet{hopwood15}, 
the SNR ratio below 600\,GHz is significantly diminished and this strongly impacts 
on the measured centroid line position (derived from $sinc$ profile fitting, see 
Section\,\ref{spire_data}) for individual transitions (see their Fig.\,16). Note 
that the line emission SNR for the point-source stellar calibrators discussed in 
the \citet{hopwood15} analysis is typically much higher than the SNR for {\it all} 
line detections discussed here (at most we achieve SNR\,$\sim$\,40).

We measured the variation of the centroid velocity position for the CO line 
emission for our sample; typical values are $\sim$\,40\,km\,s$^{-1}$ and 
$\sim$\,70\,km\,s$^{-1}$ for sources with typical SNR larger and smaller than 10, 
respectively (the median velocity is always consistent with the 
typical systemic velocity of the LMC and SMC, $\sim$\,250\,km\,s$^{-1}$ and 
$\sim$\,160\,km\,s$^{-1}$ respectively). Even for spectra with the highest SNR 
overall (N\,113\,YSO-1, SNR\,=\,22\,$-$\,41), the velocity position for the CO\,(4$-$3) 
line at 461.041\,GHz deviates by $\sim$\,3-$\sigma$ from the median centroid velocity 
for the other nine CO lines. 

The \ci\ line at 492.161\,GHz is especially affected by these uncertainties in the 
centroid line position. After careful inspection, we consider this line to be 
appropriately detected if SNR\,$\geq$\,5 and the velocity position is consistent with 
that of the nearest CO lines; this is the case for five LMC sources and one SMC source. 
The other \ci\ line at 809.342\,GHz is detected for all sources except for one SMC 
source (\#4, N\,81). The \nii\ line at 1461.13\,GHz is located in a more favourable 
part of the spectrum (better continuum SNR); this line is detected (SNR\,$\geq$\,5) for 
nine LMC and two SMC sources. These emission lines are further discussed in 
Sect.\,\ref{otherspire}.

\subsection{Total CO luminosity measured over the SPIRE range}
\label{columinosity}

\begin{figure}
\begin{center}
\includegraphics[scale=0.55]{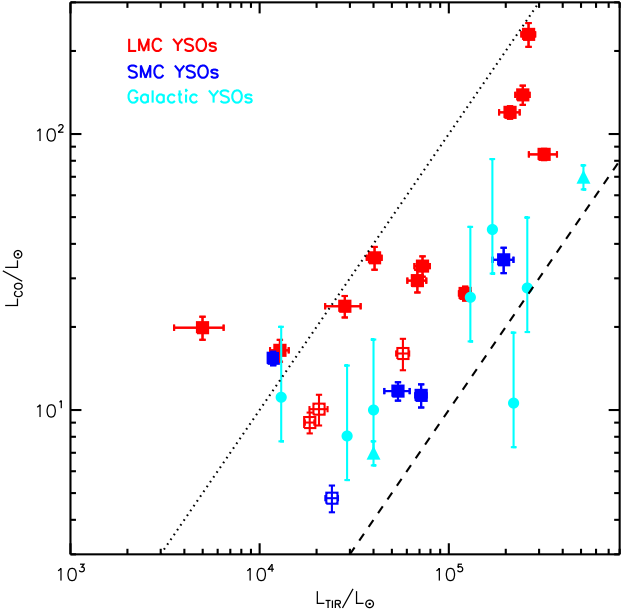}
\end{center}
\caption{$L_{\rm CO}$ measured over the SPIRE range (4\,$\leq$\,$J$\,$\leq$\,13) versus 
$L_{\rm TIR}$ for all sources. Filled squares represent the fifteen Magellanic YSOs with 
fitted CO rotational diagrams (Sect.\,\ref{co_rot}), otherwise open squares are used. 
$L_{\rm CO}$ for massive Galactic YSOs are estimated from CO\,(14$-$13) \iso\ fluxes, 
with the exception of two literature measurements \citep[triangles,][]{stock15}. Ratios 
$L_{\rm CO}/L_{\rm TIR}$ of 0.1\% (dotted line) and 0.01\% (dashed line) are shown; the
latter is the limit proposed for shock-dominated heating \citep{meijerink13}, see 
Sect.\,\ref{co_origin}.}
\label{totalco}
\end{figure}

In Fig.\,\ref{totalco} we plot the total CO luminosity $L_{\rm CO}$ {\it measured over 
the SPIRE range} -- CO\,(4$-$3) to CO\,(13$-$12) -- against $L_{\rm TIR}$; there is a 
strong correlation between the two luminosities: the Spearman's rank correlation is 
$\rho=0.74$ and the probability that the two quantities are uncorrelated is $p<0.0016$. 
Since the energy that heats the gas derives in some form from the YSO, such correlation 
is not unexpected. There is no correlation between $L_{\rm CO}$ and the dust temperature.
These findings are consistent with results for low-luminosity Galactic samples 
\citep{manoj13,manoj16,yang18}. Typically $L_{\rm CO}/L_{\rm TIR}$ is $\sim$\,0.06\% and
$\sim$\,0.02\% for the LMC and SMC sources respectively, with a large scatter. There 
is a tendency for SMC sources to be weaker CO emitters. The LMC $L_{\rm CO}/L_{\rm TIR}$ 
ratio is consistent with that found by \citet{lee16} across their SPIRE CO maps for the 
SFR N\,159W, $L_{\rm CO}/L_{\rm TIR}$\,$\sim$\,0.08\%. %The three sources highlighted 
%with large
%circles in Fig.\,\ref{totalco} are known to be protoclusters 
%(Table\,\ref{mcsample}). This could affect the estimates of $L_{\rm TIR}$ and/or 
%$L_{\rm CO}$ in ways impossible to quantify.

For the massive Galactic comparison sample observed with the \iso-LWS, CO\,(14$-$13) 
fluxes were measured for seven YSOs (Sect.\,\ref{isosample} and Table\,\ref{isosources}).
For the eight Magellanic sources with PACS CO\,(14$-$13) measurements
(Sect.\,\ref{pacs_co}), we estimate a typical fraction 
$L($CO(14$-$13))/$L_{\rm CO}$\,$\sim$\,0.12\,$\pm$\,0.06. This is consistent with a 
ratio $L($CO(14$-$13))/$L_{\rm CO}$\,$\sim$\,0.09\,$\pm$\,0.04 measured for a larger 
sample of Galactic sources \citep{green16,yang18}\footnotemark.
\footnotetext{The fluxes available in \citet{green16} have been revised according
to the procedure described in \citet{yang18}. The revised fluxes used here were 
obtained directly from the authors.} Accordingly, we adopt 
$L($CO\,(14$-$13))/$L_{\rm CO}$\,=\,0.09 to estimate $L_{\rm CO}$ for the seven Galactic 
YSOs. The $L_{\rm CO}/L_{\rm TIR}$ ratio for Galactic sources is $\sim$\,0.02\%, 
consistent with measurements for two Galactic sources \citep{stock15}. The CO 
luminosities for the Magellanic and Galactic samples are broadly consistent. However, 
the Galactic ratio is closer to that of the SMC sample, and the LMC sources tend to
exhibit higher ratios. Nevertheless, this suggests that the correlation between 
$L_{\rm CO}$ and $L_{\rm TIR}$ (or $L_{\rm bol}$) extends from low-luminosity YSOs 
\citep{manoj16,yang18} to massive YSOs, supporting a common origin for the observed CO 
emission. 

Using data available in the literature \citep[e.g.,][]{stock15,green16,yang18} we estimate
that $L_{\rm CO}$ measured over the SPIRE range (4\,$\leq$\,$J$\,$\leq$\,13) is about 
52\% of the total CO luminosity measured over the \herschel\ range ($J$\,$\geq$\,4), 
with a large scatter.

\subsection{CO rotational diagram analysis}
\label{co_rot}

When multiple rotational transitions are available, the analysis usually relies
on the so-called molecular rotational diagrams \citep[e.g.,][]{goldsmith99}, 
where the logarithm of the total number of molecules in the upper state of a 
transition $N_J$ normalised by the degeneracy of the level $g_J$ is plotted 
against the upper level excitation energy $E_J$. In this section we analyse
the CO rotational diagrams for the fifteen sources with adequate SNR (eleven in the LMC
and four in the SMC); we consider optically thin gas components in local thermodynamic 
equilibrium (LTE), and subthermally excited (non-LTE) gas. 

\subsubsection{Optically thin LTE gas models}
\label{co_rot_thin}

If the CO emission is optically thin, the total number of molecules in the upper
state of a transition $N_J$ is related to the observed line flux $F_J$ as 
follows:
\begin{equation} 
N_J= 4\pi d^2 \frac{F_J}{h \nu_J A_J}, 
\label{numbermolecules}
\end{equation}
where $\nu_J$ and $A_J$ are respectively the frequency and Einstein coefficient
for the transition, and $d$ is the distance to the source (we adopt 
$d_{\rm LMC}=50$\,kpc and $d_{\rm SMC}=60$\,kpc throughout this work). These 
population levels follow a Boltzmann distribution:
\begin{equation} 
\frac{N_J}{g_J} = \frac{N_{\rm CO}}{Q} e^{-E_J/T_{\rm rot}},
\label{boltzmann}
\end{equation}
where $N_{\rm CO}$ is the total number of molecules, $Q=kT/hcB$ is the partition
function for linear molecules ($B=192.25$\,m$^{-1}$ for CO), $E_J$ and $g_j$ are
the upper level energy (in K) and degeneracy respectively, and $T_{\rm rot}$ is the 
rotational temperature\footnotemark. \footnotetext{Basic atomic and molecular data are 
from LAMDA (Leiden Atomic and Molecular Database, \citealt{schoier05}).}
Assuming the density is high enough to thermalise all relevant energy 
levels \citep[$n \geq 10^6$\,cm$^{-3}$ for $J$\,=\,4$-$13,][]{yang10}, $T_{\rm rot}$ is 
the kinetic temperature $T$ of the gas for all transitions, and values for $T$ and 
$N_{\rm CO}$ can be simply determined from the slope and intercept of the rotational 
diagram. If the levels are not thermalised then $T_{\rm rot}$\,<\,$T$ \citep[e.g.,][]
{goldsmith99} and the true value of $T$ is underestimated; furthermore the 
calculated partition function $Q$ (a function of $T$) is too small and consequently 
$N_{\rm CO}$ is overestimated. If some transitions are optically thick, non-linear 
effects are introduced in the rotational diagram \citep[e.g.,][]{goldsmith99}. 
Non-LTE and optical depth effects are discussed subsequently.

\begin{figure*}
\begin{center}
\includegraphics[scale=1.04]{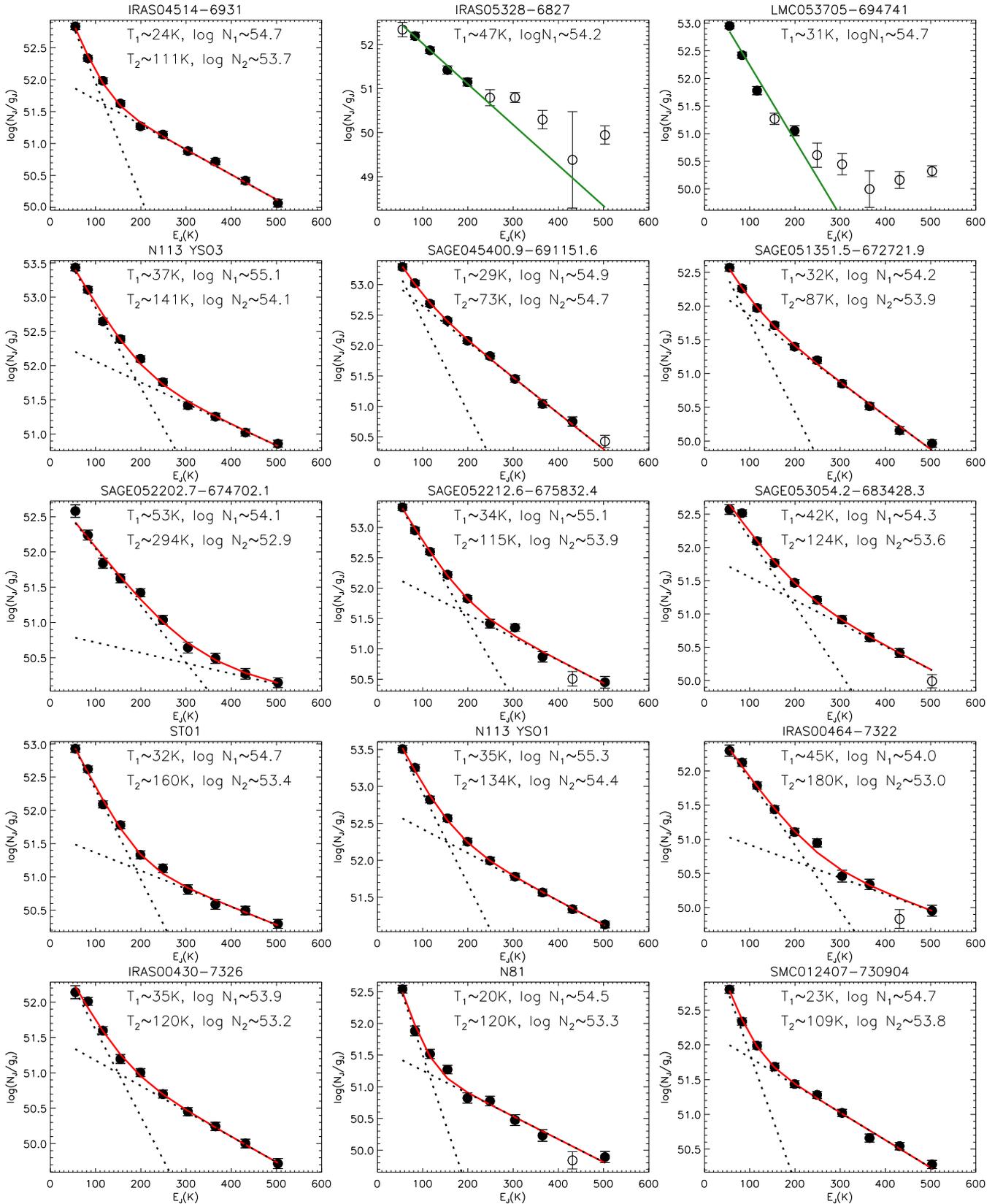}
\end{center}
\caption{CO rotational diagrams for the Magellanic YSO sample (open symbols 
indicate data points with SNR\,<\,5). The transitions measured over the SPIRE range 
are CO\,(4$-$3) to CO\,(13$-12$). The characteristic break in the rotational 
diagrams suggests the presence of two isothermal LTE components with different 
temperatures $T_{i}$ and total number of CO molecules $N_{i}$ (red lines;
the dashed lines show the contributions of the individual components); for two 
sources only a cold component is fitted for $J$\,$\leq$\,9 (green lines).
Approximate best fit values appear in each panel; fitted parameters with uncertainties
are listed in Table\,\ref{lte_pars}.}
\label{co_rotational1}
\end{figure*}

\begin{table*}
\begin{center}
\caption{\normalsize LTE fit parameters to the observed CO rotational diagrams (CO\,(4$-$3) to CO\,(13$-12$)), using two isothermal gas components with different 
temperatures $T_{i}$ and total number of CO molecules $N_{i}$. For IRAS05328$-$6827 and LMC\,053705$-$694741 a single isothermal 
component is fitted ($J$\,$\leq$\,9); for SAGE\,052202.7$-$674702.1 the properties of the warmer component are poorly constrained. Also listed are 
$\chi^2$ and number of degrees of freedom $N_{\rm f}$ for each fit.}
\label{lte_pars}
\begin{tabular}{l|c|c|c|c|c|c}
\hline
Source ID            &$T_1$&$\log N_1$&$T_2$&$\log N_2$&$\chi^2$&$N_{\rm f}$\\
                     &(K)  & (mol)    &(K)  & (mol)    &        &\\
		     \hline
\multicolumn{7}{c}{LMC YSOs}\\
\hline
       IRAS\,04514$-$6931&24.2$\pm$2.5&54.69$\pm$0.10&\al{1}11.2$\pm$6.2	   &53.68$\pm$0.06&      6.9&6\\ 
       IRAS\,05328$-$6827&47.1$\pm$5.0&54.20$\pm$0.20&			   &54.18$\pm$0.35&      1.2&2\\
     LMC\,053705$-$694741&31.9$\pm$1.7&54.70$\pm$0.10&			   &54.28$\pm$0.11&\al{1}7.8&2\\
                N113 YSO3&37.1$\pm$3.1&55.15$\pm$0.05&\al{1}41.2$\pm$17.\ar{7} &54.08$\pm$0.10&\al{1}1.0&6\\
SAGE\,045400.9$-$691151.6&29.1$\pm$8.8&54.91$\pm$0.13&	73.9$\pm$6.2	   &54.66$\pm$0.13&      2.2&5\\ 
SAGE\,051351.5$-$672721.9&32.4$\pm$7.2&54.18$\pm$0.10&	87.6$\pm$5.9	   &53.86$\pm$0.10&      6.4&6\\ 
SAGE\,052202.7$-$674702.1&53.7$\pm$6.6&54.15$\pm$0.06&\al{2}94.0$\pm$240\ar{.8}&52.89$\pm$0.23&      8.7&6\\ 
SAGE\,052212.6$-$675832.4&34.3$\pm$3.2&55.08$\pm$0.06&\al{1}15.1$\pm$18.\ar{7} &53.94$\pm$0.16&      7.4&5\\ 
SAGE\,053054.2$-$683428.3&42.8$\pm$7.0&54.34$\pm$0.06&\al{1}24.9$\pm$35.\ar{4} &53.56$\pm$0.28&      6.7&5\\ 
                   ST\,01&32.2$\pm$2.4&54.72$\pm$0.06&\al{1}60.1$\pm$25.\ar{0} &53.40$\pm$0.11&      6.5&6\\ 
               N113 YSO-1&35.1$\pm$3.5&55.26$\pm$0.06&\al{1}34.0$\pm$12.\ar{2} &54.43$\pm$0.08&      2.9&6\\ 
\hline
\multicolumn{7}{c}{SMC YSOs}\\
\hline
       IRAS\,00464$-$7322&45.4$\pm$5.1&54.04$\pm$0.06&\al{1}80.2$\pm$59.\ar{6} &52.97$\pm$0.21&      6.0&5\\ 
       IRAS\,00430$-$7326&35.3$\pm$5.2&53.95$\pm$0.09&\al{1}20.8$\pm$14.\ar{0} &53.18$\pm$0.12&      5.0&6\\ 
                    N\,81&20.3$\pm$2.8&54.51$\pm$0.15&\al{1}20.3$\pm$12.\ar{8} &53.26$\pm$0.08&      9.5&5\\ 
     SMC\,012407$-$730904&23.5$\pm$3.3&54.65$\pm$0.13&\al{1}09.6$\pm$6.6	   &53.81$\pm$0.06&      7.4&6\\ 
\hline
\end{tabular}
\end{center}
\end{table*}

Fig.\,\ref{co_rotational1} shows the rotational diagrams for the sources in our sample. 
If Eq.\,\ref{boltzmann} holds for a single isothermal CO gas 
component, the data points should lie on a straight line with the slope related 
to the gas temperature $T_{\rm rot}$. Excluding two sources (IRAS\,05328$-$6827 and
LMC053705$-$694741) for which limited data are available, only two other sources 
(SAGE\,04500.9$-$691151.6 and SAGE\,051351.5$-$672721.9) show rotational diagrams that
seem reasonably consistent with a single isothermal gas component. For the 
remainder of the sources the rotational diagrams exhibit a positive curvature as 
often is the case. This implies that the rotational temperature 
increases with $E_J$ and thus it cannot arise from a single isothermal {\it dense}
gas component. The characteristic ``break'' in the rotational diagram is usually 
interpreted as indicating multiple optically-thin LTE CO gas components of 
different temperatures. 

For each source with at least four measurements with SNR\,$\ge$\,5, we fitted 
a model with two CO gas components, with distinct temperatures $T_i$ and total
number of CO molecules $N_i$. We added in quadrature a systematic error of 10\% 
to each transition measurement error \citep{hopwood15}. The fits that minimise 
$\chi^2$ are shown in each diagram (Fig.\,\ref{co_rotational1}). The position of the 
``break'' is not the same for
all sources; typically it occurs for 6\,$\leq$\,$J$\,$\leq$\,9 
\citep[see also][]{yang17}. We note that its position is unrelated to the stitching of 
the SLW and and SSW bands described in Section\,\ref{spire_data}. For IRAS\,05328$-$6827 
and LMC053705$-$694741 a single-temperature model is fitted for $J$\,$\leq$\,9 
(top row Fig.\,\ref{co_rotational1}). Fit parameters and $\chi^2$ are listed in 
Table\,\ref{lte_pars}.

The fitted components have the following properties: $T_1$\,$\sim$\,35\,K (range 
20\,$-$\,53\,K), $N_1$\,$\sim$\,5\,$\times 10^{54}$ CO molecules (range 
0.8\,$-$\,20\,$\times 10^{54}$ molecules), $T_2$\,$\sim$\,132\,K (73\,$-$\,180\,K), 
$N_2$\,$\sim$\,4.8\,$\times 10^{53}$ CO molecules (0.8\,$-$\,50\,$\times 10^{53}$ 
molecules). Typically the colder component contributes an order of magnitude more 
CO than the (slightly) warmer component. The $T_1$ values for individual sources are 
broadly consistent with their SED-derived temperatures (Sect.\,\ref{bb}). The fitted 
rotational temperatures vary little across the sample, even though the luminosities of
the objects $L_{\rm TIR}$ vary by almost two orders of magnitude (Table\,\ref{fir_temp}); 
$T_1$ and $T_2$ values are very similar for LMC and SMC sources. The total 
number of CO molecules is typically $\sim$\,7\,$\times 10^{54}$, ranging from 
1$\,-\,$21\,$\times 10^{54}$ for individual sources. For two sources 
(SAGE\,04500.9$-$691151.6 and SAGE\,051351.5$-$672721.9, Fig.\,\ref{co_rotational1} 
second row) the temperature of the warmer component is also below 100\,K and the two 
components are characterised by a number of molecules of the same order of 
magnitude. In fact a single component would fit the observed data reasonably well 
as noted previously. For one source (SAGE\,052202.7$-$674702.1) the temperature of 
the warmer component is poorly constrained.

\citet{goicoechea12} analysed the CO ladder of a low-mass Class\,0 Galactic YSO 
and found that an LTE component of $T$\,$\sim$\,100\,K fits the observed data for 
$J$\,$\leq$\,14. Close inspection of their rotational diagram suggests that 
significant curvature is seen even over this narrow energy range. For another 
Class\,0 Galactic YSO \citet{yang17} find CO temperatures $\sim$\,43 and 
$\sim$\,197\,K over the SPIRE range. \citet{yildiz13} and \citet{yang18} analysed 
larger samples of low-luminosity Galactic YSOs; they found median temperatures of 
$\sim$\,45\,K and $\sim$\,95\,K, and $\sim$\,43\,K and $\sim$\,138\,K respectively. 
\citet{white10} performed similar analysis on a Galactic molecular cloud core and 
found that two CO components of temperatures $\sim$\,78\,K and $\sim$\,185\,K are 
required. Similarly fits to CO ladders for two HAeBe stars led to estimated 
temperatures of about $\sim$\,30\,K and $\sim$\,100\,K \citep{donaire17}. Focusing on
the \citet{yang18}, \citet{yildiz13} and Magellanic samples, the so-called cold and 
cool components have consistent temperatures $T_{\rm cold}$\,$\sim$\,40\,K and 
$T_{\rm cool}$\,$\sim$\,120\,K respectively. Despite the very different spatial
scales sampled and evolutionary stages of the sources, the derived CO temperatures
for Galactic and Magellanic YSOs over the SPIRE range are remarkably similar. 
Furthermore, these temperatures do not seem to correlate with source properties 
like $L_{\rm TIR}$ or $L_{\rm bol}$ (see also e.g.,
\citealt{manoj13,yildiz13,green13,yang18}). 

The analysis described in this section shows that the observed rotational
diagrams are only properly described by considering multiple components assuming that 
the gas is optically thin and fully thermalised for the transitions in the SPIRE 
range (see Appendix\,\ref{gas_admixture} for an alternative approach using an admixture 
of gas components). In the next section we test the validity of the LTE and optically 
thin assumptions using models by \citet{neufeld12}. 

\subsubsection{Non-LTE gas models}
\label{nonlte}

\begin{figure*}
\begin{center}
\includegraphics[scale=1.04]{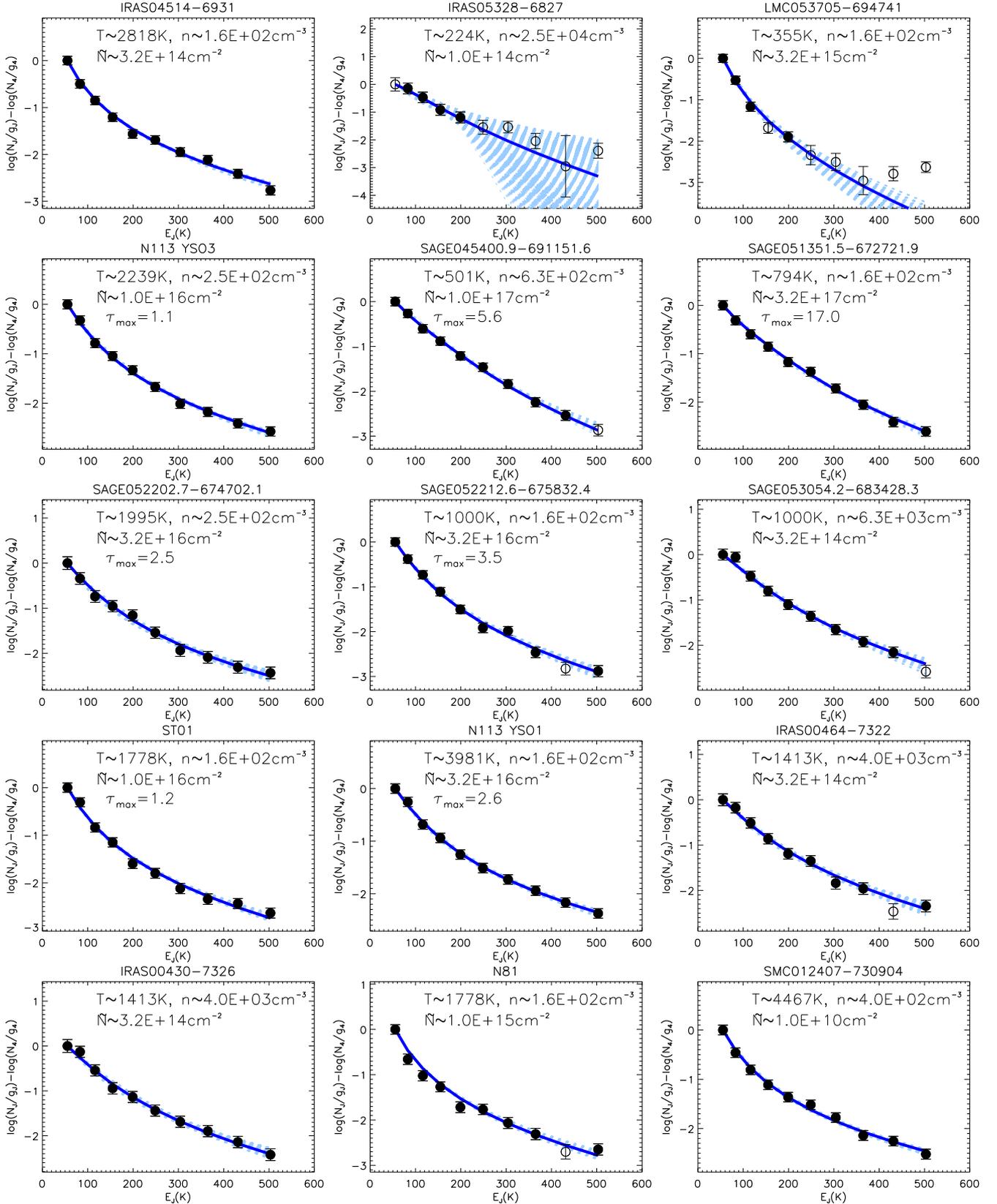}
\end{center}
\caption{CO rotational diagrams for the Magellanic YSO targets (symbols as in
Fig.\,\ref{co_rotational1}), normalised to $N_4/g_4$. The thick blue line
represents the best-fit model from $\chi^2$ minimisation; the shaded area
represents models that fulfil a 1-$\sigma$ confidence interval. The best fit 
values for gas temperature $T$, molecular hydrogen volume density $n$ and 
CO column density parameter $\tilde{N}$ are listed in each panel; fit parameters are 
given in Table\,\ref{nlte_pars}. A value for $\tau_{\rm max}$ is also provided if 
transitions become optically thick for the best fit model (see text for further details).}
\label{co_rotational3}
\end{figure*}

\begin{table*}
\begin{center}
\caption{\normalsize Fit parameters to the observed CO rotational diagrams (CO\,(4$-$3) 
to CO\,(13$-12$)), using the non-LTE model grid by \citet{neufeld12}. The model parameters are gas 
temperature $T$ (in the approximate range 10\,$-$\,5000\,K, 0.05 logarithmic steps), 
molecular hydrogen volume density $n$ (160\,$-$\,$10^{12}$\,cm$^{-3}$, 0.2 logarithmic 
steps) and CO column density parameter ${\tilde N}_{CO}$ 
($10^{10}$\,$-$\,$10^{18}$\,cm$^{-2}$ per km\,s$^{-1}$, 0.5 logarithmic steps). 
The best value is given for each parameter and the two limits provide
the 1-$\sigma$ confidence interval (i.e. range for which $(\chi^2-\chi^2_{\rm min})$\,$\leq$\,3.53).
For IRAS05328$-$6827 the confidence interval includes the full parameter space. 
Values for maximum optical depth $\tau_{\rm max}$ and number of optically thick transitions $N(\tau >1)$,
as well as $\chi^2$ values are also provided.}
\label{nlte_pars}
\begin{tabular}{l|c|c|c|c|c|c|c|c|c|c|c|c}
\hline
Source ID            &$T$&$T_{\rm min}$&$T_{\rm max}$&$n$&$n_{\rm min}$&$n_{\rm max}$&$\tilde{N}$&$\tilde{N}_{\rm min}$&$\tilde{N}_{\rm max}$&$\tau_{\rm max}$&$N(\tau>1)$&$\chi^2$\\
                     &\multicolumn{3}{c}{(K)}&\multicolumn{3}{c}{(m$^{-3}$)}&\multicolumn{3}{c}{cm$^{-2}$ per km\,s$^{-2}$} &&&        \\
		     \hline
\multicolumn{13}{c}{LMC YSOs}\\
\hline
       IRAS\,04514$-$6931&\al{2}818&\al{1}778&3162&      160&160&       630&3.2$\times$\,10$^{14}$&1.0$\times$\,10$^{10}$&1.0$\times$\,10$^{16}$&         & &8.6\\
       IRAS\,05328$-$6827&      224&         &    &\al{5}000&   &          &1.0$\times$\,10$^{14}$&       &       &    & &   \\
     LMC\,053705$-$694741&      355&      178& 501&      160&160& \al{4}000&3.2$\times$\,10$^{15}$&1.0$\times$\,10$^{10}$&3.2$\times$\,10$^{16}$&         & &1.8\\
                N113 YSO3&\al{2}239&\al{1}122&3162&      250&160& \al{2}500&1.0$\times$\,10$^{16}$&1.0$\times$\,10$^{10}$&3.2$\times$\,10$^{16}$&    1.1&1&4.0\\
SAGE\,045400.9$-$691151.6&      501&      355& 891&      630&160&\al{16}000&1.0$\times$\,10$^{17}$&1.0$\times$\,10$^{10}$&3.2$\times$\,10$^{17}$&      5.6&5&2.0\\
SAGE\,051351.5$-$672721.9&      794&      501&1259&      160&160&\al{10}000&3.2$\times$\,10$^{17}$&1.0$\times$\,10$^{10}$&3.2$\times$\,10$^{17}$&\al{1}7.0&6&3.3\\
SAGE\,052202.7$-$674702.1&\al{1}995&\al{1}000&5012&      250&160& \al{4}000&3.2$\times$\,10$^{16}$&1.0$\times$\,10$^{10}$&1.0$\times$\,10$^{17}$&      2.5&3&3.4\\
SAGE\,052212.6$-$675832.4&\al{1}000&      562&1413&      160&160& \al{4}000&3.2$\times$\,10$^{16}$&1.0$\times$\,10$^{10}$&1.0$\times$\,10$^{17}$&      3.5&2&5.4\\
SAGE\,053054.2$-$683428.3&\al{1}000&      447&1995&\al{6}300&160&\al{16}000&3.2$\times$\,10$^{14}$&1.0$\times$\,10$^{10}$&3.2$\times$\,10$^{17}$&         & &4.0\\
                   ST\,01&\al{1}778&\al{1}259&2818&      160&160& \al{1}000&1.0$\times$\,10$^{16}$&1.0$\times$\,10$^{10}$&1.0$\times$\,10$^{16}$&      1.2&1&8.6\\
               N113 YSO-1&\al{3}981&\al{1}778&5012&      160&160& \al{2}500&3.2$\times$\,10$^{16}$&1.0$\times$\,10$^{10}$&3.2$\times$\,10$^{16}$&       2.6&3&1.9\\
\hline
\multicolumn{13}{c}{SMC YSOs}\\
\hline
       IRAS\,00464$-$7322&\al{1}413&      794&5012&\al{4}000&160& \al{6}300&3.2$\times$\,10$^{14}$&1.0$\times$\,10$^{10}$&3.2$\times$\,10$^{17}$&          & & 3.8\\
       IRAS\,00430$-$7326&\al{1}413&      891&5012&\al{4}000&160& \al{6}300&3.2$\times$\,10$^{14}$&1.0$\times$\,10$^{10}$&3.2$\times$\,10$^{17}$&          & & 2.5\\
                    N\,81&\al{1}778&\al{1}413&2512&      160&160&       400&1.0$\times$\,10$^{15}$&1.0$\times$\,10$^{10}$&3.2$\times$\,10$^{15}$&          & &12.5\\
     SMC\,012407$-$730904&\al{4}467&\al{2}239&5012&      400&160& \al{1}000&1.0$\times$\,10$^{10}$&1.0$\times$\,10$^{10}$&1.0$\times$\,10$^{16}$&          & & 5.1\\
\hline
\end{tabular}
\end{center}
\end{table*}

\citet{neufeld12} showed that CO rotational diagrams can exhibit curvature,
i.e. the rotational diagram changes monotonically with the upper level energy $E_J$
(mathematically $T_{\rm rot} \equiv -\left(k\,d\ln (N_J/g_J)/dE_J\right)^{-1}$), for a 
single isothermal gas component if the gas is not thermalised. If the gas is
sub-thermal, a positive curvature arises in the lower density regime, while for
high but non-thermal densities a negative curvature results. In other words,
a ``break'' in the rotational diagram does not necessarily imply multiple gas
components. 

\citet{neufeld12} solved the equations of statistical equilibrium for CO gas,
assuming a uniform density and temperature. Radiative transfer was treated using 
the escape probability method and a large velocity gradient in a single direction
was assumed \citep[for more details refer to][and references therein]{neufeld12}. 
A pre-computed grid of solutions to the equations of statistical equilibrium for 
CO (David Neufeld, private communication) was fitted to the observed rotational 
diagrams, assuming a single uniform CO component.\footnotemark 
\footnotetext{\citet{neufeld12} discussed CO transitions in the PACS range 
($J$\,$\ge$\,14), but the grid of models includes transitions in the SPIRE range as 
well ($J$\,<\,14).} The grid parameters are gas temperature $T$ (in the approximate range 
10\,$-$\,5000\,K, 0.05 logarithmic steps), molecular hydrogen volume density $n$ 
(160\,$-$\,$10^{12}$\,cm$^{-3}$, 0.2 logarithmic steps) and CO column density parameter 
${\tilde N}_{CO}$ ($10^{10}$\,$-$\,$10^{18}$\,cm$^{-2}$ per km\,s$^{-1}$, 0.5 logarithmic
steps; see \citealt{neufeld12} for a full description). 

The rotational diagrams shown in Fig.\,\ref{co_rotational3} are normalised to
$N_4/g_4$. For each object the best-fit model is shown in blue, and the dashed 
area indicates the range of models that correspond to a 1-$\sigma$ confidence 
interval for three free parameters ($\chi^2-\chi^2_{\rm min}$\,$\le$\,3.53). 
{Fit parameters are given in Table\,\ref{nlte_pars}; the maximum and minimum limits 
provide the 1-$\sigma$ confidence interval (note that often the best-fit value is at a 
limit of the grid). Values for $\chi^2$ and $\tau_{\rm max}$ (defined as the maximum 
value for the optical depth) are also provided for the best fit model (the latter only 
if transitions become optically thick). 

Even though the parameters are not well constrained, a few facts are clear from 
the fits in Fig.\,\ref{co_rotational3} (we discuss only the thirteen objects 
with more than four transitions detected with good SNR). With the exception of 
two objects (SAGE\,04500.9$-$691151.6 and SAGE\,051351.5$-$672721.9) the best fit 
temperatures are high ($\sim$\,1000\,$-$\,4500\,K) and the 
molecular hydrogen densities are rather low ($\sim$\,160\,$-$\,6300\,cm$^{-3}$). As an 
independent check, we also compared our rotational diagrams to the grid of 
Radex models \citep{vandertak07} used by \citet{lee16}; similarly the 
parameters are poorly constrained but the $\chi^2$ fitting also favours high 
temperature, low density solutions. In fact, no model with density 
$n$\,$\gsim$\,1.6\,$\times 10^4$\,cm$^{-3}$ is able to fit the data; at such low 
densities none of the energy levels are thermalised even at high temperature. 
Consequently, non-LTE models clearly favour sub-thermal excitation conditions. 

The column density parameter is not constrained at all, but for seven out of 
thirteen sources the best solution sees one or more lower-$J$ transitions become 
optically thick; even in such cases there are optically thin models with 
slightly higher densities (but still $n$\,$\lsim$\,1.6\,$\times 10^4$\,cm$^{-3}$) 
that provide acceptable fits. We note that no $^{13}$CO transitions were detected, and 
the upper limits do not place meaningful limits on the gas optical depth. 

Purely in terms of $\chi^2$, the non-LTE fits are better than the 
LTE optically thin two-component fits (with fewer free parameters).

\subsection{Properties of CO emitting gas: physical conditions and origin}
\label{co_origin}

As already pointed out, $L_{\rm{CO}}$ correlates strongly with a measure of YSO 
luminosity ($L_{\rm bol}$ or $L_{\rm TIR}$), for any YSO sample. There seems to be no 
correlation between parameters derived from the LTE analysis of the CO rotational 
diagrams and YSO  properties, even taking into account uncertainties. The derived
rotational temperatures are remarkably constant (variations are at most a factor two), 
while the YSO luminosity ($L_{\rm bol}$ or $L_{\rm TIR}$) changes by as much as five 
orders of magnitude across all samples. Therefore this suggests that while the amount of 
excited CO gas is related to the source luminosity the conditions of the excited gas are 
not.

However, the analysis of the CO rotational diagrams observed for Magellanic YSOs does not
unambiguously pinpoint the gas conditions. A single high-temperature isothermal 
($T$\,$\gsim$\,1000\,K) gas component can fit the observations, but the gas 
densities would be rather low ($n$\,$\lsim$\,$10^4$\,cm$^{-2}$), i.e. the hot gas 
would be very clearly subthermal. On the other hand, cold ($T$\,$\lsim$\,200\,K) LTE
gas provides a good match for the observed CO intensities and ratios, but 
multiple components are required. \citet{manoj13} arrive at very similar
conclusions, for their sample of low-luminosity 
($L_{\rm bol}$\,$\lsim$\,200\,L$_{\odot}$) Galactic YSOs. They fitted CO rotational 
diagrams in the PACS range ($J$\,$\geq$\,14), meaning that the temperatures derived are 
higher. Nevertheless, the ambiguity between the two regimes mentioned above is also seen.
 \citet{manoj13} argue that it is more difficult to conceive the existence of
multiple LTE components with temperatures essentially insensitive to YSO 
luminosity. On the other hand, in the low-density regime, a large range in gas 
temperatures results in a narrow range in rotational temperatures: for a high 
luminosity YSO the CO gas may be hotter but the resulting CO emission is not 
significantly enhanced. This can also be seen in Fig.\,\ref{co_rotational3}, where
temperatures between 1000 and 5000\,K result in models that are indistinguishable.
Even though \citet{manoj13} favour the high temperature sub-thermal solutions, the
fitted rotational diagrams alone cannot discriminate between the two sets of 
emitting gas conditions.

\begin{figure*}
\begin{center}
\includegraphics[scale=0.58]{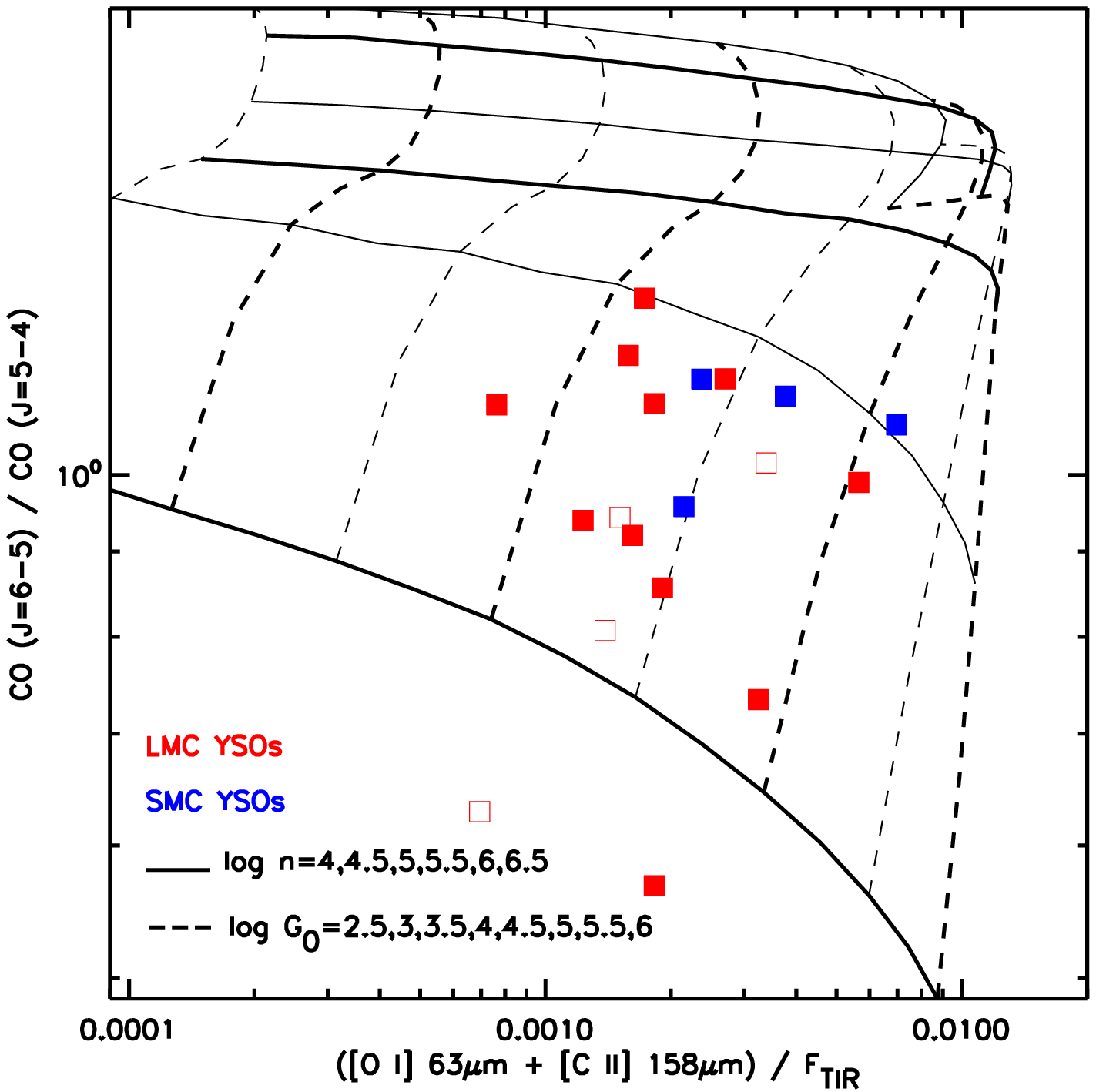}
\includegraphics[scale=0.58]{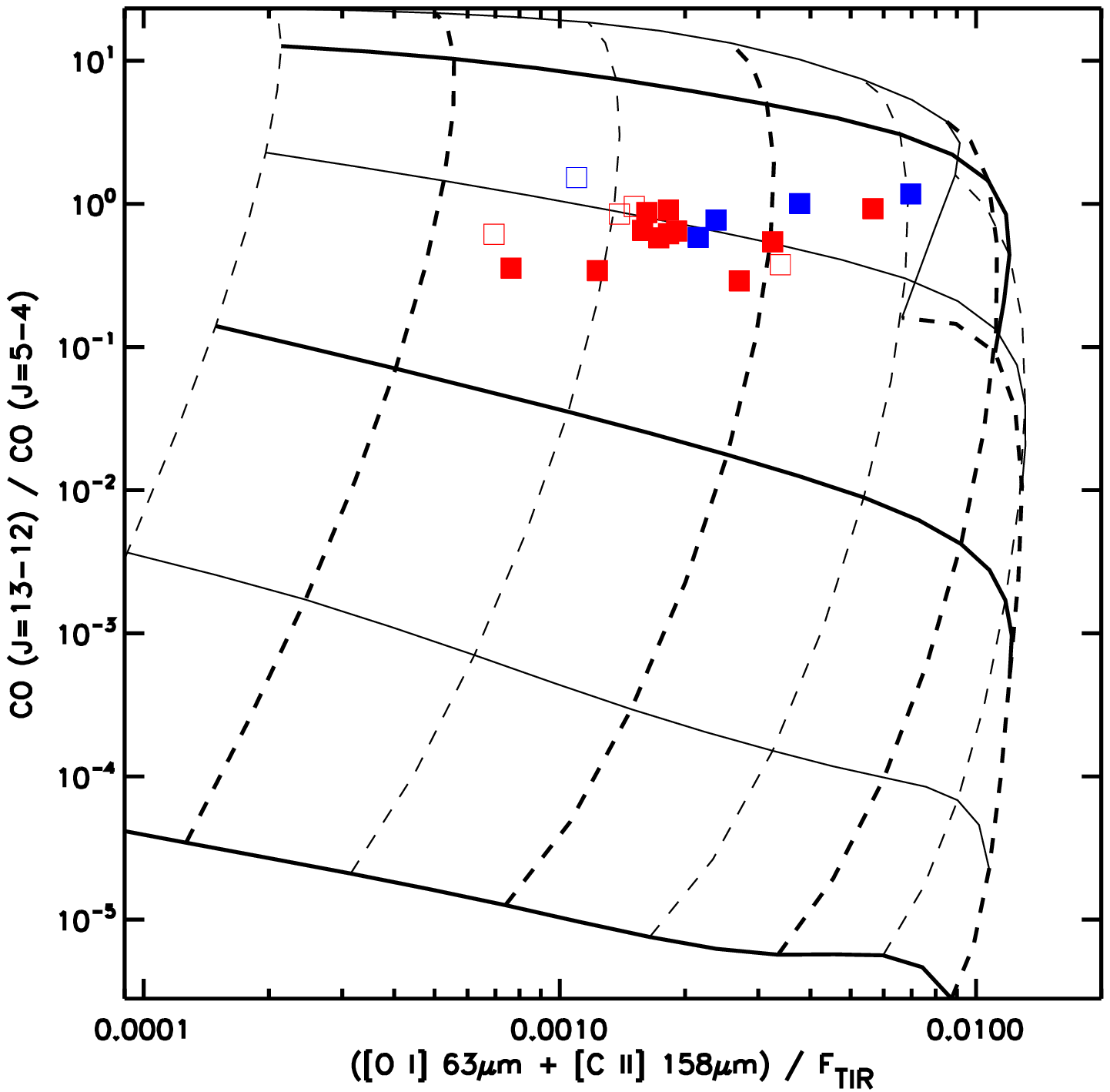}
\end{center}
\caption{Observed atomic and CO emission compared to model predictions from the PDR 
Toolbox \citep{kaufman99,kaufman06}. Cloud density $n$ increases towards the top of 
the diagrams, and strength of the FUV radiation field $G_0$ increases towards the 
left. Symbols have the same meaning as in Fig.\,\ref{totalco}. The grid of PDR models 
represented in this figure covers a different range in parameter values (i.e. higher 
densities) than the grid in Fig.\,\ref{pdr_diagnostic}. Also note that 
$n$\,$\gsim$\,$10^4$\,cm$^{-3}$ is required when considering the ratio
CO\,(6$-$5)/CO\,(5$-$4), while $n$\,$\gsim$\,$10^5$\,cm$^{-3}$ is appropriate for 
CO\,(13$-$12)/CO\,(5$-$4).}
\label{pdr_co}
\end{figure*}

\begin{figure*}
\begin{center}
\includegraphics[scale=0.58]{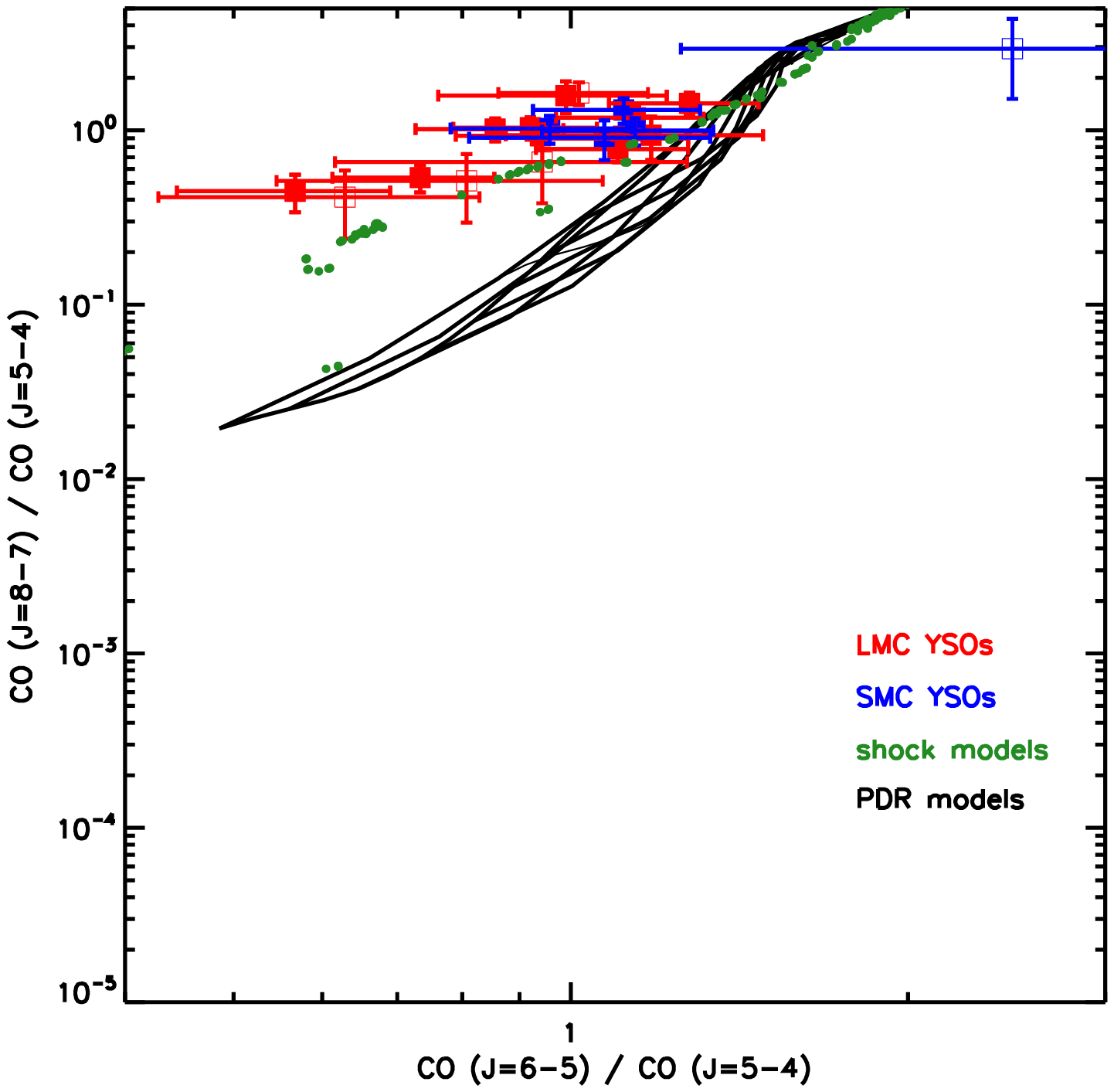}
\includegraphics[scale=0.58]{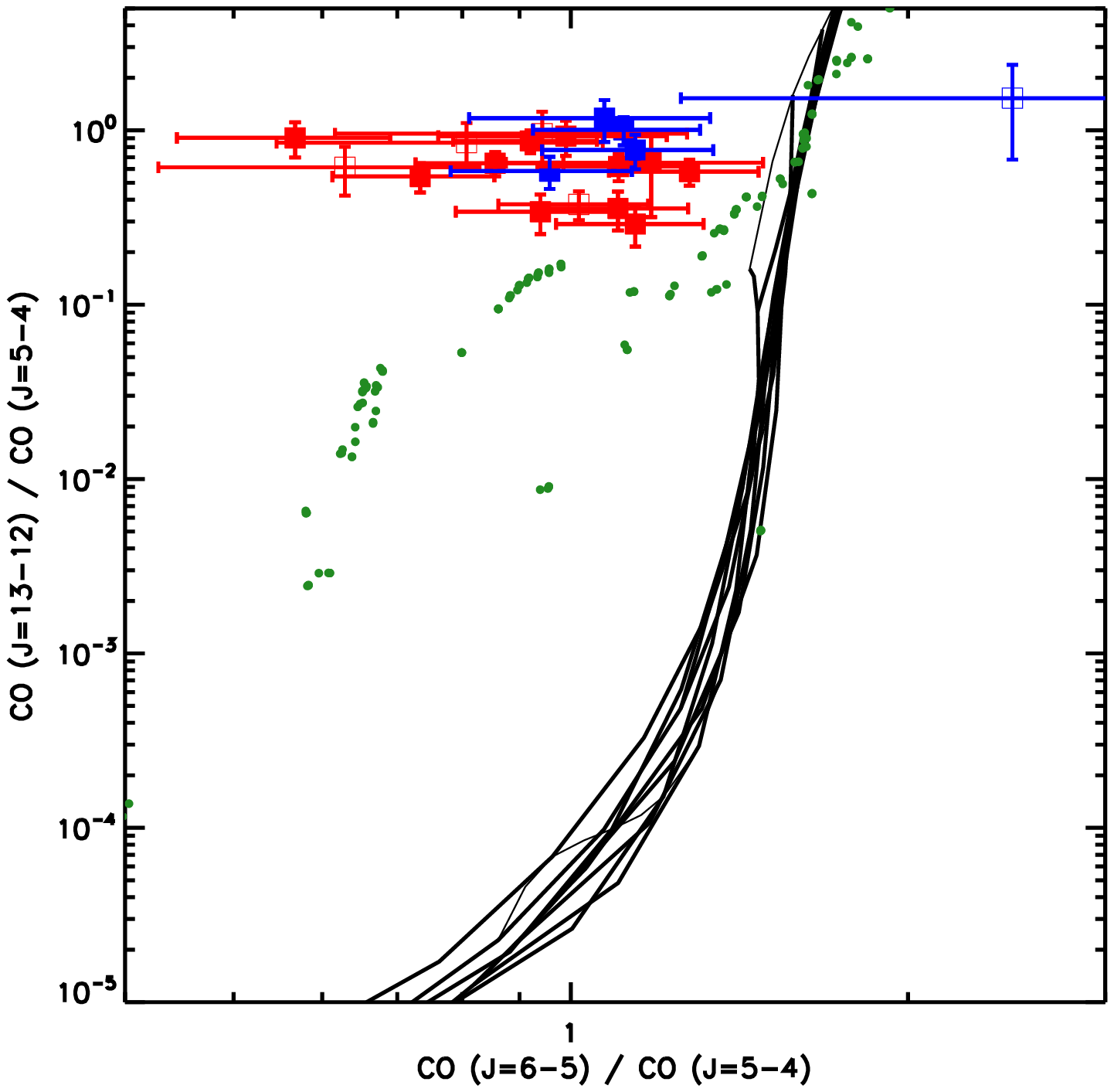}
\end{center}
\caption{Observed CO emission ratios compared to model predictions from the PDR 
Toolbox \citep{kaufman99,kaufman06} and shock models \citep{flower15}. PDR 
models (full lines) predict a very narrow range in CO line ratios, and a much 
steeper decrease (by several orders of magnitude) in fluxes for higher energy 
transitions, that is not observed. Green symbols show the ratios predicted by a selection 
of shock models \citep[see text and also][]{lee16}. Shock heating is a more likely 
excitation mechanism for the observed CO emission.}
\label{pdr_co_only}
\end{figure*}

The origin of the CO excitation can either be UV heating in PDRs or shock heating 
in molecular outflows. The constancy of the fitted rotational temperatures (over 
five orders of magnitude in luminosity, as shown in this work and Galactic samples 
-- \citealt{manoj13,green16,yang18}), argues against UV photons as the main heating 
mechanism. To further explore this issue, we have compared CO emission line ratios 
to those predicted by the PDR models \citep{kaufman99,kaufman06} used in 
Sect.\,\ref{pacs_emission} to analyse the \oi\ and \cii\ emission. 
Figure\,\ref{pdr_co} shows that the PDR conditions responsible for the atomic line 
emission (namely $n$\,$\lsim$\,$10^4$\,cm$^{-3}$, Fig.\,\ref{pdr_diagnostic}) are not 
consistent with the observed CO emission ratios for our sample: for the same 
photoelectric heating efficiency the required densities are higher 
($n$\,$\gsim$\,$10^4$\,cm$^{-3}$). Therefore the same PDR gas is unlikely to be 
responsible for the atomic and CO emission \citep[see also][for two Galactic 
examples for which a similar conclusion was reached]{stock15}. Note that for 70\%
of the sources represented in Fig.\,\ref{pdr_co}, CO emission is optically thin
for the transitions plotted (see Sect.\,\ref{nonlte}). The diagram also suggests 
that the model density range that best reproduces the observed ratios progressively 
increases when higher-J CO transitions are considered:  
$n$\,$\gsim$\,$10^4$\,cm$^{-3}$ is required when considering the ratio
CO\,(6$-$5)/CO\,(5$-$4), while $n$\,$\gsim$\,$10^5$\,cm$^{-3}$ is appropriate for 
CO\,(13$-$12)/CO\,(5$-$4). This indicates that a single PDR gas component cannot give 
rise to the observed CO emission across this energy range.
 
In Fig.\,\ref{pdr_co_only} we further explore the behaviour of pairs of CO ratios.
For CO gas excited in PDR conditions the predicted ratios occupy an extremely narrow
range of values; for a given CO\,(6$-$5)/CO\,(5$-$4) ratio the PDR models predict a 
much steeper decrease (by five orders of magnitude) in CO\,(13$-$12) fluxes than 
what is actually observed (right panel). This is still the case even when only the CO 
cold component contribution (CO\,(8$-$7)/CO\,(5$-$4) ratio, left panel) is considered. 
We conclude that the CO emission observed towards Magellanic YSOs is very unlikely to be 
excited in PDRs, even when multiple gas components are considered. This is consistent 
with the analysis by \citet{lee16} who found that PDR heating cannot be responsible for 
the excitation of CO gas in the Magellanic SFR N\,159W \citep[see also][for a similar
analysis in the 30\,Doradus region]{lee19}. 

Figure\,\ref{pdr_co_only} also shows the grid of shock models used by
\citet[][computed using the Paris$-$Durham shock code, \citealt{flower15}]{lee16}
for a range of representative conditions\footnotemark\footnotetext{For these input
parameters the shock grid contains only C-type shocks \citep{lee16}.}: pre-shock 
density $10^4$\,$-$\,$10^6$\,cm$^{-3}$, unity magnetic field strength parameter (this 
parameter reflects the strength of magnetic field transverse to the shock 
propagation direction), and shock velocity 4\,$-$\,20\,km\,s$^{-1}$. This range of 
velocities is consistent with CO outflow velocities observed in the LMC 
\citep{fukui15,shimonish16}. These shock models are for solar metallicity and do 
not include grain$-$grain interactions; none of these effects significantly impact 
CO line ratios over the J-range discussed here (see Appendix in \citealt{anderl13} 
and \citealt{lee16}). Figure\,\ref{pdr_co_only} (left) suggests that shock 
models are reasonably consistent with the observed CO emission ratios for Magellanic
YSOs. However these models are still not able to fully reproduce the observed 
ratios over the whole J-range, i.e when higher-J transitions are included 
(Fig.\,\ref{pdr_co_only} right). This suggests that if the CO emission originates 
from shocks, multiple gas components are required. 

\citet{meijerink13} proposed that shock-heated gas both in SFRs and active 
galactic nuclei (AGN) can yield much larger $L_{\rm CO}/L_{\rm TIR}$ ratios than PDRs and 
XDRs (X-ray dominated regions). This is due to the fact that while PDRs and XDRs 
heat both the gas and the dust, shocks in typical conditions heat the dust less
effectively, leading to larger ratios. Their models predict that a ratio 
$L_{\rm CO}/L_{\rm TIR}$\,$\gsim$\,0.01\% is the threshold diagnostic for shock heating. 
Most Galactic and all Magellanic objects in Fig.\,\ref{totalco} exhibit ratios above 
this limit, suggesting that in massive YSO environments, CO is heated by shocks. Despite 
the very different spatial scales probed, both on SFR-wide scales and on the scales of 
individual massive YSOs shock-heated gas seems to be the origin of most CO emission.

High spectral resolution \herschel-HIFI observations of high-J CO and \hho\ 
transitions in low-mass Galactic YSOs support the presence of at least three 
kinematically distinct gas components of different physical conditions within 
outflows \citep[][see their Fig.\,10, and references therein]{kristensen17}: the 
cool component ($\lsim$\,100\,K, observed with SPIRE) originates from the entrained 
(i.e. swept-up) envelope gas, the warm component ($\sim$\,300\,K, observed with PACS)
comes from cavity shocks (gas directly interacting with the shocks in the walls of 
the outflow cavity) or from the disc wind, while the hot component ($\gsim$\,600\,K, 
observed with HIFI) emerges from distinct ``spot shocks'' near the base of the outflow 
or jet. The temperature of the ubiquitous warm $\sim$\,300\,K component is the result from
the change in the dominant cooling molecules from H$_2$ to CO, and the hot component 
arises in gas prior to the onset of H$_2$ formation. Both hotter components 
contribute to the excitation of \hho\ emission \citep[e.g.,][]{mottram14}. As 
described in Sect.\,\ref{columinosity} the fact that the correlation between 
YSO luminosity and total CO luminosity extends from low-luminosity YSOs to 
massive YSOs supports a common origin for the observed CO emission across the mass 
range.

In summary, our analysis of the CO emission for massive YSOs in the Magellanic
Clouds supports an origin in shocked gas. Multiple shock components are required to 
explain the observations (akin to what is observed for low-luminosity Galactic YSOs),
suggesting the observed SPIRE emission is more likely to arise from LTE 
low-temperature gas (typically $T_{\rm cold} \sim 35$\,K and $T_{\rm cool} \sim 132$\,K
for the Magellanic YSOs), rather than a single subthermal hot gas component.

\subsection{Other lines in the SPIRE range}
\label{otherspire}

In this section we discuss atomic transitions detected in the SPIRE range. As 
already mentioned \nii\ emission traces the ionised gas that also contributes to 
the \cii\ emission, while \ci\ emission traces regions where the gas is shielded 
enough so that carbon is found in a mixture of atomic C and CO. 

\subsubsection{\nii\ emission}
\label{cii_nii}

Out of the nineteen sources observed with SPIRE, nine LMC and two SMC sources show 
\nii\ emission at 205\,\micron. We use the \nii\ emission to assess 
how much of the \cii\ emission originates from the ionised gas. Following the method
proposed by \citet{oberst11}, theoretical line intensity \cii/\nii\ ratios
expected from ionised gas are used to estimate the ionised gas contribution to the
total \cii\ emission. We calculate the fine structure level abundances of N$^+$ and 
C$^+$ as a function of electron density using theoretical collisional rates; a 
correction factor due to the ionic abundance fraction [(C$^+$/C)/(N$^+$/N)]\,=\,1.26 
is applied \citep[determined using the MAPPINGS III photoionisation grid,][]
{sunderland13}. Finally, we adopt a C/N ratio of 5.9 appropriate for \hii\ regions 
in the Galaxy and the Magellanic Clouds \citep[][and references therein]
{carlos-reyes15}. For the typical low electron density 
$n_{\rm e}$\,$\sim$\,100\,cm$^{-3}$ that produces \cii\ emission 
\citep{cormier15,chevance16}, the theoretical ratios are 
(\cii/\nii\,122\,\micron)$_{\rm T}$\,$\sim$\,2.4 and 
(\cii/\nii\,205\,\micron)$_{\rm T}$\,$\sim$\,7.7\footnotemark.
\footnotetext{Such ratios depend only 
very slightly on the ionised gas temperature in the range 5000\,$-$\,15000\,K.}
For the Magellanic sample, the observed ratios are 
(\cii/\nii\,205\,\micron)$_{\rm O}$\,>\,20, calculated using the fluxes in 
Tables\,\ref{linefluxes_pacs} and \ref{linefluxes_spire}. The \cii\ PDR emission 
fraction is $1-($\cii/\nii$)_{\rm T}/($\cii/\nii$)_{\rm O}$; between  
61\% and 96\% (median $\sim$\,82\%) of the \cii\ emission originates from the PDR, 
rather than from low density ionised gas (individual fractions are listed in 
Table\,\ref{linefluxes_pacs}). For resolved star forming regions in the LMC and SMC,
\citet{cormier15} estimated this contribution to be at least 75\% (see also 
\citealt{lebouteiller12}; \citealt{chevance16}), while \citet{jameson18} estimated a 
contribution of $\gsim$\,95\% for five SMC SFRs. Given the uncertainties, these 
estimates are reasonably consistent. Theoretical and observed \cii/\nii\,122\,\micron\,
ratios are used to calculate the PDR gas fraction for the Galactic sample 
(Appendix\,\ref{iso_appendix}); a median of $\sim$\,60\% of the \cii\ emission 
originates from the PDR gas component, broadly consistent with other estimates
\citep{oberst11,bernard-salas12}. Despite the large scatter for both Galactic and
Magellanic samples, these fractions are also consistent with an increased contribution
from the ionised gas component for higher metallicity environments 
\citep[e.g.,][and references therein]{cormier19}.

\subsubsection{\ci\ emission}
\label{cii_ci}

For five LMC objects and one SMC object we have detected \ci\ emission at 370 and
609\,\micron; a further eight sources exhibit emission at 370\,\micron\ only. The 
ratio \ci$_{370}$/\ci$_{609}$ is sensitive to density and it can be used to infer 
the excitation temperature $T_{\rm ex}$ for optically thin LTE gas. Under such 
conditions the population levels follow a Boltzmann distribution
\citep[see also][]{spinoglio12}: 
\begin{equation} 
T_{\rm ex}=\Delta E\,\left[\ln \frac{N_1\,g_2}{N_2\,g_1}\right]^{-1},
\end{equation}
\noindent where $\Delta E$ is the energy difference between the two levels (in K). 
Other quantities have their usual meanings; see Eq.\,\ref{numbermolecules} and 
\ref{boltzmann} for their definition and the relation to observed line 
fluxes\addtocounter{footnote}{-4}\footnotemark.\addtocounter{footnote}{+3}
For the Magellanic sample the \ci\ line ratio varies in the range 1.2\,$-$\,1.8,
suggesting the emission is optically thin \citep[e.g.,][]{kramer04}. The inferred LTE 
excitation temperatures $T_{\rm ex}$ are in the narrow range 18\,$-$\,24\,K. If the 
C-emitting gas were not thermalised \citep[$n\,\lsim\,10^{4}$\,cm$^{-3}$, e.g.,][]
{tielens85}, the observed ratios would favour higher temperature gas 
($T_{\rm ex}$\,$\gsim$\,30\,K, see Fig.\,9 in \citealt{pereira13}).

The two sources with the lowest LTE temperature ($\sim$\,18\,K) are 
SAGE\,04500.9$-$691151.6 and SAGE\,051351.5$-$672721.9; referring back to 
Sect.\,\ref{co_rot}, these are the two sources for which the CO emission was closest 
to that predicted by a single isothermal gas component, and for which the 
two-component LTE fit predicted the lowest temperatures for the CO cool component 
(<\,100\,K).

We compared the ratios \ci$_{370}$/\ci$_{609}$ and \ci$_{370}$/CO (CO measured over 
the SPIRE range) with the ratios predicted by shock models \citep{flower15} -- C-type
shock: pre-shock density $n$\,=\,$10^3$\,$-$\,$10^6$\,cm$^{-3}$, shock velocity 
$v$\,=\,10\,$-$\,40\,km\,s$^{-1}$, magnetic field strength parameter $b$\,=\,1; 
J-type: $n$\,=\,$10^3$\,$-$\,$10^5$\,cm$^{-3}$, $v$\,=\,10\,$-$\,35\,km\,s$^{-1}$, 
$b$\,=\,0.1. For the \ci$_{370}$/\ci$_{609}$ ratio the predictions are $\gsim 2$ and 
$\gsim 3$ respectively for C- and J-type shocks. The \ci$_{370}$/CO ratios are 
predicted to be $\lsim 0.02$ and $\lsim 0.036$, respectively. As already mentioned, 
for the Magellanic sample \ci$_{370}$/\ci$_{609}$\,=\,1.2$-$1.8 (for six sources), 
and \ci$_{370}$/CO\,=\,0.035$-$0.12 (for fourteen sources). The observed ratios are 
therefore not consistent with the \ci\ emission originating in the same shocked gas 
that also is likely to produce the observed CO emission \citep[see also][]{lee16}.

We also investigated whether the observed \ci\ emission could originate from the PDR
gas responsible for the \oi\ and \cii\ emission, by looking at the 
\ci$_{370}$/\ci$_{609}$ ratio in conjunction with both the \oi/\cii\ and the 
(\oi+\cii)/$F_{\rm TIR}$ ratios. We find that the \ci\ emission is consistent with a low
$G_0/n$ PDR regime, even if suggesting slightly lower $n$ and $G_0$ conditions than 
those derived from Fig.\,\ref{pdr_diagnostic}. As described in \citet{rollig07} the 
predicted intensities for \ci\ are very model dependent, affecting the \ci\ line 
ratio and the comparison with other atomic lines. Furthermore, \ci\ emission may have
a smaller filling factor and may trace different layers of the PDR compared to the
\cii\ emission. Nevertheless, our analysis suggests that the \ci\ emission in these 
sources originates from PDR gas.

\section{Discussion} 

\subsection{Line flux correlations}
\label{linecorrelations}

As already discussed in Sect.\,\ref{pacs_emission} there is a strong correlation 
between the \oi\ and \cii\ emission, supporting their common origin in PDR gas.
There is no correlation between total CO emission and these two atomic lines,
consistent with a different origin for CO (likely shocked gas). We now discuss the 
relationships between the fluxes for other species in this study.

There is no correlation between \hho\ emission at 179\,\micron\ and \oi\ or \cii\
emission, supporting an origin in distinct gas components 
\citep[see also][]{karska18}. Since \hho\ emission is thought to originate from the
warm and hot components ($T \gsim 300$\,K) of the outflow system in young stars 
(see discussion in Sect.\,\ref{co_origin}) and \oi\ or \cii\ emission originates 
from the PDR, this is to be expected. The CO emission measured over the SPIRE range 
originates from the distinct cool and cold component in the outflow system 
($T_{\rm cool}$\,$\sim$\,120\,K and $T_{\rm cold}$\,$\sim$\,40\,K respectively);
accordingly no correlation is found between \hho\ and total CO emission. 

OH emission is also thought to originate within the outflow 
\citep[e.g.,][]{wampfler13}, even though it has been difficult to consistently 
reconcile observed molecular emission with shock models \citep[see discussion in][]
{karska18}. We do not find any correlation between OH and \hho\ emission; there 
is however a larger spread in the observed \hho\ fluxes, compared to \oh\ fluxes. 

Most of the known \hho\ maser sources exhibit \hho\ emission in the PACS range; \oh\ 
emission is detected towards one of two \oh\ maser sources in the LMC. 

\subsection{Far-IR line cooling}
\label{coolingbudget}

In this section we discuss how the different atomic and molecular species contribute
to the total line emission, an essential component of the cooling budget of the
YSO environment.

\subsubsection{Total emission line luminosities}
\label{totallines}

For \oi, \hho\ and OH we only have observations for a single line or doublet, 
and for CO we only observe over the SPIRE range, therefore we use correction factors
to convert line fluxes to total luminosities; these are estimated from a variety of
sources from the literature and described in previous sections. The corrections 
(i.e. multiplication factors) applied to the observed luminosities to obtain total 
luminosities are as follows: 1.09 for \oi\ at 63\,\micron\ 
\citep[adopted from][]{karska18}, 1.92 for CO observed in the SPIRE range 
\citep[computed using the large dataset from][]{yang18}, 12.5 for \hho\ observed at 
179\,\micron, 4.17 for OH observed at 79\,\micron\ and finally 5.26 for OH observed 
at 84.4\,\micron\ only \citep[estimated using data from][]
{karska13,wampfler13,karska18}; no correction is needed to obtain \cii\ 
luminosities. We restrict our discussion of the gas cooling budget to the thirteen 
sources (nine LMC and four SMC) with good CO measurements across the SPIRE range. 
All these sources also have \hho\ and OH measurements or 3-$\sigma$ upper limits. 
The total line luminosity $L_{\rm LIR}$ correlates moderately with the total IR 
luminosity $L_{\rm TIR}$ (Spearman's parameters $\rho$\,$\sim$\,0.7, $p$\,$\sim$\,0.01);
their ratio is $L_{\rm LIR}/L_{\rm TIR}$\,$\sim$\,0.4\% (range 0.2\,$-$\,0.9\%).
Table\,\ref{pies} lists line luminosities for each species as fractions of 
$L_{\rm LIR}$, as well as $L_{\rm LIR}/L_{\rm TIR}$.

\begin{table}
\begin{center}
\caption{\normalsize Line luminosities as fractions of the total IR line luminosity
$L_{\rm LIR}$ for \oi, \cii, CO, and \hho\ and \oh\ when detected. The last column gives
the ratio of total IR line luminosity to total IR luminosity $L_{\rm
LIR}$/$L_{\rm TIR}$.}
\label{pies}
\begin{tabular}{l|@{\hspace{-1mm}}c|@{\hspace{-1mm}}c|@{\hspace{-1mm}}c|@{\hspace{-1mm}}c|@{\hspace{-1mm}}c|@{\hspace{-1mm}}c}
\hline
Source ID               &$L_{\rm [OI]}$&$L_{\rm [CII]}$&$L_{\rm CO}$&$L_{\rm H_2O}$&$L_{\rm OH}$&$L_{\rm LIR}$\\
                        &($L_{\rm LIR}$)&($L_{\rm LIR}$)&($L_{\rm LIR}$)  &($L_{\rm LIR}$)     &($L_{\rm LIR}$)  & ($L_{\rm TIR}$)       \\
\hline
\multicolumn{7}{c}{LMC}\\
\hline
IRAS04514$-$6931        &0.43 &0.22 &0.26 &0.06  &0.03         &0.003\\
N\,113 YSO3             &0.47 &0.26 &0.24 &0.02  &0.01         &0.005\\
SAGE045400.9$-$691151.6 &0.38 &0.39 &0.19 &0.04  &0.00\rlap{7} &0.006\\
SAGE051351.5$-$672721.9 &0.42 &0.39 &0.15 &      &             &0.003\\
SAGE052202.7$-$674702.1 &0.37 &0.36 &0.17 &      &             &0.009\\
SAGE052212.6$-$675832.4 &0.41 &0.33 &0.24 &      &             &0.002\\
SAGE053054.2$-$683428.3 &0.25 &0.33 &0.31 &      &             &0.003\\
ST01                    &0.30 &0.32 &0.25 &      &             &0.007\\
N\,113 YSO1             &0.36 &0.25 &0.36 &0.03  &0.00\rlap{4} &0.005\\
\hline
\multicolumn{7}{c}{SMC}\\
\hline
IRAS00464$-$7322        &0.26 &0.32 &0.28 &      &             &0.009\\
IRAS00430$-$7326        &0.51 &0.30 &0.09 &0.05  &             &0.003\\
N\,81                   &0.75 &0.10 &0.05 &      &             &0.009\\
SMC012407-730904        &0.73 &0.17 &0.07 &      &             &0.005\\
\hline
\end{tabular}
\end{center}
\end{table}

\subsubsection{Contributions from \oi, \cii\ and CO to gas cooling}
\label{cooling_top3}

\begin{figure}
\begin{center}
\includegraphics[scale=0.42]{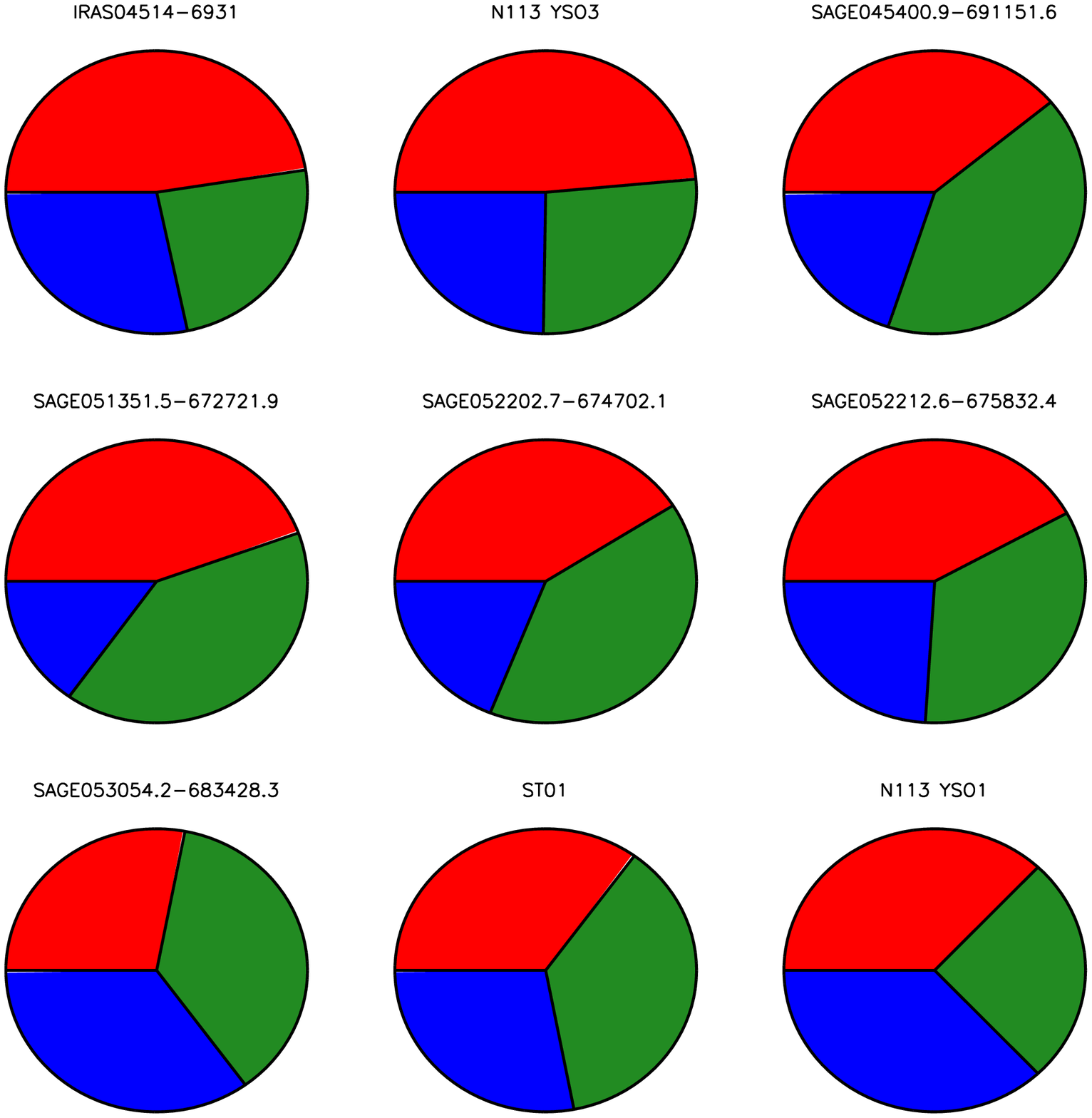}
\includegraphics[scale=0.42]{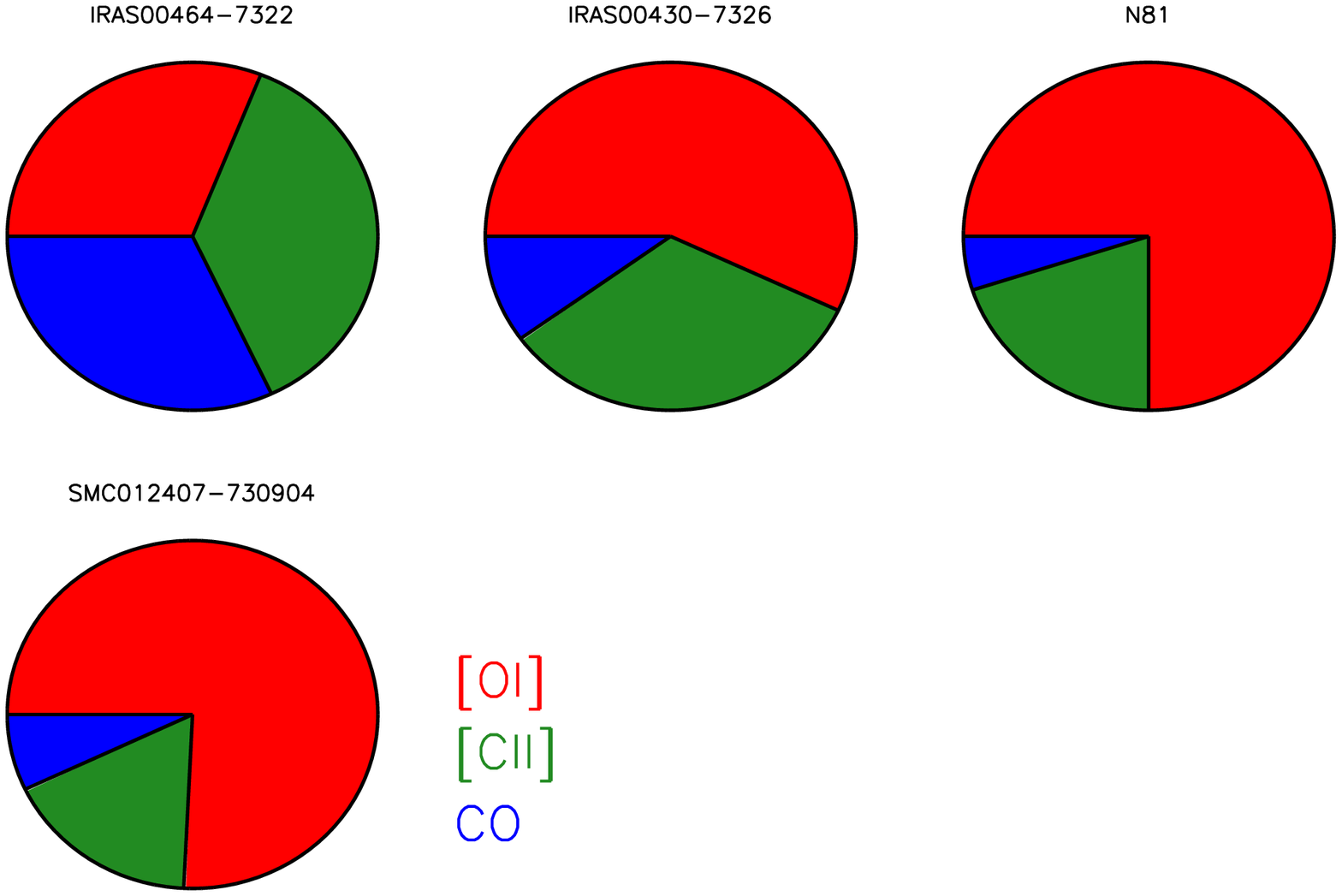}
\end{center}
\caption{Fractions of gas cooling for LMC and SMC YSOs, for the main gas species 
discussed in this work: \oi\ (red), \cii\ (green) and CO (blue). Note that the 
fractions displayed here are those from Table\,\ref{pies} renormalised to 
$L_{(\rm [OI]+[CII]+CO)}$ for  sources with a contribution from \hho\ and \oh.}
\label{cooling}
\end{figure}

Figure\,\ref{cooling} shows the relative contributions of \oi, \cii\ and CO that
given their ubiquitous nature are expected to be the main contributors to far-IR 
line cooling in YSOs \citep[e.g.,][and references therein]
{karska13,karska14,karska18}. For the majority of the sources (8/13), the strongest
line emission, and thus the main gas coolant, is \oi; for four sources \cii\ is the
main contributor, while for the last source CO and \oi\ contribute in equal
measure (with a smaller \cii\ contribution). Note that for many sources the \oi\ and
\cii\ contributions are very comparable. Median fractions normalised to their 
summed contributions are 42\%, 34\% and 24\%, respectively for \oi, \cii\ and CO  
(source-to-source variations are significant). Luminosity fractions for \oi\ and \cii\ 
are computed using line fluxes uncorrected for the extended emission contribution 
discussed in Sect.\,\ref{linemorphology}. This is to ensure that they can be compared to 
Galactic sources observed with \iso\ for which such corrections are 
unavailable. For completeness, if these corrections were applied to the Magellanic 
sources the fractions would become 46\%, 23\% and 31\%, respectively for \oi, \cii\ and 
CO. 

\begin{figure}
\begin{center}
\includegraphics[scale=0.42]{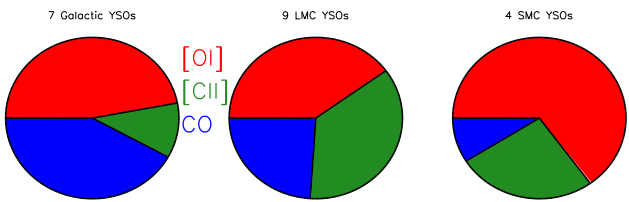}
\end{center}
\caption{Median gas cooling contributions for the main gas species for the 
Galactic, LMC and SMC massive YSO samples (the number of objects in each sample is 
listed). The {\it renormalised} cooling fractions \oi:\cii:CO are: 
47:11:42 (Galaxy), 40:36:24 (LMC) and 65:26:9 (SMC).}
\label{cooling_comparison}
\end{figure}

\begin{figure}
\begin{center}
\includegraphics[scale=0.55]{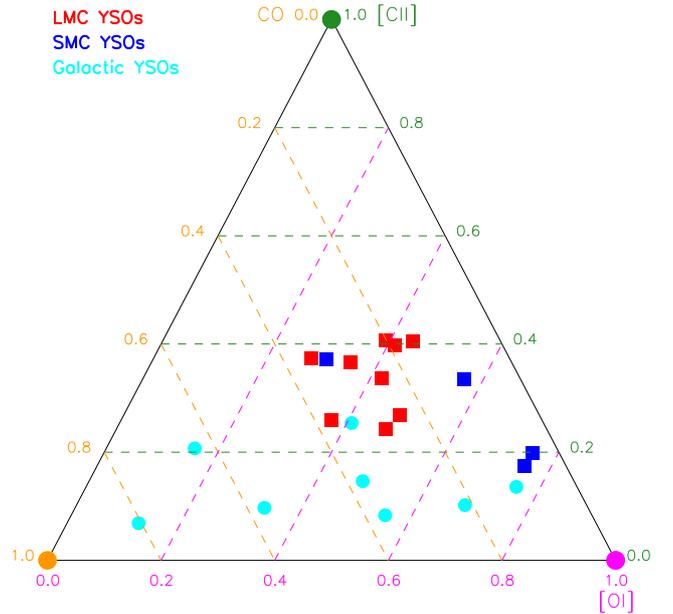}
\end{center}
\caption{Ternary diagram showing the cooling fractions for the main gas species:
\oi, \cii\ and CO. An object's fractions should be read by following the 
colour-coded dashed lines (\oi\ magenta, \cii\ green and CO orange) back to the 
relevant axes. Note that fractions displayed here are those from Table\,\ref{pies} 
renormalised to $L_{(\rm [OI]+[CII]+CO)}$.}
\label{ternary}
\end{figure}

We compared the luminosity fractions for massive Magellanic YSOs and Galactic
massive YSOs observed with \iso\ LWS. As described in Sect.\,\ref{isosample}, we do 
not directly compare our Magellanic sample to the massive YSO sample described in 
\citet{karska14} due to the different spatial scales sampled. Furthermore, \cii\ 
emission is often saturated for those sources; when reliably detected, it accounts 
for between 10$-$25\% of the total far-IR line luminosity. Individual gas cooling 
fractions for the Galactic comparison sources can be found in 
Fig.\,\ref{iso_cooling_pies} and Table\,\ref{iso_pies}. The median contributions for
the Galactic sample are 41\%, 10\% and 37\%, respectively for \oi, \cii\ and CO. It is 
noticeable from Fig.\,\ref{iso_cooling_pies} that for three out of seven sources CO is 
the main contributor to line emission while for the remaining four sources \oi\ is the 
main contributor.

The renormalised median contributions for the LMC, SMC and Galactic massive YSO 
samples are shown in Fig.\,\ref{cooling_comparison}. The figure suggests that the 
contribution from \cii\ emission is reduced in the Galactic sample compared to the 
Magellanic sample. Furthermore, while in the Galaxy the median contributions of 
\oi\ and CO are comparable, in the Magellanic sample \oi\ emission is more 
dominant, particularly in the SMC. It is also noticeable that the CO relative
contribution decreases from the Galaxy to the SMC, i.e. from higher to lower
metallicity environments. Given that source-to-source variations are 
significant, the fractions for the individual sources from the different samples 
are shown collectively in the ternary diagram in Fig.\,\ref{ternary}. For the 
Magellanic sample, the \cii\ fractions are consistently between $\sim$\,20\% and 
$\sim$40\%. The \oi\ fraction is larger than $\sim$\,30\%, while the CO fraction is 
always less than 40\%, in other words \oi\ emission is dominant. For the Galactic 
sample, the \cii\ fraction is at most 25\%, consistently lower than for the 
Magellanic sample. The range of the \oi\ and CO contributions are comparable, 
but the diagram strongly suggests that there is more variation in terms of \oi/CO 
ratio in the Galactic sample compared to the Magellanic sample. 

We investigated in more detail the samples properties, to check for any trends and
potential evolutionary effects. We constructed equivalent diagrams to that in 
Fig.\,\ref{ternary}, but separating sources with/without: silicate absorption, PAH 
and fine structure emission, and finally maser and radio emission (diagrams not 
shown). For more evolved sources in the UCH\,{\sc ii} and later stages, radio 
emission is typically strong due to free$-$free emission in the ionised gas 
\citep[e.g.,][]{hoare05}. PAH emission and fine structure emission can also identify 
a more evolved source (with the caveat that the wider YSO environment can contaminate
the \spitzer-IRS spectrum), while silicate absorption suggests a more embedded 
source \citep[e.g.,][]{seale09,woods11}. We found no statistically significant 
differences, either between the Galactic and Magellanic samples nor between samples 
with and without any of these indicators. Both the Galactic and Magellanic samples 
are made up of a very eclectic mix of sources, but there is no indication of any
evolutionary effects. 

\citet{karska14} found that for their sample of Galactic high-mass YSOs observed with 
PACS, line emission (and thus cooling) is dominated by CO and \oi, with a tendency 
for more evolved sources to have more significant \oi\ contributions 
\citep[as seen also for lower luminosity Galactic YSOs,][]{karska18}. In our analysis
we find similarly large source-to-source variations but we find no evidence for any 
evolutionary trend. Nevertheless, while evolutionary effects cannot be completely 
ruled out, there seems to be a shift in cooling dominance from CO to \oi\ from the 
Galaxy to the Magellanic Clouds (Figs.\,\ref{cooling_comparison} and \ref{ternary}). 
Furthermore, the predominance of \oi\ cooling seems to increase further for the SMC 
YSOs. Potentially, this could be a result of the lower CO gas-phase abundances at 
low metallicity, particularly in the SMC \citep[e.g.,][]{leroy07}. 

We also analysed the behaviour of the \oi/\cii\ and \oi/CO ratios. Median \oi/CO 
ratios are 1.2, 1.7 and 8 and \oi/\cii\ ratios are 2.1, 1.1 and 2.7 respectively for 
Galactic, LMC and SMC samples. Kolmogorov$-$Smirnov (K$-$S) tests reveal that the 
three sets of \oi/\cii\ measurements are statistically indistinguishable; for the 
\oi/CO ratios the K$-$S tests between Magellanic and Galactic sources, and LMC and 
SMC YSOs reveal that these sets of measurements are distinct, with null hypothesis 
probabilities 0.08 and 0.04 respectively. We investigated the behaviour of these 
ratios as a function of several measurable quantities: $L_{\rm TIR}$ (or $L_{\rm bol}$ 
for the Galactic sources), CO$_2$ ice column density ($N({\rm CO_2})$, another potential 
indicator of a cooler environment; column densities are compiled from
\citealt{gibb04,shimonishi10,oliveira09,oliveira11,seale11}), \oiii$/L_{\rm TIR}$ or 
\oiii$/L_{\rm bol}$ fraction, and finally mid-IR flux ratio ($F60/F100$ from \iras\ 
photometry for Galactic sources and $F70/F100$ from \spitzer\ MIPS and \herschel\ 
PACS photometry for Magellanic sources, Tables\,\ref{isosources}
and \ref{targetphotometry} respectively). There are no significant correlations
between the \oi/\cii\ or \oi/CO ratios with $L_{\rm TIR}$ or $L_{\rm bol}$, 
$N({\rm CO_2})$, and \oiii$/L_{\rm TIR}$ or \oiii$/L_{\rm bol}$. However 
Fig.\,\ref{cooling_ratios} reveals that not only there seems to be a correlation between 
\oi/CO ratio plotted against mid-IR flux ratio (albeit weak, Spearman $\rho$\,=\,0.53, 
$p$\,=\,0.034), the Galactic and Magellanic samples are different, as predicted by the 
K$-$S tests above. While being mindful of the small sample sizes, the mid-IR flux ratios 
are lower for Galactic sources; this suggests that the SED peaks at longer wavelengths, 
implying lower typical dust temperatures. The SMC sources tend to have higher ratios than
the LMC sources, consistent with higher dust temperatures in the low-metallicity SMC 
environment \citep[see Sect.\,\ref{bb}, and e.g.,][]{vanloon10a,vanloon10b}. Even 
though YSOs are generally associated with CO peaks \citep[e.g.,][]{sewilo13}, the 
observed CO emission is reduced for sources with higher dust temperature. This is 
consistent with the expectation that in general CO abundance is reduced and atomic 
species are more dominant in the less UV-shielded environments predominant at lower 
metallicity \citep[e.g.,][]{israel11}. 

\begin{figure}
\begin{center}
\includegraphics[scale=0.6]{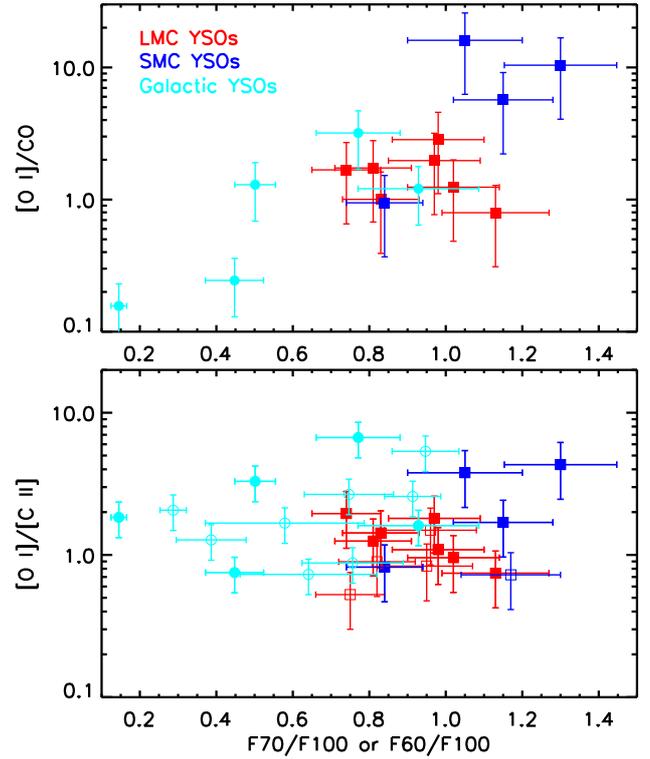}
\end{center}
\caption{Cooling line ratios against mid-IR flux ratio: $F60/F100$ derived from 
{\it IRAS} fluxes for Galactic sources and $F70/F100$ derived from \spitzer\ MIPS and 
\herschel\ PACS fluxes for Magellanic sources. The uncertainties are
dominated by the uncertainties resulting from the process to infer total line 
luminosities (Sect.\,\ref{totallines}). Filled symbols represent sources with CO 
measurements; sources shown with open symbols (lower panel) only have reliable \oi\ 
and \cii\ measurements.}
\label{cooling_ratios}
\end{figure}

\begin{figure}
\begin{center}
\includegraphics[scale=0.42]{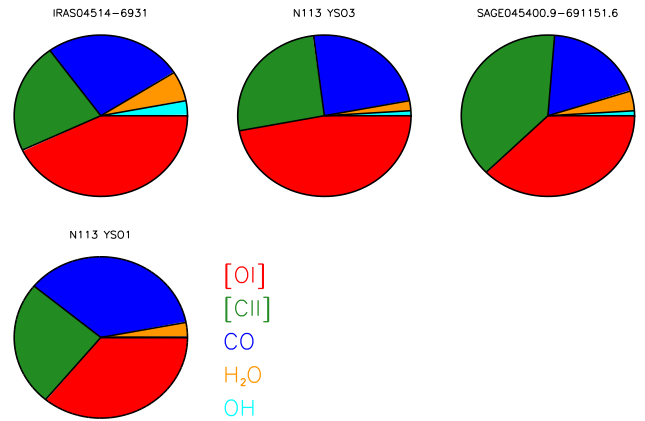}
\end{center}
\caption{Cooling fractions for LMC YSOs, for those objects with \hho\ (orange) and
\oh\ (cyan) detections. Note that for N\,113-YSO1 the \oh\ fraction is less than 1\%.
None of the SMC YSOs in our sample exhibit \oh\ emission.}
\label{cooling_all}
\end{figure}

\subsubsection{Contributions from \hho\ and \oh\ to gas cooling}

For four out of thirteen sources both \hho\ and OH measurements are 
available (in the LMC). For these sources \hho\ and OH account for at most 6\% 
and 3\% of the total line cooling respectively, or at most 9\% 
collectively (Fig.\,\ref{cooling_all}). If 3-$\sigma$ upper limits are considered, 
\hho\ and \oh\ could conservatively contribute up to 12\% and 6\% respectively. This
indeed confirms that \oi, \cii\ and CO are definitively the main cooling agents 
($\gsim$\,90\% of the total line cooling).

There is a clear trend of decreasing contribution of \hho\ and \oh\ cooling from low
luminosity sources \citep{karska18} to massive YSOs in the Galaxy \citep{karska14}.
Our analysis suggests this trend extends to massive YSOs in the Magellanic Clouds.
\hho\ and \oh\ molecules can be destroyed by numerous processes 
\citep*[see review by][and references therein]{vandishoeck13}, including 
photodissociation that would be more dominant in massive YSO environments. 
There could also be an evolutionary effect since \hho\ emission is predominantly 
seen for evolved massive YSOs \citep{vandertak13}. However no such trend with 
evolutionary class is found in \citet{karska18}. 

\section{Summary and Conclusions}

We have presented \herschel\ PACS and SPIRE FTS spectroscopy of a sample of massive YSOs
in the Magellanic Clouds. Our analysis focused on a variety of emission lines: \oi\,
\oiii, \cii, CO, \hho\ and \oh\ observed with PACS, and CO, \ci\ and \nii\ observed
with SPIRE. We have compared the properties of the Magellanic sample with those of a
Galactic YSO sample observed with the \iso\ LWS; this is to ensure that the spatial 
scales probed by the two samples are in principle comparable. We summarise here our 
main findings:
\vspace{-2.5mm}
\begin{itemize}\item {\it Bright atomic emission line morphology:} \oi\ and \cii\ 
emission is detected for all sources. By making use of the 25 PACS spaxels, we show 
that the emission exhibits a point source contribution superposed on more extended 
diffuse emission; we are able to successfully separate the YSO contribution. The 
morphology of the \oiii\ emission is very varied: the YSO is detected for nine targets, 
while for four targets the emission at the YSO position is consistent with ambient 
contamination; for a further six targets no \oiii\ emission is detected.

\item {\it Atomic emission line diagnostics and PDR emission:} We find that \oi\ and 
\cii\ emission (after removing the diffuse contribution (see above) and the ionised 
gas contribution to \cii) is tightly correlated, and it is also correlated with the 
total IR luminosity. The measured ratio \oi/\cii\ is relatively low suggesting that 
shocks are not important contributors to \oi\ emission. We conclude that both \oi\ 
and \cii\ emission originate from the PDR associated with the massive YSOs. The line 
emission is also more dominant (with respect to the dust emission) in Magellanic YSOs 
when compared to the Galactic massive sample. 

The efficiency of photoelectric gas heating (inferred from the ratio 
(\oi+\cii)/$F_{\rm TIR}$) shows a large source-to-source scatter, however it is clearly 
higher for Magellanic compared to Galactic YSOs (medians respectively 0.25\% and 
0.1\%). This can be understood as a consequence of the reduced grain charge in low 
metallicity environments. When compared to PDR models, our results are consistent
with a lower $G_0/n$ ratio for the Magellanic sources, that can be interpreted as 
evidence of the changed properties in the low-metallicity ISM (porous and clumpy), even 
on the smaller spatial scales of individual YSOs.

\nii\ emission at 205\,\micron\ is detected for nine LMC and two SMC YSOs. For these 
sources we are able to constrain the ionised gas contribution to the \cii\ emission. 
We conclude that most of the \cii\ emission originates from the PDR ($\sim$\,82\% 
median). \ci\ emission is detected at 370 and 609\,\micron\ for six sources. This
emission originates from PDR gas, with a temperature of 18\,$-$\,24\,K for LTE gas or 
$\ge$\,30\,K for non-LTE gas. 

\item {\it Properties and origin of the CO emission:}
The observed CO emission is consistent with shock excitation. We analysed the CO 
rotational diagrams for fifteen sources. We find that these are broadly consistent 
with originating from either two low-temperature LTE gas components or a single
isothermal (hot) low-density component. Based on the fact that CO ratios suggest the
presence of multiple shock components, and on what is known from Galactic sources,
the former seems the most likely scenario. The temperatures of the LTE components
($T_{\rm cold}$\,$\sim$\,40\,K and $T_{\rm cool}$\,$\sim$\,120\,K) are remarkably
consistent across the luminosity range, and from Galactic to Magellanic environments.

\item {\it Additional molecular detections:} \hho\ emission is detected towards six
LMC sources and tentatively towards one SMC source. \oh\ emission is detected towards
five sources, and absorption is detected towards two sources, all LMC YSOs. The
emission is generally weak, as observed towards massive Galactic YSOs, signalling a
very modest contribution to far-IR cooling (see below).

\item {\it YSO cooling budget:} Our observations include the most prominent emission
lines in YSO environments, that are thus the main contributors to the far-IR cooling
of the YSO envelope, both in Galactic and Magellanic sources. We find that \oi\ and 
\cii\ are the main contributors for Magellanic YSOs in contrast to Galactic YSOs for 
which CO and \oi\ contribute by a similar amount, with a decreased \cii\ 
contribution. Furthermore, there seems to be a trend for a reduced fraction 
attributed to CO emission from the Galaxy $\rightarrow$ LMC $\rightarrow$ SMC, 
suggesting a link to metallicity. We find that the Magellanic YSOs not only have higher 
\oi/CO ratios, they also have mid-IR fluxes consistent with higher dust temperatures. 
This is consistent with a reduction in CO emission in environments where the dust is 
warmer due to reduced UV-shielding. 
\end{itemize}
Our analysis shows evidence for metallicity effects in the properties of the ISM in the
near-environment of massive YSOs in the Magellanic Clouds. The interaction of the 
LMC and the SMC has potentially led to metallicity gradients in both galaxies
\citep[e.g.,][]{fukui17,choudhury18,tsuge19}. It will be interesting to place our results
in the context of these metal abundance variations, when such detailed large scale maps 
become available. 

\section*{Acknowledgements}
We thank Agata Karska and Yao-Lun Yang for sharing their line measurements and useful 
discussions. Our analysis benefited from support from \herschel\ PACS and SPIRE experts,
namely: Rosalind Hopwood, Ivan Valtchanov, Edward Polehampton, Katrina Exter and Elena 
Puga. This work makes use of data collected by the \herschel\ Space Observatory. HIPE is 
a joint development by the \herschel\ Science Ground Segment 
Consortium, consisting of ESA, the NASA \herschel\ Science Center, and the HIFI, 
PACS and SPIRE consortia. This work is based in part on observations made with 
the \spitzer\ Space Telescope, which is operated by the Jet Propulsion Laboratory,
California Institute of Technology under a contract with NASA. We acknowledge 
financial support from the NASA Herschel Science Center, JPL contracts 
\#1381522, \#1381650 \& \#1350371. The material is based upon work supported by 
NASA under award number 80GSFC17M0002. M.-Y.L. was partially funded through the 
sub-project A6 of the Collaborative Research Council 956, funded by the 
Deutsche Forschungsgemeinschaft (DFG). We thank the anonymous referee for a 
constructive report.
%%%%%%%%%%%%%%%%%%%%%%%%%%%%%%%%%%%%%%%%%%%%%%%%%%



%\bibliographystyle{mnras}


\vspace*{10mm}

\newpage

%\clearpage

%%%%%%%%%%%%%%%%%%%%%%%%%%%%%%%%%%%%%%%%%%%%%%%%%%

%%%%%%%%%%%%%%%%% APPENDICES %%%%%%%%%%%%%%%%%%%%%

\appendix

\section{Target {\bf \it SPITZER} and {\bf \it HERSCHEL} photometry}
\label{phot_appendix}

Table\,\ref{targetphotometry} lists the photometry for the Magellanic targets. 
Aperture photometry is performed for the \spitzer\ IRAC 3.6$-$8\,\micron\ and MIPS 
24 and 70\,\micron\ bands (labelled I1$-$4 and M1$-$2): target centring, aperture 
radius and sky annulus radii are optimised for each target; a minimum 10\% flux 
error is adopted. PSF band-merged catalogue photometry (\citealt{seale14}, see also
\citealt{meixner13}) is used where available for the \herschel\ PACS (100 and
160\,\micron, P2$-$3) and SPIRE (250, 350 and 500\,\micron, S1$-$3) bands; for a 
single target (\#4, N\,81) $^*$ indicates that aperture photometry was used instead.
We did perform aperture photometry for those sources for which only \herschel\ 
S1$-$S3 flux limits are available; however we do not consider those measurements 
reliable given the bright environmental emission and large beam size.

\begin{landscape}
\begin{center}
\begin{table}
\begin{center}
\caption{\normalsize \spitzer\ and \herschel\ photometry for the sample of
Magellanic YSOs discussed here (flux densities in mJy). Photometric band identifications are as follows: I1$-$4: IRAC bands 
(3.6, 4.5, 5.8 and 8.0\,\micron); M1$-$2: MIPS bands (24 and 70\,\micron); P2$-$3: PACS 
bands (100 and 160\,\micron); S1$-$S3: SPIRE bands (250, 350 and 500\,\micron). We measure 
new aperture photometry for the I1$-$I4 and M1$-$M2 bands; a flux uncertainty of at least 10\% 
is adopted for these measurements. We list catalogue (PSF) photometry for P2$-$P3 and S1$-$S3 from 
\citet{seale14} unless $^*$ indicates otherwise (aperture photometry is used
for \#4, N\,81). Sources 
\#7A, B are not resolved at all in any \herschel\ photometric bands, but two \spitzer\ sources
are identified separated by $\sim 6$\arcsec; \#7C is $\sim 20$\arcsec\ from \#7A, B and it is 
thus only resolved in PACS P2$-$3 (see Fig.\,\ref{n113_fov}, bottom). Sources
\#10, 11 and 12 are the three 
prominent YSOs in N\,113 (see Fig.\,\ref{n113_fov}, top); they are spatially resolved in all \herschel\ 
bands (\#10 and \#12 are respectively $\sim 20$\arcsec\ and $\sim 26$\arcsec\ from \#11), however
only marginally so in the SPIRE photometric bands; thus it is not possible to extract meaningful 
individual fluxes in S2$-$3.}
\label{targetphotometry}
\begin{tabular}{l|l|c@{\hspace{5mm}}|c@{\hspace{5mm}}|c@{\hspace{5mm}}|c@{\hspace{5mm}}|c@{\hspace{5mm}}|c@{\hspace{5mm}}|c@{\hspace{5mm}}|c@{\hspace{5mm}}|c@{\hspace{5mm}}|c@{\hspace{5mm}}|c}
\hline
\#&Source ID               &I1               &I2                       &\,I3	                &\,I4	                 &\,\,\,M1	            &\,\,M2		           &\,\,\,P2		       &\,\,\,P3 	           &\,\,\,S1		       &\,\,\,S2		   &\,\,\,S3 	        \\
  &                        &3.6\,\micron     &4.5\,\micron             &5.8\,\micron            &8\,\micron              &24\,\micron               &70\,\micron                   &100\,\micron               &160\,\micron               &250\,\micron               &350\,\micron               &500\,\micron \\
\hline
\multicolumn{13}{c}{SMC YSOs}\\
\hline
 1 &IRAS\,00430$-$7326	    &\al{3}4.8$\pm$3.5& \al{7}0.1$\pm$7.0      &\al{12}3.4$\pm$12.\ar{3}&\al{14}4.3$\pm$14.\ar{4}&\al{1}515$\pm$15\ar{2}    &      7798$\pm$780      &      6783$\pm$354         &      4467$\pm$305	   & \al{1}368$\pm$61          &       720$\pm$43          &       255$\pm$24   \\%H3		   
 2 &IRAS\,00464$-$7322	    &      3.9$\pm$0.4&       8.9$\pm$0.9      & \al{1}5.7$\pm$1.6      & \al{2}1.1$\pm$2.1      &      120$\pm$12          &      1924$\pm$192      &      2279$\pm$127         &      2104$\pm$163	   &       996$\pm$45          & $\geq$234                 &\null	        \\%H4		
 3 &S3MC\,00541$-$7319	    &\al{1}0.9$\pm$1.1& \al{3}2.9$\pm$3.3      & \al{6}7.7$\pm$6.8      & \al{8}6.3$\pm$8.6      &      551$\pm$55          &      2591$\pm$259      &      2208$\pm$118         &      1475$\pm$113	   & $\geq$631	               & $\geq$469	           & $\geq$254	        \\%H13	       
 4 &N\,81 		    &\al{1}0.5$\pm$1.1& \al{1}4.3$\pm$1.4      & \al{2}4.9$\pm$2.5      & \al{6}7.6$\pm$6.8      &  \al{1}296$\pm$13\ar{0}  &      9495$\pm$950      &      9049$\pm$905\ar{$^*$}&      7063$\pm$706\ar{$^*$}& \al{2}742$\pm$27\ar{4$^*$}& \al{1}343$\pm$13\ar{4$^*$}&       644$\pm$64\ar{$^*$}\\%H2		
 5 &SMC\,012407$-$73090 (N\,88A)&\al{3}5.5$\pm$3.5& \al{7}2.3$\pm$7.2      &\al{13}8.8$\pm$13.\ar{9}&\al{42}9.9$\pm$43.\ar{0}&  \al{7}294$\pm$72\ar{9}  &\al{2}1720$\pm$217\ar{2}&\al{1}6390$\pm$875         &      8997$\pm$613	   & $\geq$2735	               &$\geq$1340	           & $\geq$706	        \\%H1	       
\hline
\multicolumn{13}{c}{LMC YSOs}\\
\hline
 6 &IRAS\,04514$-$6931	    &      6.8$\pm$0.7& \al{1}7.2$\pm$1.7      & \al{6}1.8$\pm$6.2      &\al{11}8.9$\pm$11.\ar{9}&  \al{1}696$\pm$17\ar{0}  &\al{1}3167$\pm$131\ar{7}&\al{1}7810$\pm$116\ar{4}   &\al{1}2200$\pm$750	   & \al{3}484$\pm$21\ar{8}    &$\geq$2445                 & \al{1}404$\pm$10\ar{1}\\%H11		   
 7A&SAGE\,045400.2$-$691155.4 &      8.7$\pm$0.9& \al{2}5.5$\pm$2.5      & \al{5}8.8$\pm$5.9      & \al{9}1.7$\pm$9.2      &\multirow{2}{*}{\al{2}617$\pm$26\ar{2}} &\multirow{2}{*}{$\leq$50597}&\multirow{2}{*}{\al{1}1830$\pm$114\ar{1}}&\multirow{2}{*}{\al{2}1180$\pm$134\ar{2}}&\multirow{3}{*}{\al{10}300$\pm$65\ar{5}} &\multirow{3}{*}{$\geq$3758}&\multirow{3}{*}{$\geq$3802}\\%H6A		  
 7B&SAGE\,045400.9$-$691151.6 &      8.6$\pm$0.9& \al{1}0.5$\pm$2.8      & \al{3}7.6$\pm$4.9      & \al{8}3.5$\pm$8.3      &                                            &			           & 			                     & 			                       & 			                 &			          &			       \\%H6	       
 7C&SAGE\,045403.0$-$691139.7 &      4.7$\pm$0.5&       5.9$\pm$0.6      & \al{2}0.4$\pm$2.0      & \al{4}5.4$\pm$4.5      &\al{1}456$\pm$15\ar{5}    &	          $\leq$48865& 	  8862$\pm$108\ar{5}   &\al{1}1720$\pm$841         & 			       &			                 &			       \\%H6B		      
 8 &IRAS\,05011$-$6815	    &      1.0$\pm$0.2&       1.7$\pm$0.3      &       3.1$\pm$0.4      &       5.4$\pm$0.7      &      445$\pm$44          &      4762$\pm$476        &      5026$\pm$333         &      3994$\pm$258         & $\geq$1756                &\al{1}212$\pm$65           &       687$\pm$50  \\%H17	      
 9 &SAGE\,051024.1$-$701406.5 &      8.8$\pm$0.9& \al{3}1.2$\pm$3.1      & \al{7}3.4$\pm$7.3      &\al{11}6.0$\pm$11.\ar{6}&      588$\pm$59          &      2470$\pm$247        &      3279$\pm$220         &      3108$\pm$204	   & \al{1}770$\pm$10\ar{8}    &      940$\pm$49           &       372$\pm$28  \\%H16	    
10 &N\,113\,YSO-1	            &\al{1}4.6$\pm$1.5& \al{2}1.6$\pm$2.2      & \al{8}5.7$\pm$8.6      &\al{23}4.0$\pm$23.\ar{4}&\al{5}130$\pm$51\ar{3}    &\al{4}9959$\pm$499\ar{6}&\al{6}0100$\pm$405\ar{6}   &\al{5}4880$\pm$334\ar{5}   &\al{20}540$\pm$12\ar{84}   &\multirow{3}{*}{$\geq$3074}&\multirow{3}{*}{$\geq$5855}\\%H5	   
11 &N\,113\,YSO-4	            &\al{1}7.5$\pm$1.7& \al{2}7.6$\pm$2.8      & \al{8}4.2$\pm$8.4      &\al{20}7.2$\pm$20.\ar{7}&\al{4}797$\pm$48\ar{0}    &\al{1}7469$\pm$174\ar{7}&\al{1}8200$\pm$150\ar{2}   &\al{1}9820$\pm$127\ar{4}   & \al{5}705$\pm$37\ar{9}    &   	                   &                   \\%H7A(mid)        
12 &N\,113\,YSO-3	            &\al{1}7.3$\pm$1.7& \al{2}5.7$\pm$2.6      & \al{8}2.3$\pm$8.2      &\al{22}6.2$\pm$22.\ar{6}&\al{4}809$\pm$48\ar{1}    &\al{4}3832$\pm$438\ar{3}&\al{4}5210$\pm$298\ar{9}   &\al{3}0460$\pm$187\ar{1}   &\al{11}480$\pm$71\ar{4}    &	                   &       	       \\%H7(maser)  
13 &SAGE\,051351.5$-$672721.9 &\al{1}1.6$\pm$1.2& \al{1}7.6$\pm$1.8      & \al{7}2.6$\pm$7.3      &\al{18}9.6$\pm$19.\ar{0}&\al{3}870$\pm$38\ar{7}    &\al{2}0677$\pm$206\ar{8}&\al{2}1060$\pm$138\ar{8}   &\al{1}5700$\pm$961	   & \al{6}405$\pm$39\ar{2}    &\al{2}721$\pm$14\ar{7}     &       985$\pm$77  \\%H10		  
14 &SAGE\,052202.7$-$674702.1 &      4.3$\pm$0.4&       3.9$\pm$0.4      & \al{1}9.2$\pm$1.9      & \al{4}9.6$\pm$5.0      &      869$\pm$87          &       $\leq$8437       &      7020$\pm$474         &      7201$\pm$453         & \al{4}015$\pm$24\ar{6}    &\al{1}914$\pm$10\ar{7}     &       969$\pm$76  \\%H19		  
15 &SAGE\,052212.6$-$675832.4 &\al{4}1.7$\pm$4.2& \al{9}5.1$\pm$9.5      &\al{23}1.0$\pm$23.\ar{1}&\al{37}7.3$\pm$37.\ar{8}&\al{7}935$\pm$79\ar{4}    &\al{4}3135$\pm$431\ar{4}&\al{5}3050$\pm$348\ar{8}   &\al{3}1420$\pm$192\ar{4}   & \al{8}222$\pm$52\ar{4}    &\al{3}354$\pm$21\ar{6}     &$\geq$4127         \\%H8	       
16 &SAGE\,052350.0$-$675719.6 &      2.7$\pm$0.3&       4.7$\pm$0.5      & \al{1}8.6$\pm$1.9      & \al{5}2.0$\pm$5.2      &\al{1}400$\pm$14\ar{0}    &\al{1}0694$\pm$107\ar{0}&\al{1}3100$\pm$860         &      9640$\pm$598         & \al{3}921$\pm$24\ar{1}    &\al{1}897$\pm$10\ar{6}     &$\geq$ 821         \\%H14	       
17 &SAGE\,053054.2$-$683428.3 &\al{3}1.6$\pm$3.2& \al{7}0.8$\pm$7.1      &\al{14}2.4$\pm$14.\ar{2}&\al{19}2.9$\pm$19.\ar{4}&\al{1}731$\pm$17\ar{3}    &\al{1}4416$\pm$144\ar{2}&\al{1}2780$\pm$853         &\al{1}0180$\pm$636         & $\geq$3551	               &$\geq$1823	           &$\geq$1436         \\%H9	    
18 &IRAS\,05328$-$6827	    &\al{6}1.1$\pm$6.1&\al{10}1.9$\pm$10.\ar{2}&\al{16}8.0$\pm$16.\ar{8}&\al{18}5.8$\pm$18.\ar{6}&      551$\pm$55        &      $\leq$4794          &      2556$\pm$210         &      3581$\pm$253         & $\geq$1884                &$\geq$1133                 &$\geq$ 779         \\%H15  	     
19 &LMC\,053705$-$694741      &      2.0$\pm$0.2&       1.4$\pm$0.1      &       9.9$\pm$1.0      & \al{2}5.0$\pm$2.5      &          $\geq$71.5    &      $\leq$4351          &      1428$\pm$180         &      1865$\pm$197         &       835$\pm$89          &\null                      &$\geq$1041         \\%H12  	     
20 &ST\,01		    &\al{1}8.4$\pm$1.8& \al{3}5.5$\pm$3.6      & \al{8}3.3$\pm$8.3      &\al{13}7.5$\pm$13.\ar{8}&      888$\pm$89        &        9129$\pm$913      &      8967$\pm$602         &      7793$\pm$492         & $\geq$4134                &$\geq$1861                 &$\geq$1315         \\%H18  	     
\hline
\end{tabular}
\end{center}
\end{table}
\end{center}
\end{landscape}

\section{Selected SED fit examples}
\label{sed_appendix}

Figure\,\ref{sed_examples} shows examples of the two component modified blackbody fits 
to the SEDs of Magellanic YSOs, as described in Sect.\,\ref{bb}. The fit parameters for 
individual objects are tabulated in Table\,\ref{fir_temp}. The integrated
luminosity $L($\,>\,10\,\micron) is adopted as the object's total IR luminosity 
$L_{\rm TIR}$. Example spectra for the same two sources are shown in 
Fig.\,\ref{spec_examples}.

\begin{figure}
\begin{center}
\includegraphics[scale=0.55]{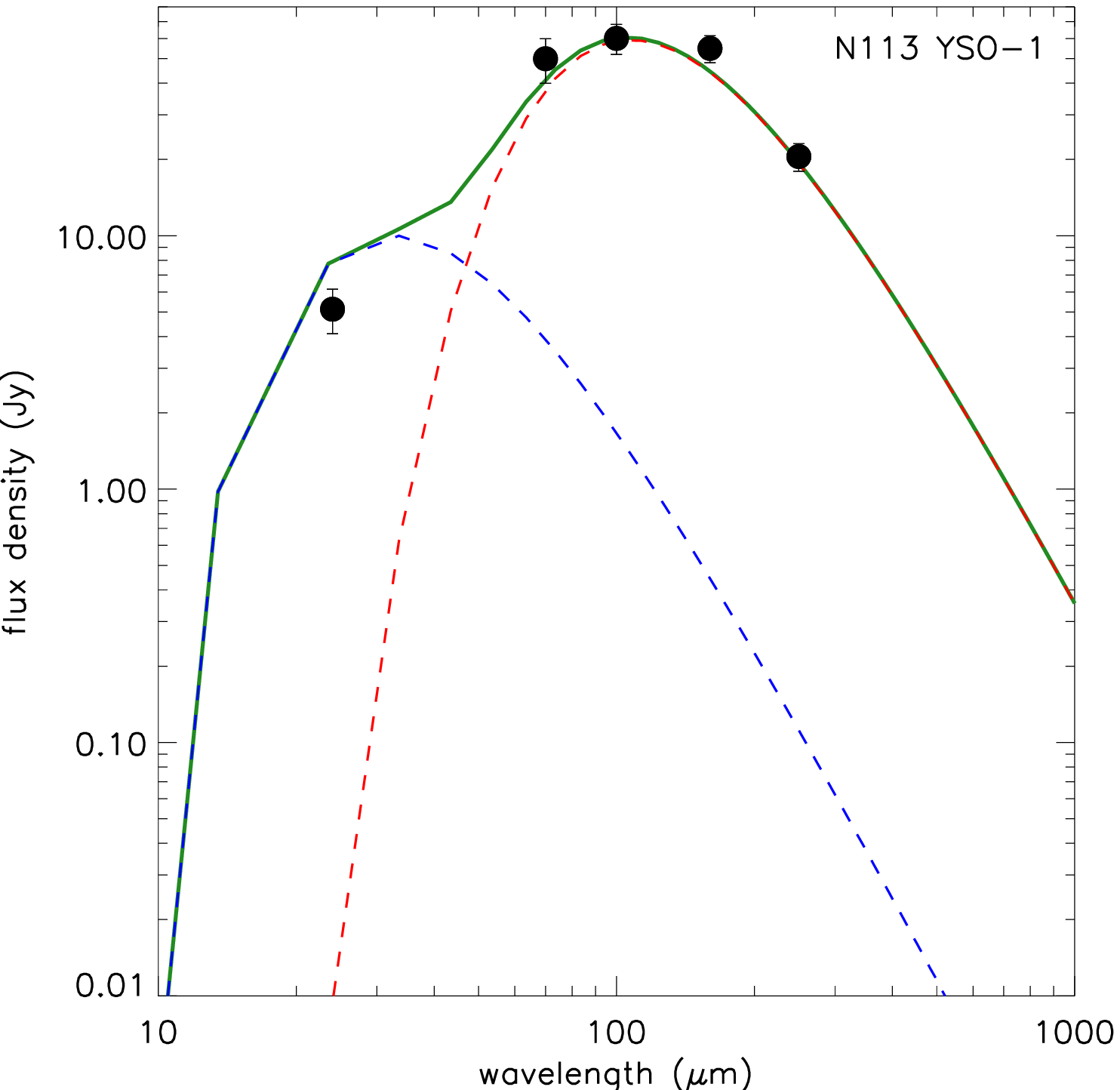}
\includegraphics[scale=0.52]{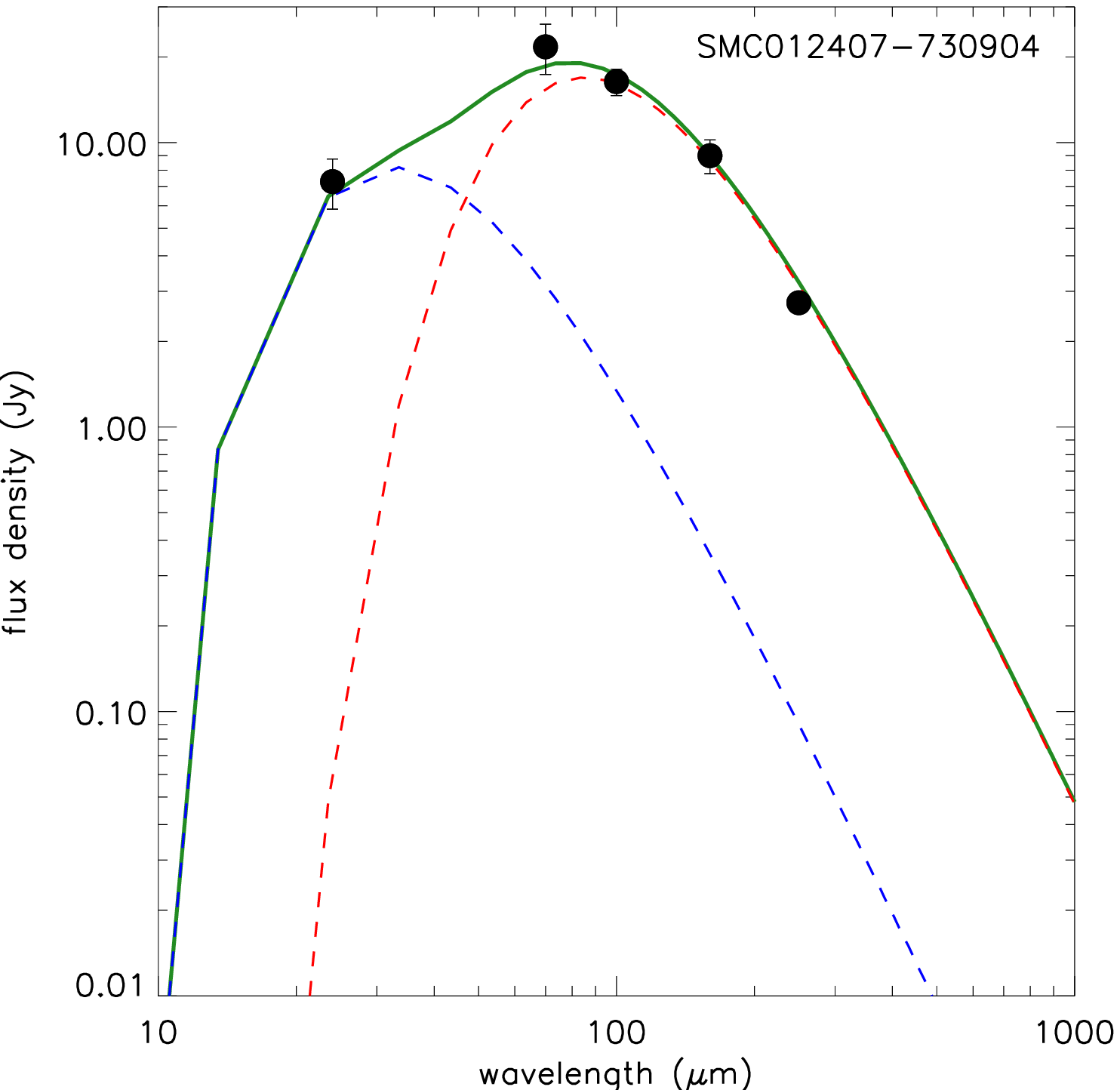}
\end{center}
\caption{Examples of two component modified blackbody SED fits for N\,113 YSO-1 (in the 
LMC) and SMC\,012407$-$730904 (N\,88A). The total fit (green line) is shown, as well as
the two temperature components: the temperature of the warmer components (blue dashed
line) is not well constrained; the cooler component (red line) has temperature of
$\sim$\,34\,K and $\sim$\,38\,K respectively for these two examples.}
\label{sed_examples}
\end{figure}

\newpage

\clearpage

\newpage

\section{Selected example spectra}
\label{spec_appendix}

Figure\,\ref{spec_examples} shows examples of the PACS and SPIRE spectra for an LMC and 
an SMC YSO respectively. They are typical of our sample for \oi, \oiii\ and \cii; for
CO they exemplify the highest and the lowest SNR achieved. The SPIRE spectra are
representative of those sources with SNR\,$\ge$\,5 for which the analysis of the 
rotational diagram was performed. This SNR limit was chosen somewhat empirically; 
as discussed in Sect.\,\ref{spire_int_vel}, for lower SNR the centroid line 
positions vary widely and line identifications become unreliable. It is consistent with 
limits set in other works \citep{lee16}. Note that for the SMC source 
SMC\,012407$-$730904, emission for \ci\ at 609\,\micron\ and \nii\ at 205\,\micron\ are 
not detected.

In Figs.\,\ref{hho_spec} and \ref{oh_spec} we show \hho\ and \oh\ spectra for 
those sources for which the transitions are detected. N\,113-YSO3 is the only source
for which both the 179.5 and 108\,\micron\ transitions were detected; the detection
of \hho\ at 108\,\micron\ for IRAS\,00430$-$7326 is very tentative. For the \oh\
doublet at 84\,\micron\ the bluest component is detected in emission for two sources.
For \oh\ at 79.1\,\micron\ emission is detected for three sources, for N\,113-YSO1
only the blue component is detected and for SAGE\,052212.6$-$675832.4 and 
SAGE\,052350.0$-$675719.6 weak absorption is detected. No \oh\ transitions are detected
towards any SMC source. Line fluxes are tabulated in the next section.

\begin{figure*}
\begin{center}
\includegraphics[scale=0.3]{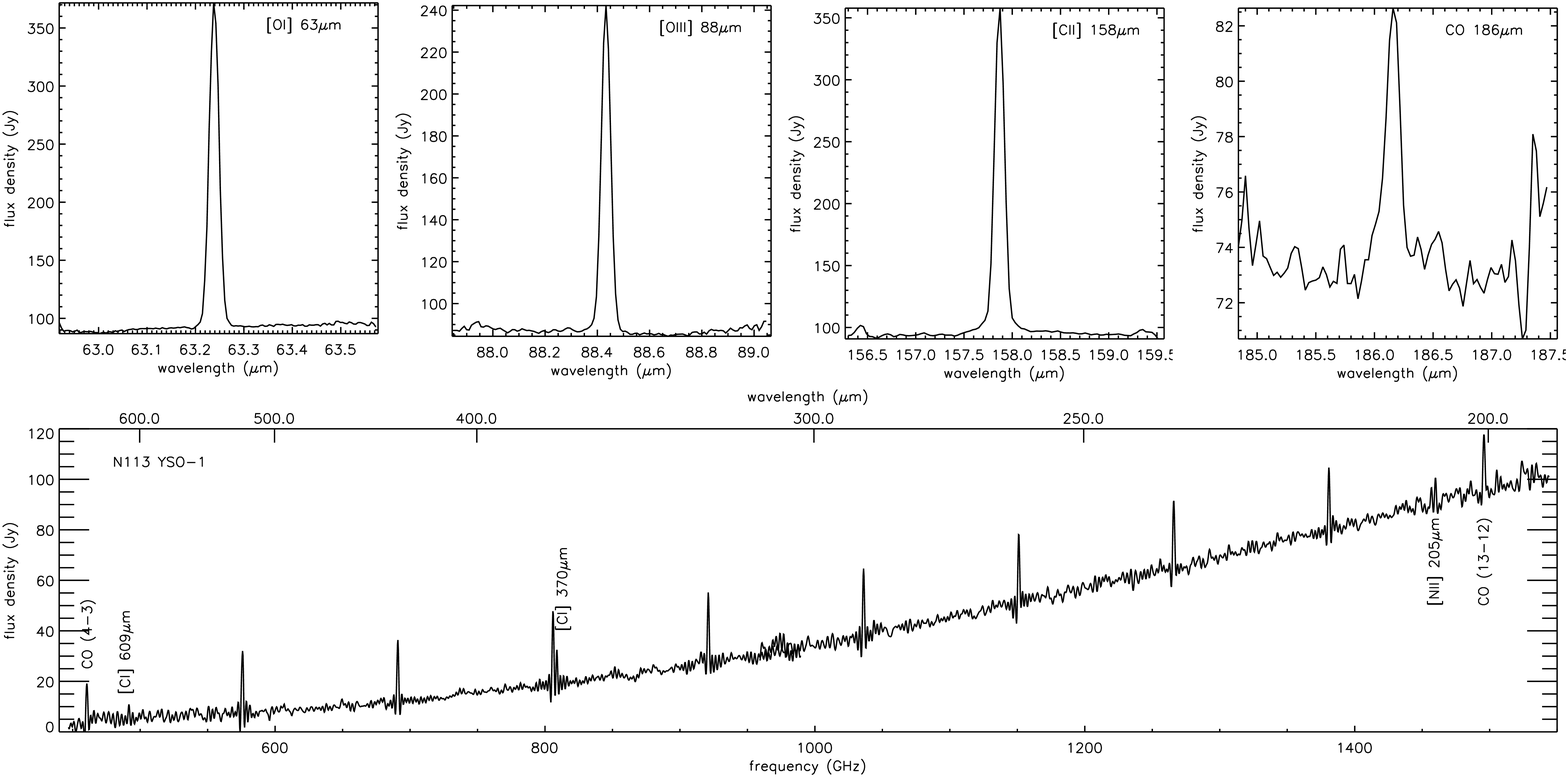}
\includegraphics[scale=0.3]{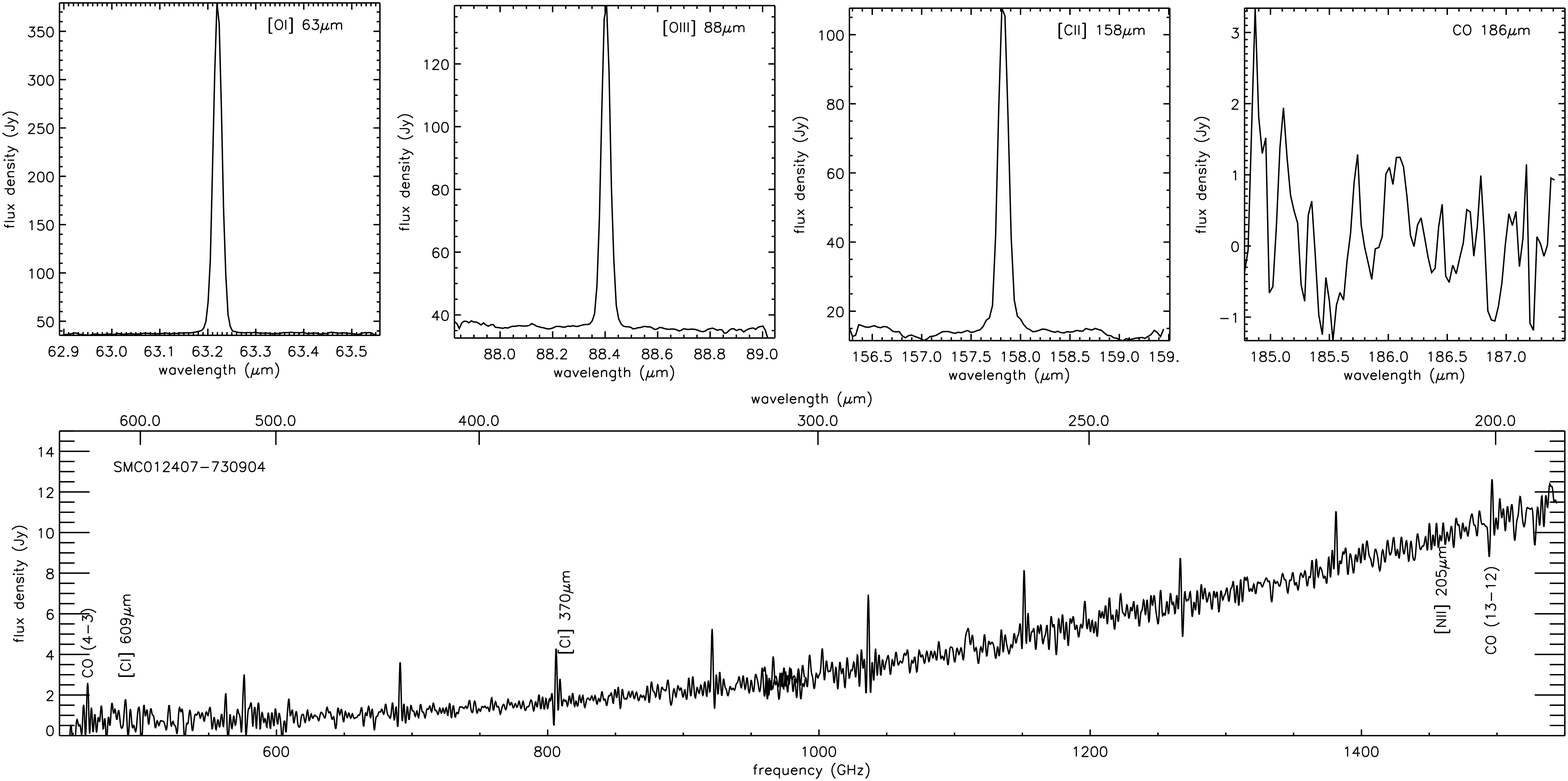}
\end{center}
\caption{PACS and SPIRE spectra for N\,113 YSO-1 (in the LMC) and SMC\,012407$-$730904 
(N\,88A).}
\label{spec_examples}
\end{figure*}

\begin{figure*}
\begin{center}
\includegraphics[scale=0.57]{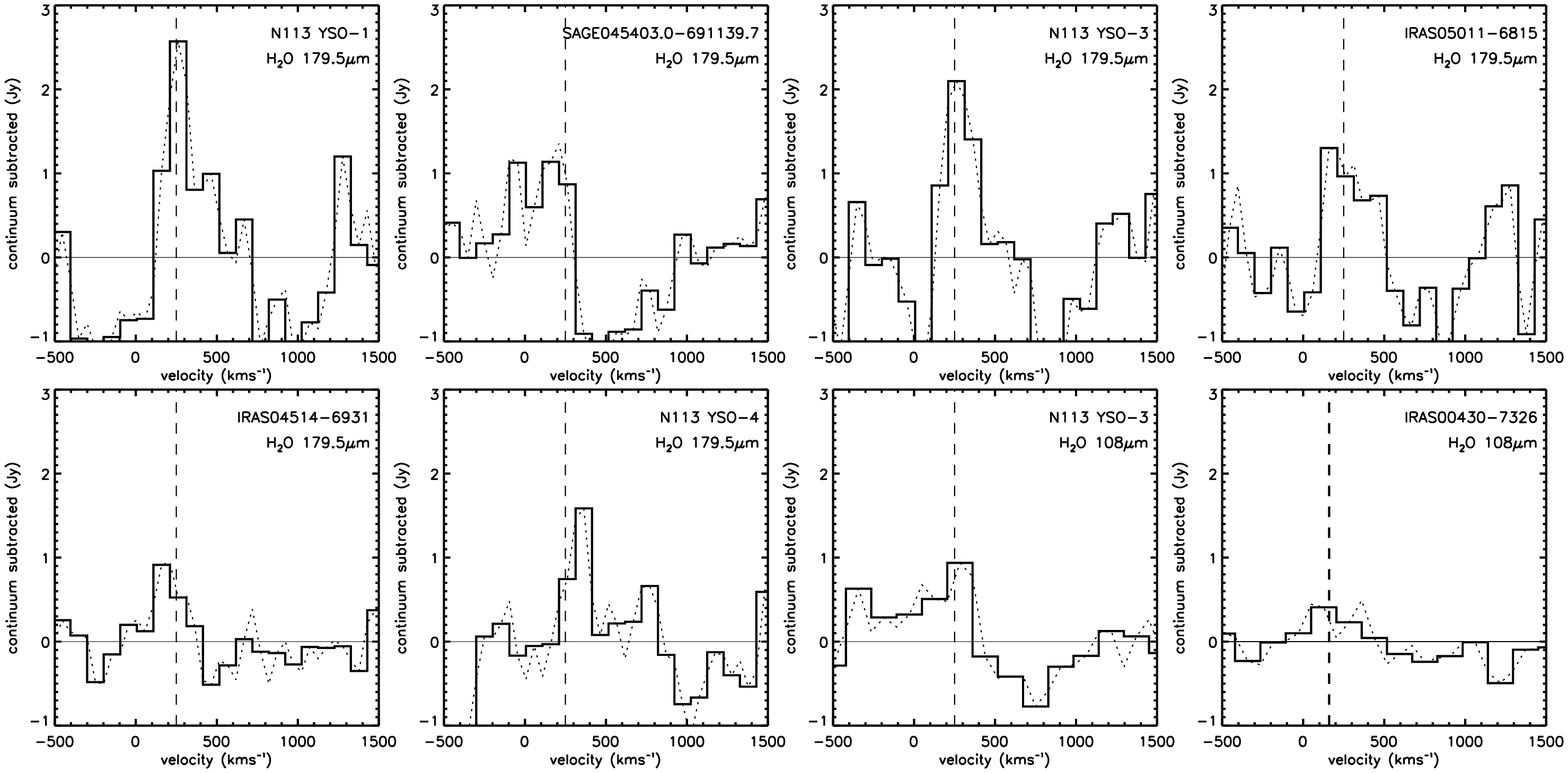}
\end{center}
\caption{\hho\ detections for Magellanic YSOs. Note that N\,113 YSO-3 is the only source
for which both the 179.5 and 108\,\micron\ transitions were detected. For the
SMC source IRAS\,00430$-$7326 there is a tentative detection at 108\,\micron\ only. The
vertical dashed line represent the galaxies systemic velocity.}
\label{hho_spec}
\end{figure*}

\begin{figure*}
\begin{center}
\includegraphics[scale=0.57]{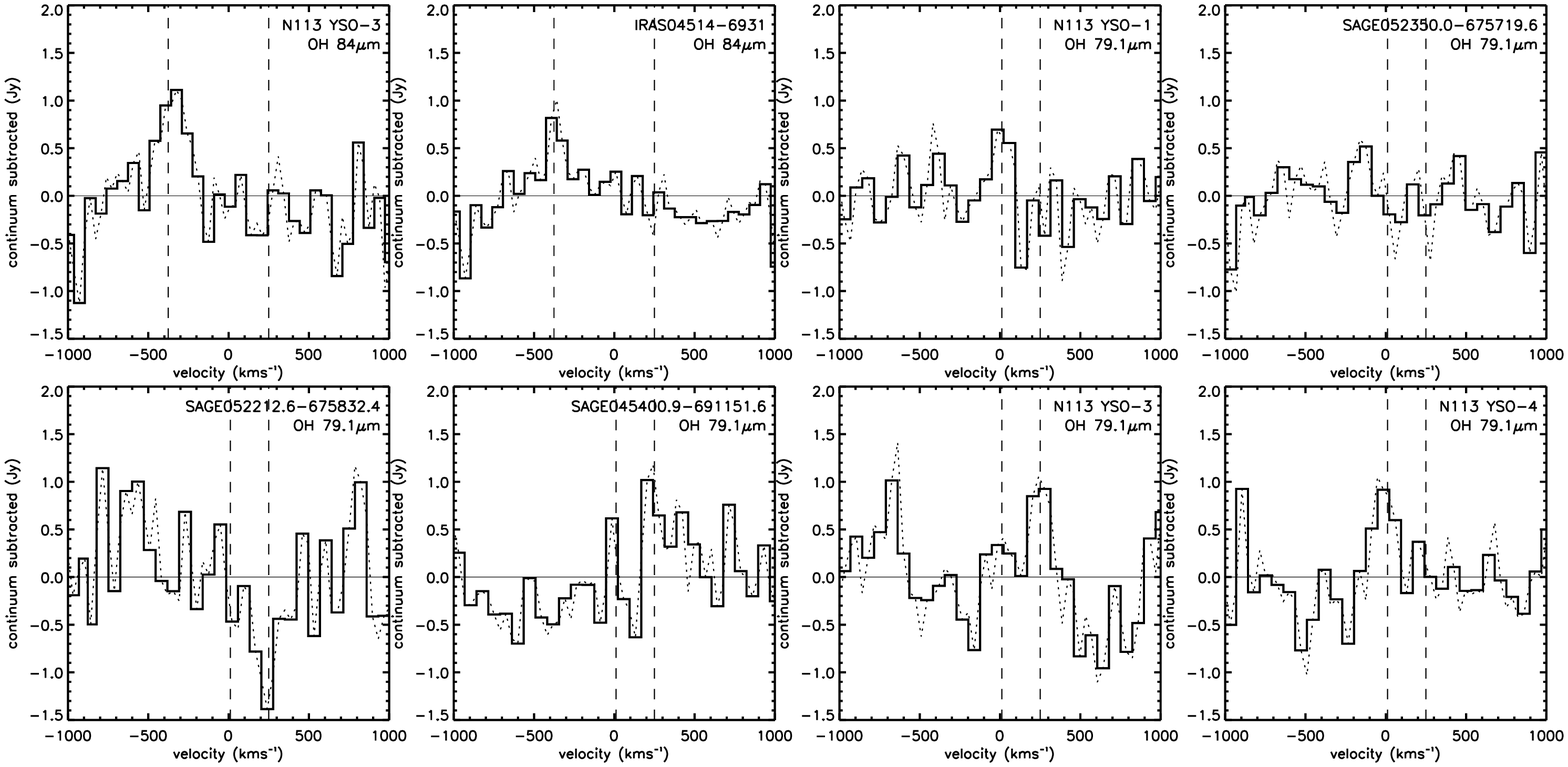}
\end{center}
\caption{OH detections for Magellanic YSOs.  Most sources shown exhibit \oh\ emission
for at least one doublet component; for SAGE\,052212.6$-$675832.4 and 
SAGE\,052350.0$-$675719.6 weak absorption is detected. The redmost doublet component is
displayed at the galaxies systemic velocity.}
\label{oh_spec}
\end{figure*}

\clearpage

\section{Line fluxes for Magellanic YSOs}
\label{linefluxes}

In this section we present measured line fluxes and 3-$\sigma$ upper limits for 
individual Magellanic YSOs. These are shown in Tables\,\ref{linefluxes_pacs} and 
\ref{linefluxes_spire} respectively for lines observed with PACS and SPIRE FTS.

\begin{landscape}
\begin{center}
\begin{table}
\begin{center}
\caption{Measured fluxes for Magellanic YSOs for lines in the PACS
range ($\times 10^{-18}$\,W\,m$^{-2}$). Upper limits are given at 3-$\sigma$. For 
\oi, \cii\ and \oiii\ both measured fluxes and fluxes corrected for the extended 
environmental emission (Sect\,\ref{linemorphology}) are given; for \oiii\ if the latter is not listed the emission is 
considered to be mostly ambient to the YSO wider environment. The fraction of \cii\ 
emission that originates from the PDR gas is also given (see
Sect\,\ref{cii_nii}). Sources \#7A and B are not resolved at
PACS wavelengths.}
\label{linefluxes_pacs}
\begin{tabular}{l@{\hspace{0mm}}|c@{\hspace{0mm}}|c@{\hspace{0mm}}|c@{\hspace{0mm}}|c@{\hspace{0mm}}|c@{\hspace{0mm}}|c@{\hspace{0mm}}|c@{\hspace{0mm}}|c@{\hspace{0mm}}|c@{\hspace{0mm}}|c@{\hspace{0mm}}|c@{\hspace{0mm}}|c@{\hspace{0mm}}|c@{\hspace{0mm}}|c@{\hspace{0mm}}|c@{\hspace{0mm}}|c@{\hspace{0mm}}|c@{\hspace{0mm}}|c@{\hspace{0mm}}|}
\hline
\#&SourceID&\multicolumn{2}{c}{\oi\,63\,\micron}&\multicolumn{3}{c}{\cii\,158\,\micron}&\multicolumn{2}{c}{\oiii\,88\,\micron}&CO&\multicolumn{2}{c}{\hho}&\multicolumn{4}{c}{\oh}\\
  &         &Measured&Corrected&Measured&Corrected&PDR&Measured&Corrected&186\,\micron&108\,\micron&179.5\,\micron&79.11\,\micron&79.18\,\micron&84.4\,\micron&84.6\,\micron\\
\hline
\multicolumn{16}{c}{SMC YSOs}\\
\hline
 1&       IRAS\,00430$-$7326&1010.0$\pm$ 6.8& 900.0$\pm$ 6.8& 648.1$\pm$ 6.5& 471.0$\pm$ 6.7&0.96& 116.8$\pm$ 3.0& 116.7$\pm$ 3.0&
$\leq$16.8    &
 11.3$\pm$ 3.3&$\leq$16.8    &
$\leq$19.9    &$\leq$19.9    &              &                  \\
 2&       IRAS\,00464$-$7322& 228.8$\pm$ 4.1& 210.0$\pm$ 4.1& 303.0$\pm$ 3.9& 150.0$\pm$ 4.0&0.74&  	     &  	     &
              &
              &$\leq$ 7.4    &
$\leq$11.0    &$\leq$11.0    &              &                  \\
 3&       S3MC\,00541$-$7319& 115.9$\pm$ 5.1& 115.0$\pm$ 5.0& 173.9$\pm$ 3.4& 121.0$\pm$ 3.8&1.00&$\leq$  17.3   &  	     &
 16.9$\pm$ 3.1&
$\leq$11.8    &$\leq$ 8.1    &
$\leq$11.1    &$\leq$11.1    &$\leq$10.3    &$\leq$10.3        \\
 4&                  N\,81&2955.8$\pm$18.4&2670.0$\pm$18.4& 848.3$\pm$ 6.8& 680.0$\pm$ 6.8&1.00&4771.0$\pm$12.8&4290.0$\pm$ 8.8&
$\leq$11.0    &
$\leq$20.8    &$\leq$14.1    &
              &              &$\leq$10.7    &$\leq$10.7    &                  \\
 5&     SMC\,012407$-$730904&5729.6$\pm$16.6&5400.0$\pm$16.6&1442.4$\pm$11.6&1160.0$\pm$11.8&1.00&1671.9$\pm$ 7.0&1580.0$\pm$ 7.0&
 18.4$\pm$ 4.0&
$\leq$24.4    &$\leq$14.0    &
$\leq$21.0    &$\leq$21.0    &$\leq$13.5    &$\leq$13.5        \\
\hline
\multicolumn{16}{c}{LMC YSOs}\\
\hline
 6&       IRAS\,04514-6931&1118.1$\pm$ 6.2&1100.0$\pm$16.0& 621.7$\pm$ 5.1& 500.0$\pm$ 5.0&1.00&$\leq$  19.4   &  	     &
$\leq$25.2    &
$\leq$17.7    & 14.0$\pm$ 2.0&
              &              &16.8$\pm$ 3.8&$\leq$13.0          \\
% 7&SAGE045400.9-691151.6&5314.4$\pm$36.3&3800.0$\pm$30.2&5972.3$\pm$31.8&3510.0$\pm$33.6&0.87&9401.8$\pm$17.1&  	     &
%$\leq$32.5    &
%$\leq$53.3    & 44.7$\pm$11.9&
%  4.8$\pm$ 2.9& 19.1$\pm$ 3.7&$\leq$24.1    &$\leq$24.1        \\
7A&SAGE\,045400.2$-$691155.4&\multirow{2}{*}{3757.4$\pm$26.4}&\multirow{2}{*}{3000.0$\pm$26.9}&\multirow{2}{*}{3082.3$\pm$19.8}&\multirow{2}{*}{1850.0$\pm$21.3}&\multirow{2}{*}{0.74}&\multirow{2}{*}{5061.7$\pm$17.1}&&
 \multirow{2}{*}{$\leq$32.5    }&
 \multirow{2}{*}{$\leq$32.8    }&\multirow{2}{*}{$\leq$14.0} &
 \multirow{2}{*}{  4.8$\pm$ 2.9}& \multirow{2}{*}{19.1$\pm$ 3.7}&\multirow{2}{*}{$\leq$24.7}    &\multirow{2}{*}{$\leq$24.7}        \\
7B&SAGE\,045400.9$-$691151.6 &&&&&&&&&&&&&&&\\
7C&SAGE\,045403.0$-$691139.7 &1557$\pm$12.9& 800$\pm$13.9&2890$\pm$24.9&1660$\pm$26.1&0.72&4340$\pm$24.6&&
$\leq$35.3&
$\leq$53.2&44.7$\pm$11.9&
$\leq$27.6&$\leq$27.6&$\leq$24.1&$\leq$24.1\\
 8&       IRAS\,05011$-$6815&  91.4$\pm$16.2&  90.0$\pm$ 0.2& 119.1$\pm$ 8.1&  94.0$\pm$ 8.9&1.00&$\leq$  21.8   &  	     &
$\leq$20.9    &
$\leq$34.0    & 23.2$\pm$ 5.5&
$\leq$18.4    &$\leq$18.4    &$\leq$14.5    &$\leq$14.5        \\
 9&SAGE\,051024.1$-$701406.5& 140.1$\pm$ 9.1& 140.0$\pm$ 9.1& 289.4$\pm$ 7.6& 217.0$\pm$18.0&1.00&$\leq$  24.9   &  	     &
 12.9$\pm$ 5.7&
$\leq$19.0    &$\leq$22.2    &
$\leq$16.8    &$\leq$16.8    &$\leq$22.1    &$\leq$22.1        \\
10&            N\,113 YSO1&5218.3$\pm$22.2&3500.0$\pm$23.3&3974.4$\pm$22.3&1990.0$\pm$24.8&0.78&2635.8$\pm$10.1&2310.0$\pm$10.0&
121.4$\pm$ 7.4&
$\leq$44.1    & 39.6$\pm$15.7&
 13.9$\pm$ 3.7&$\leq$20.1    &$\leq$32.8    &$\leq$32.8        \\
11&            N\,113 YSO4&5186.2$\pm$27.6&3470.0$\pm$28.4&3777.5$\pm$19.8&1796.0$\pm$22.7&    &6459.7$\pm$18.0&6130.0$\pm$18.0&
 26.4$\pm$ 9.7&
$\leq$44.6    & 16.1$\pm$ 2.8&
 22.0$\pm$ 4.3&  5.8$\pm$ 3.3&$\leq$20.9    &$\leq$20.9        \\
12&            N\,113 YSO3&6192.8$\pm$26.8&4333.0$\pm$18.7&3723.1$\pm$25.7&1700.0$\pm$28.0&0.91& 208.0$\pm$ 5.9&  	     &
 45.9$\pm$ 5.1&
 28.0$\pm$ 8.5& 29.2$\pm$ 5.6&
  6.5$\pm$ 6.3& 19.6$\pm$ 5.3& 37.7$\pm$ 5.4&$\leq$16.3        \\
13&SAGE\,051351.5$-$672721.9&1701.6$\pm$ 7.5&1530.0$\pm$ 7.5&1700.2$\pm$13.5&1180.0$\pm$14.0&0.91& 164.6$\pm$ 3.9&  	     &
 22.0$\pm$ 6.2&
$\leq$27.9    &$\leq$16.3    &
$\leq$15.4    &$\leq$15.4    &             &                   \\
14&SAGE\,052202.7$-$674702.1&1173.1$\pm$11.9&1050.0$\pm$11.8&1226.1$\pm$ 9.4&1000.0$\pm$ 9.5&0.86&$\leq$  38.3   &  	     &
$\leq$25.8    &
$\leq$32.9    &$\leq$17.5    &
$\leq$25.9    &$\leq$25.9    &$\leq$22.2    &$\leq$22.2        \\
15&SAGE\,052212.6$-$675832.4&3314.4$\pm$16.2&2330.0$\pm$11.2&2881.4$\pm$17.3& 800.0$\pm$ 5.0&0.81&8227.4$\pm$29.0&  	     &
$\leq$29.5    &
$\leq$34.6    &$\leq$18.2    &
$-$10.8$\pm$ 4.0&$-$30.3$\pm$ 0.7&$\leq$38.5    &$\leq$38.5        \\
16&SAGE\,052350.0$-$675719.6& 588.7$\pm$ 8.4& 522.0$\pm$ 8.4& 714.4$\pm$ 6.3& 500.0$\pm$ 6.4&1.00&  17.3$\pm$ 5.0&  17.3$\pm$ 5.0&
$\leq$18.1    &
$\leq$25.3    &$\leq$20.3    &
$-$6.4$\pm$ 1.5&$-$5.3$\pm$ 1.0&$\leq$22.1    &$\leq$22.1        \\
17&SAGE\,053054.2$-$683428.3& 598.3$\pm$ 6.7& 539.0$\pm$ 6.7& 873.9$\pm$ 9.3& 610.0$\pm$ 9.6&0.76&1712.1$\pm$ 8.0&  78.0$\pm$13.0&
$\leq$15.1    &
$\leq$34.1    &$\leq$22.8    &
$\leq$12.2    &$\leq$12.2    &             &                   \\
18&       IRAS\,05328$-$6827& 108.3$\pm$ 9.1& 100.0$\pm$ 9.0& 539.0$\pm$ 7.7& 160.0$\pm$ 9.4&0.81&  13.9$\pm$ 3.7&  14.0$\pm$ 3.7&
 13.1$\pm$ 5.6&
$\leq$32.8    &$\leq$27.6    &
$\leq$21.2    &$\leq$21.2    &$\leq$20.9    &$\leq$20.9        \\
19&     LMC\,053705$-$694741&  80.6$\pm$ 3.9&  25.0$\pm$ 4.0& 615.8$\pm$ 6.1&  92.0$\pm$ 8.9&0.62&  36.7$\pm$ 4.7&  37.0$\pm$ 4.7&
$\leq$23.2    &
$\leq$30.2    &$\leq$24.2    &
$\leq$12.9    &$\leq$12.9    &             &                   \\
20&                 ST\,01&1000.5$\pm$ 9.9& 900.0$\pm$22.4&1137.0$\pm$ 7.5& 790.0$\pm$ 8.3&1.00&$\leq$  27.9   &  	     &
$\leq$25.1    &
$\leq$19.6    &$\leq$26.4    &
$\leq$17.2    &$\leq$17.2    &$\leq$16.9    &$\leq$16.9        \\
\hline
\end{tabular}
\end{center}
\end{table}
\end{center}
\end{landscape}
% 7&SAGE045400.9-691151.6&5314.4$\pm$36.3&3800.0$\pm$30.2&5972.3$\pm$31.8&3510.0$\pm$33.6&0.87&9401.8$\pm$17.1&  	     &
%$\leq$32.5    &
%$\leq$53.3    & 44.7$\pm$11.9&
%  4.8$\pm$ 2.9& 19.1$\pm$ 3.7&$\leq$24.1    &$\leq$24.1        \\
% 7A&SAGE045400.2$-$691155.4&\multirow{2}{*}{3757.4$\pm$26.4}&\multirow{2}{*}{3000.0$\pm$26.9}&\multirow{2}{*}{3082.3$\pm$19.8}&\multirow{2}{*}{0.74}&\multirow{2}{*}{1850.0$\pm$21.3}&\multirow{2}{*}{5061.7$\pm$17.1}&&
% \multirow{2}{*}{$\leq$32.5    }&
% \multirow{2}{*}{$\leq$32.8    }&\multirow{2}{*}{$\leq$14.0} &
% \multirow{2}{*}{  4.8$\pm$ 2.9}& \multirow{2}{*}{19.1$\pm$ 3.7}&\multirow{2}{*}{$\leq$24.7}    &\multirow{2}{*}{$\leq$24.7}        \\
% 7B&SAGE045400.9$-$691151.6 &&&&&&&&&&&&&&&\\
% 7C&SAGE045403.0$-$691139.7 &1557$\pm$12.9& 800$\PM$13.9&2890$\pm$24.9&1660$\pm$26.1&4340$\pm$\24.6&&
%$\leq$35.3&
%$\leq$53.2&44.7$\pm$11.9&
%$\leq$27.6&$\leq$27.6&$\leq$24.1&$\leq$24.1\\

\begin{landscape}
\begin{center}
\begin{table}
\begin{center}
\caption{Measured fluxes for Magellanic YSOs for lines in the 
SPIRE FTS range ($\times 10^{-18}$\,W\,m$^{-2}$). Upper limits are given at 3-$\sigma$.
Sources \#7A, B and C are not resolved in these SPIRE observations; source \#11
has no SPIRE spectrum. For the sources marked $*$, the SNR ratio is not
sufficient for the analysis of the CO rotational diagram.
}
\label{linefluxes_spire}
\begin{tabular}{l|l|c@{\hspace{0mm}}|c@{\hspace{0mm}}|c@{\hspace{0mm}}|c@{\hspace{0mm}}|c@{\hspace{0mm}}|c@{\hspace{0mm}}|c@{\hspace{0mm}}|c@{\hspace{0mm}}|c@{\hspace{0mm}}|c@{\hspace{0mm}}|c@{\hspace{0mm}}|c@{\hspace{0mm}}|c@{\hspace{0mm}}|c@{\hspace{0mm}}|c@{\hspace{0mm}}|c@{\hspace{0mm}}|}
\hline
 &                          &\multicolumn{10}{c}{CO}&\multicolumn{2}{c}{\ci}&\nii\\
 &                          &4$-$3&5$-$4&6$-$5&7$-$6&8$-$7&9$-$8&10$-$9&11$-$10&12$-$11&13$-$12&370\,\micron&609\,\micron&205\,\micron\\
\hline
\multicolumn{15}{c}{SMC YSOs}\\
\hline
 1&           IRAS\,00430$-$7326&   5.4$\pm$ 1.0&  12.2$\pm$ 1.0&  11.7$\pm$ 1.0&  10.0$\pm$ 1.0&  12.5$\pm$ 1.0&  11.1$\pm$ 0.9&  10.6$\pm$ 0.9&  10.5$\pm$ 0.9&   9.2$\pm$ 0.9&   7.1$\pm$ 0.9&
  5.1$\pm$ 1.0&$\leq$ 3.1    &
  3.4$\pm$ 0.9\\
 2&           IRAS\,00464$-$7322&   7.8$\pm$ 1.2&  15.8$\pm$ 1.2&  18.1$\pm$ 1.2&  17.4$\pm$ 1.2&  15.8$\pm$ 1.2&  19.6$\pm$ 1.9&  10.8$\pm$ 1.9&  13.1$\pm$ 1.9&   6.3$\pm$ 1.9&  12.2$\pm$ 1.9&
 12.1$\pm$ 1.2&  7.7$\pm$ 1.2&
 10.3$\pm$ 1.9\\
 3&       S3MC\,00541$-$7319$^*$&$\leq$ 4.4     &   3.4$\pm$ 1.5&   8.3$\pm$ 1.5&   6.0$\pm$ 1.5&   9.8$\pm$ 1.5&   5.6$\pm$ 1.6&   1.5$\pm$ 1.6&   2.8$\pm$ 1.6&$\leq$ 4.7     &   5.1$\pm$ 1.6&
$\leq$ 4.5    &$\leq$ 4.4    &
$\leq$ 4.7    \\
 4&                      N\,81&  13.5$\pm$ 1.3&   9.1$\pm$ 1.3&   9.7$\pm$ 1.3&  11.9$\pm$ 1.4&   8.2$\pm$ 1.3&  13.2$\pm$ 1.9&  11.2$\pm$ 1.9&  10.2$\pm$ 1.9&   6.4$\pm$ 1.9&  10.6$\pm$ 1.9&
$\leq$ 4.1    &$\leq$ 4.0    &
$\leq$ 5.7    \\
 5&         SMC\,012407$-$730904&  24.6$\pm$ 1.8&  25.8$\pm$ 1.8&  28.8$\pm$ 1.8&  30.9$\pm$ 1.8&  33.8$\pm$ 1.8&  42.4$\pm$ 2.4&  39.3$\pm$ 2.4&  27.4$\pm$ 2.4&  32.2$\pm$ 2.4&  26.0$\pm$ 2.4&
 10.0$\pm$ 1.8&$\leq$ 5.5    &
$\leq$ 7.3    \\
\hline
\multicolumn{15}{c}{LMC YSOs}\\
\hline
 6&           IRAS\,04514$-$6931&  38.3$\pm$ 1.7&  37.1$\pm$ 1.7&  40.9$\pm$ 1.7&  38.8$\pm$ 1.7&  33.2$\pm$ 1.7&  44.2$\pm$ 2.2&  41.1$\pm$ 2.2&  45.1$\pm$ 2.2&  34.9$\pm$ 2.2&  23.0$\pm$ 2.2&
 19.4$\pm$ 1.7&$\leq$ 5.1    &
$\leq$ 6.5    \\
7A&    SAGE\,045400.2$-$691155.4& \multirow{3}{*}{109.7$\pm$ 7.0}& \multirow{3}{*}{180.5$\pm$ 6.9}& \multirow{3}{*}{206.1$\pm$ 6.9}& \multirow{3}{*}{234.9$\pm$ 7.1}& \multirow{3}{*}{213.0$\pm$ 6.9}& \multirow{3}{*}{214.5$\pm$11.1}& \multirow{3}{*}{152.6$\pm$11.1}&  \multirow{3}{*}{95.2$\pm$11.1}&  \multirow{3}{*}{74.7$\pm$11.1}&  \multirow{3}{*}{52.4$\pm$11.2}&
 \multirow{3}{*}{71.2$\pm$ 7.1}& \multirow{3}{*}{63.2$\pm$ 7.0}&
\multirow{3}{*}{102.7$\pm$10.9}\\
7B&SAGE\,045400.9$-$691151.6 &&&&&&&&&&&&&&\\
7C&SAGE\,045403.0$-$691139.7 &&&&&&&&&&&&&&\\
 8&       IRAS\,05011$-$6815$^*$&  19.2$\pm$ 3.8&  24.7$\pm$ 3.7&  15.5$\pm$ 3.7&  12.3$\pm$ 3.8&  10.2$\pm$ 3.7&  10.2$\pm$ 3.5&   6.7$\pm$ 3.5&   2.8$\pm$ 3.5&  12.2$\pm$ 3.5&  15.1$\pm$ 3.5&
$\leq$11.4    &$\leq$11.4    &
$\leq$10.6    \\
 9&SAGE\,051024.1$-$701406.5$^*$&  20.9$\pm$ 3.4&  15.6$\pm$ 3.4&  14.7$\pm$ 3.4&  15.3$\pm$ 3.4&  10.2$\pm$ 3.4&   9.6$\pm$ 3.1&   8.2$\pm$ 3.1&   4.2$\pm$ 3.1&   1.7$\pm$ 3.1&  14.9$\pm$ 3.1&
$\leq$10.2    &$\leq$10.3    &
$\leq$ 9.4    \\
10&                N\,113 YSO1& 180.8$\pm$ 8.2& 307.0$\pm$ 8.1& 282.5$\pm$ 8.1& 338.7$\pm$ 8.2& 317.8$\pm$ 8.1& 315.9$\pm$10.1& 323.0$\pm$10.1& 317.3$\pm$10.1& 289.5$\pm$10.1& 266.8$\pm$10.2&
124.0$\pm$ 8.2& 65.8$\pm$ 8.1&
112.1$\pm$10.1\\
11&                N\,113 YSO4&               &               &               &               &               &               &               &               &               &
              &              &
              \\
12&                N\,113 YSO3& 152.8$\pm$ 7.7& 220.7$\pm$ 7.6& 188.9$\pm$ 7.6& 223.9$\pm$ 7.7& 224.2$\pm$ 7.6& 183.9$\pm$ 8.9& 141.8$\pm$ 8.9& 155.4$\pm$ 8.9& 140.1$\pm$ 8.9& 142.8$\pm$ 8.9&
 75.9$\pm$ 7.7& 42.9$\pm$ 7.6&
 43.4$\pm$ 8.8\\
13&    SAGE\,051351.5$-$672721.9&  20.9$\pm$ 1.4&  31.3$\pm$ 1.4&  39.9$\pm$ 1.4&  47.7$\pm$ 1.4&  44.6$\pm$ 1.4&  50.4$\pm$ 1.5&  38.2$\pm$ 1.5&  28.4$\pm$ 1.5&  19.1$\pm$ 1.5&  18.1$\pm$ 1.5&
 11.8$\pm$ 1.4& 10.3$\pm$ 1.4&
 19.5$\pm$ 1.5\\
14&    SAGE\,052202.7$-$674702.1&  21.4$\pm$ 3.9&  29.8$\pm$ 3.8&  29.5$\pm$ 3.8&  38.6$\pm$ 3.9&  47.1$\pm$ 3.8&  34.9$\pm$ 3.5&  23.8$\pm$ 3.5&  26.8$\pm$ 3.5&  24.8$\pm$ 3.5&  27.6$\pm$ 3.5&
 28.2$\pm$ 3.9&$\leq$11.7    &
 22.4$\pm$ 3.5\\
15&    SAGE\,052212.6$-$675832.4& 120.6$\pm$ 7.3& 153.4$\pm$ 7.2& 169.0$\pm$ 7.2& 152.6$\pm$ 7.4& 119.8$\pm$ 7.2&  83.2$\pm$11.0& 120.7$\pm$11.0&  63.6$\pm$11.0&  42.6$\pm$11.0&  54.6$\pm$11.1&
 98.9$\pm$ 7.4&$\leq$22.1    &
 69.5$\pm$10.9\\
16&SAGE\,052350.0$-$675719.6$^*$&$\leq$20.3     &  36.6$\pm$ 6.7&  29.5$\pm$ 6.7&  28.1$\pm$ 6.8&  18.8$\pm$ 6.7&  21.1$\pm$ 5.7&$\leq$17.2     &  21.0$\pm$ 5.7&  18.9$\pm$ 5.7&  31.1$\pm$ 5.7&
$\leq$20.4    &$\leq$20.3    &
$\leq$17.2    \\
17&    SAGE\,053054.2$-$683428.3&  21.0$\pm$ 2.9&  56.4$\pm$ 2.9&  53.0$\pm$ 2.9&  53.6$\pm$ 2.9&  52.4$\pm$ 2.9&  52.0$\pm$ 4.0&  44.5$\pm$ 4.0&  38.5$\pm$ 4.0&  34.6$\pm$ 4.0&  19.2$\pm$ 4.0&
 27.5$\pm$ 2.9& 15.5$\pm$ 2.9&
 26.8$\pm$ 4.0\\
18&           IRAS\,05328$-$6827&  12.3$\pm$ 4.5&  26.6$\pm$ 4.4&  31.4$\pm$ 4.4&  24.2$\pm$ 4.5&  25.0$\pm$ 4.4&  19.8$\pm$ 8.0&  33.8$\pm$ 8.0&  17.0$\pm$ 8.0&   3.2$\pm$ 8.0&  17.4$\pm$ 8.1&
 23.9$\pm$ 4.5&$\leq$13.4    &
 74.3$\pm$ 8.0\\
19&         LMC\,053705$-$694741&  50.2$\pm$ 3.7&  45.1$\pm$ 3.6&  25.6$\pm$ 3.6&  17.0$\pm$ 3.7&  20.2$\pm$ 3.6&  13.0$\pm$ 6.5&  15.0$\pm$ 6.5&   8.4$\pm$ 6.5&  19.0$\pm$ 6.5&  40.8$\pm$ 6.5&
 22.4$\pm$ 3.7&$\leq$11.0    &
 30.6$\pm$ 6.5\\
20&                     ST\,01&  47.6$\pm$ 3.6&  71.3$\pm$ 3.6&  52.3$\pm$ 3.6&  54.9$\pm$ 3.6&  38.2$\pm$ 3.6&  43.0$\pm$ 4.5&  34.8$\pm$ 4.5&  33.3$\pm$ 4.5&  41.5$\pm$ 4.5&  38.7$\pm$ 4.5&
 32.4$\pm$ 3.6&$\leq$10.9    &
$\leq$13.6    \\
\hline
\end{tabular}
\end{center}
\end{table}
\end{center}
\end{landscape}

\section{Galactic massive YSO sample observed with {\bf \it ISO}}
\label{iso_appendix}

\newcommand{\mtc}{\multicolumn}
\begin{table*}
\begin{center}
\caption{Galactic massive YSOs observed with \iso\ and used as a Galaxy comparison sample.
TDT number is the unique identifier for \iso\ observations.
References for source bolometric luminosities and distances are as follows: 
1: \citet{vandertak13}; 2: \citet{stock15}; 3: \citet{guzman12};
4: \citet{lumsden12}; 5: \citet{dewit09}; 6: \citet{navarete15}; 7: \citet{boley13}
8: \citet{dewit11}. The three sources that do not show \oi\ in emission at 63\,\micron\ 
(source ID between parentheses, ``a'' and ``u'' in the \oi\ notes column respectively for line in absorption and undetected)
are listed for completeness but are not included in further analysis. $F60$ and
$F100$ are fluxes from IRAS catalogues (uncertainties <\,20\%). 
For the sources used in the cooling budget discussion in 
Sect\,\ref{cooling_top3} additional properties are provided in the last column: 
ice absorption, 9: \citet{gibb04}; \hho\ maser emission, 10: \citet{palagi93}; 
radio emission, 11: \citet{wilson03}, 12: \citet*{wendker91}, 
13: \citet{momose01}.}
\label{isosources}
\begin{tabular}{lcccccllccl}
\hline
             Source ID  &RA (J2000)        &Dec&TDT \#& $L_{\rm bol}$&$d$    &Ref.&\oi  &$F60$&$F100$&Source\\
                        &(h:m:s)          &($\degr:^\prime:^{\prime\prime}$)   &
			     & (10$^3$\lsun)&(kpc)&    &notes&
			(10$^3$\,Jy)& (10$^3$\,Jy) &properties\\
\hline
             W3\,IRS5&02:25:40.6&+62:05:51  &47301305&\al{1}70  & 2.0 &1& & & &ice$^9$, maser$^{10}$, radio$^{11}$\\ 
           AFGL\,4029&03:01:29.2&+60:29:12  &86300969&  20  & 2.2 &5&\\	    
            AFGL\,490&03:27:38.7&+58:47:01  &62300962&   4\ar{.4}& 1.4 &6&&0.71&0.78&\\	    
        Mon\,R2\,IRS3&06:07:47.8&$-$06:22:55&71002509&  12\ar{.6}& 0.8\ar{3}&7&&13.1&20.2&\\	    
  G298.2234$-$00.3393&12:10:01.1&$-$62:49:54&06300265&\al{46}30  &\al{1}1.3&3&&10.7&11.3&\\	    
           AFGL\,4176&13:43:01.7&$-$62:08:56&11702012&  25\ar{.6}& 5.3 &3&&2.7&3.57&\\	    
           NGC\,6334I&17:20:53.3&$-$35:47:00&32400422&\al{2}60  & 1.7 &1&&11.4&22.0&\\  
  G000.3138$-$00.2000&17:47:09.7&$-$28:46:28&69601311&\al{4}80  & 8.0 &3&&2.05&7.14&\\	    
         G5.89$-$0.39&18:00:30.4&$-$24:04:02&12500924&  51  & 1.3 &1&&&&\\	    
                M8\,E&18:04:53.3&$-$24:26:42&47700355&  20  & 1.2\ar{5}&4&&1.61&2.78&\\	    
	(G10.47$+$0.03)&18:08:38.2&$-$19:51:50&70300101&\al{3}70  & 5.8 &1&a&&&\\
	       (W33\,A)&18:14:39.1&$-$17:52:07&51500805&  44  & 2.4 &1&u&&&\\
          GGD27$-$ILL&18:19:12.0&$-$20:47:31&32901362&  25  & 1.9 &3&&2.77&3.71&\\	    
	   (AFGL\,2136)&18:22:26.6&$-$13:30:16&34000203&  70  & 2.0 &8&u&&&\\
          AFGL\,7009S&18:34:20.9&$-$05:59:42&47801136&  29  & 3.1 &3&&0.95&2.14&ice$^9$\\	    
        G29.96$-$0.02&18:46:03.8&$-$02:39:22&15201381&\al{3}50  & 6.0 &1&&7.50&11.7&\\  
        G31.41$+$0.31&18:47:34.3&$-$01:12:46&70500104&\al{2}30  & 7.9 &1&&1.09&2.82&\\	    
          W51\,N$-$e1&19:23:43.8&+14:30:26  &52501399&\al{1}00  & 5.1 &1&&&&\\	    
           S106\,IRS4&20:27:26.6&+37:22:48  &86000201&  40  & 1.5 &2&&10.1&13.1&\\	    
           AFGL\,2591&20:29:24.7&+40:11:19  &86100216&\al{2}20  & 3.3 &1&&5.31&5.72&ice$^9$, maser$^{10}$, radio$^{12}$\\	    
           DR\,21(OH)&20:39:00.8&+42:22:48  &34700439&  13  & 1.5 &1&&2.43&1.68&maser$^{10}$\\  
      NGC\,7538\,IRS1&23:13:45.3&+61:28:10  &22001650&\al{1}30  & 2.7&1&&7.07 &14.1&ice$^9$, maser$^{10}$, radio$^{13}$\\	
\hline    
\end{tabular}
\end{center}
\end{table*}

\addtocounter{footnote}{-8}

Table\,\ref{isosources} lists relevant source properties for the Galactic comparison
sample observed with the \iso\ LWS; all these observations were performed with the 
medium-resolution wavelength range (AOT L01) mode \citep[see ][ for details]{gry03}; 
spectra were recovered from the IDA\footnotemark. The relatively low SNR of the 
spectra, especially at shorter wavelengths, meant we were only able to measure 
fluxes for the strongest emission lines: \cii\ at 158\,\micron, \oi\ at 63 and 
145\,\micron, \oiii\ at 88\,\micron, \nii\ at 122\,\micron\ and CO at 186\,\micron. 
We measured all line fluxes from archival spectra rather than using published 
measurements, since those spectra have been re-reduced with the same calibration 
version.
\addtocounter{footnote}{+7}

All sources exhibit emission for \cii\ and most exhibit \oi\ emission. For the 
three sources with \oi\ at 63\,\micron\ undetected or in absorption, 
\oi\ at 145\,\micron\ is undetected. %W33\,A exhibits \oi\ 63\,\micron\ in absorption
% and %145\,\micron\ in emission in the PACS observations of \citet{karska14}.
For the remaining 19 sources in the ISO sample \oi\ at 63 and 145\,\micron\ are 
both in emission. Three sources (DR\,21(OH), NGC\,6334\,I and AFGL\,2591) show 
P-Cygni or inverse P-Cygni profiles for \oi\ 63\,\micron\ (and emission at 
145\,\micron) in the PACS spectra presented by \citet{karska14}; for W51N-e1 \oi\ 
63\,\micron\ absorption is detected \citep{karska14}. We believe this discrepancy in 
line profiles is due to the fact that the \iso\ observations sample a much larger 
fraction of the more extended YSO environment, i.e. it is a result of the vastly 
different beam sizes \citep[see also discussion in ][]{stock15}.

For this sample \oi\ emission is always stronger at 63\,\micron\ than at 
145\,\micron, typically by a factor $\sim\,7$, with a large scatter; the 
63\,\micron\ contributes $\sim$\,0.87\,$\pm$\,0.1 to the total \oi\ emission, 
consistent with what is observed for the sources with simple emission profiles in 
the \citet{karska14} sample \citep[see also][]{karska18}. For optically thin 
emission, the ratio of 63\,\micron\ to 145\,\micron\ emission is $\gsim$\,10 
\citep{tielens85}; for all but one of theses \iso\ sources the observed ratio is 
<\,10, implying that one or both lines may be optically thick. We use these ratios 
to estimate total \oi\ luminosities for the Magellanic sample in 
Sect.\ref{coolingbudget}, but are mindful that \oi\ luminosities could in fact be higher.
 
We detect \nii\ emission at 122\,\micron\ for seven sources (5-$\sigma$\,detections).
The theoretical \cii/\nii\,122\,\micron\, ratio is used to calculated the fraction 
of \cii\ emission that originates from the ionised gas rather than the PDR, using the 
method described in Sect.\,\ref{cii_nii}. The predicted theoretical ratio is 
(\cii/\nii\,122\,\micron)$_{\rm T}$\,$\sim$\,2.4, while observed ratios are 
(\cii/\nii\,122\,\micron)$_{\rm O}$\,>\,2.7. We estimate that between 11 and 76\% of the 
\cii\ emission originates from the PDR, with a median value $\sim$\,60\%.  Estimates 
available in the literature are typically higher, with $\gsim$\,70\% of the emission
originating from the ionised gas \citep[e.g.,][]{oberst11,bernard-salas12}. This fraction
is dependent on the C/N ratio. \citet{oberst11} adopt C/N\,=\,1.8, appropriate for the 
diffuse ISM; for Galactic \hii\ regions C/N is significantly larger \citep[e.g.,]
[and references therein]{carlos-reyes15}, consistent with our adopted value 
C/N\,=\,5.9. This in turn leads to lower derived PDR gas fractions. 

Out of the 19 sources, twelve and seven sources exhibit emission for \oiii\ at
88\,\micron\ and CO\,(14$-$13) at 186\,\micron\ respectively. No \hho\ emission is 
detected at 179.5, 180.5 or 108\,\micron. The spectral resolution of the \iso\ 
detectors SW4 and SW5 is too low for any reliable OH detections at 79 and 
84\,\micron.

We used the seven Galactic massive sources with measured \oi, \cii\ and CO line 
luminosities as a comparison sample to the Magellanic YSOs for the line cooling 
analysis (Sect.\,\ref{cooling_top3}). The last column in Table\,\ref{isosources} lists 
relevant source properties for this subsample. The luminosity of CO\,(14$-$13) at 
186\,\micron\ accounts for $\sim$\,3\% of the total CO luminosity measured across the 
SPIRE and PACS ranges \citep[$J \geq 4$, estimated using data from][]{green16,yang18}. 
Cooling fractions for these seven Galactic sources are shown in 
Fig.\,\ref{iso_cooling_pies} and listed in Table\,\ref{iso_pies}.

\begin{figure}
\begin{center}
\includegraphics[scale=0.42]{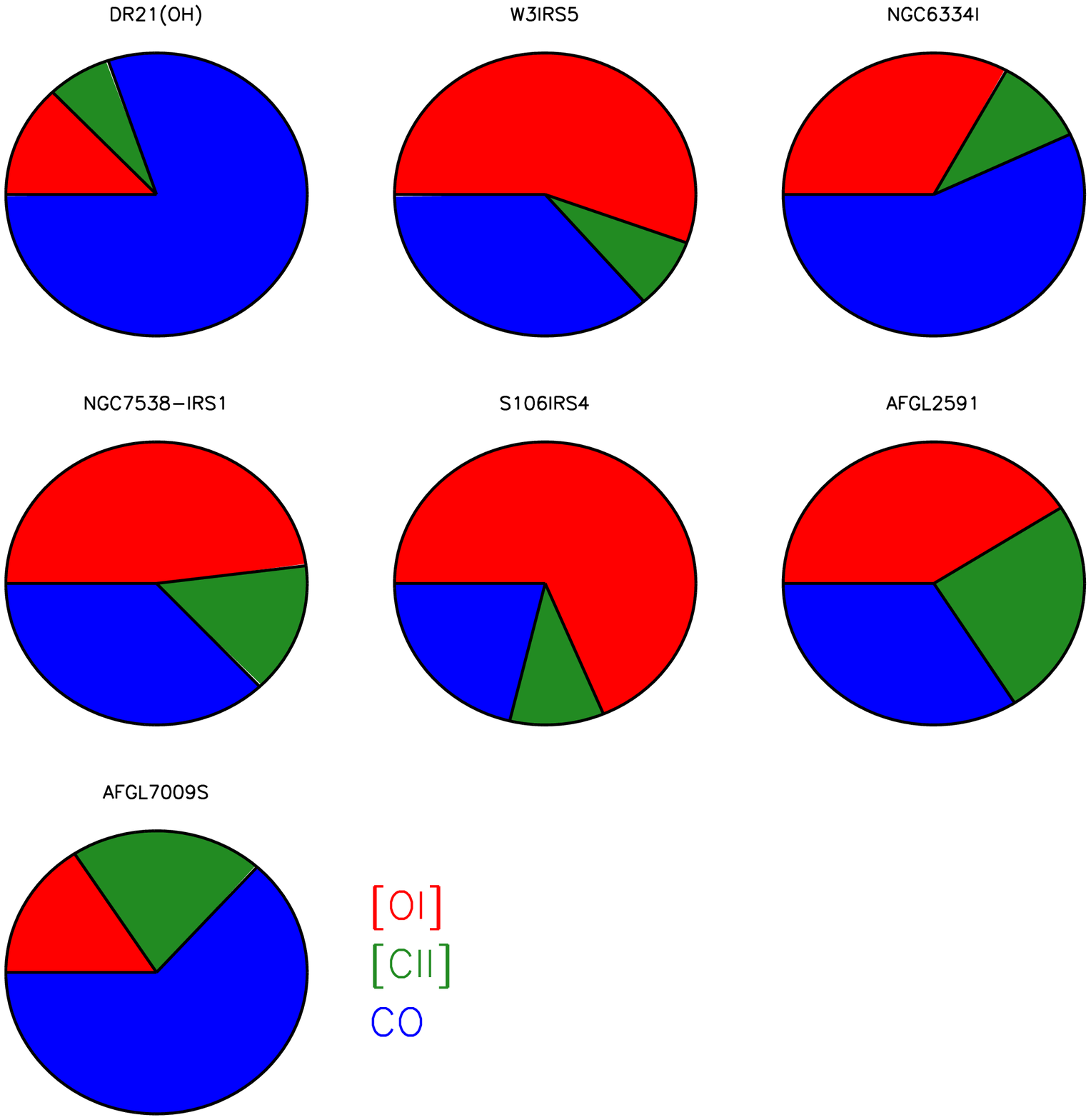}
\end{center}
\caption{Cooling fractions for a sample of Galactic YSOs, for the main gas species 
discussed in this work: \oi\ (red), \cii\ (green) and CO (blue). The fractions are 
listed in Table\,\ref{iso_pies}.}
\label{iso_cooling_pies}
\end{figure}

\begin{table}
\begin{center}
\caption{\normalsize Line luminosities as fractions of the total IR line luminosity
$L_{\rm LIR}$ for \oi, \cii\ and CO. The last column gives the ratio of total IR line luminosity
to total IR luminosity $L_{\rm LIR}$/$L_{\rm TIR}$.}
\label{iso_pies}
\begin{tabular}{l|@{\hspace{-1mm}}c|@{\hspace{-1mm}}c|@{\hspace{-1mm}}c|@{\hspace{-1mm}}c}
\hline
Source ID               &$L_{\rm [OI]}$      &$L_{\rm [CII]}$&$L_{\rm CO}$&$L_{\rm LIR}$\\
                        &($L_{\rm LIR}$)     &($L_{\rm LIR}$)&($L_{\rm LIR}$)  & ($L_{\rm TIR}$)       \\
\hline
DR\,21(OH)              &0.13     &0.07      &0.81    &0.003\\
W3\,IRS5                &0.55     &0.08      &0.36    &0.002\\
NGC\,6334I              &0.33     &0.10      &0.57    &0.000\rlap{5}\\
NGC\,7538\,IRS1         &0.48     &0.15      &0.37    &0.001\\
S\,106\,IRS4            &0.68     &0.10      &0.21    &0.003\\
AFGL\,2591              &0.41     &0.25      &0.34    &0.000\rlap{4}\\
AFGL\,7009S             &0.16     &0.21      &0.64    &0.001\\
\hline
\end{tabular}
\end{center}
\end{table}

\clearpage

\section{CO rotational diagrams fitted with admixture of LTE gas components}
\label{gas_admixture}

Another empirical description of CO rotational diagrams invokes an
admixture of gas components with temperatures described by $d\,N/d\,T$\,=\,$aT^{-b}$, 
over a temperature interval $T_{\rm min}$\,<\,T\,<\,$T_{\rm max}$; such power-law 
distribution of temperatures has been used to successfully model rotational 
diagrams for both H$_2$ \citep{neufeld08} and CO \citep{neufeld12,manoj13}. 
In the limit of {\it optically thin LTE gas} and following \citet{neufeld08}
and \citet{neufeld12}:
\begin{equation} 
\frac{N_J}{g_J} = - \frac{N_{\rm CO}}{k/hcB}
\left[\frac{b-1}{T^{1-b}_{\rm min}-T^{1-b}_{\rm max}}\right]
\left[\frac{\Gamma(b,z_1)-\Gamma(b,z_2)}{E_J}\right],
\label{powerlaw}
\end{equation}
where $b$ is the power law index, $z_1 \equiv E_J/kT_{\rm min}$, 
$z_2 \equiv E_J/kT_{\rm max}$ and $\Gamma(b,z_i)$ is the upper incomplete gamma 
function\footnotemark
\footnotetext{$\Gamma(b,z_i) \equiv \int^{\infty}_{z_i} t^{b-1} e^{-t} dt.$}. 
The function in Eq.\,\ref{powerlaw} can be fitted to the observed rotational 
diagrams, with free parameters $b$, $N_{\rm CO}$, $T_{\rm min}$ and $T_{\rm max}$. Given 
that for the SPIRE CO transitions 55.3\,$\leq E_J({\rm K})\leq$\,503.1, our fit 
solutions are insensitive to the value of $T_{\rm max}$, which is thus kept fixed at 
$T_{\rm max}$\,=\,5000\,K; they are however sensitive to the value of $T_{\rm min}$ 
(lower limit $T_{\rm min}$\,=\,4\,K).

Figure\,\ref{co_rotational2} shows the temperature power-law fits to the
rotational diagrams for the YSO sample; fit parameters are listed in
Table\,\ref{tdist_pars}. The power-lax index is typically
 $b$\,$\sim$\,3.2 (range 2.6\,$-$\,4.4), consistent with literature 
values \citep{neufeld12,manoj13}. For the other two parameters we obtain: 
$N_{\rm CO}$\,$\sim$\,9.5\,$\times 10^{54}$ molecules (1\,$-$\,23\,$\times 10^{54}$ 
molecules) and $T_{\rm min}$\,$\sim$\,23\,K (10\,$-$\,42\,K). The total amount of CO is 
similar to that derived from the two-temperature model (Sect.\,\ref{co_rot_thin}). 
For one object the minimum temperature is not well constrained 
(N\,81, Fig.\,\ref{co_rotational2} bottom row). For 10 out of 13 objects, a 
two-temperature model provides a better fit in terms of minimised $\chi^2$, likely a 
reflection of the larger number of free parameters.

While gas with temperatures over a continuum range may seem a reasonable premise,
the competing processes of gas cooling and heating are likely to favour and give rise
to gas with more restricted conditions. This is in fact what leads to the predominance
of the different ISM phases. Therefore, the CO emission is perhaps more likely to be
dominated by a few components with a temperature spread over a relatively small range.

\begin{table}
\begin{center}
\caption{\normalsize Fit parameters to the observed CO rotational diagrams (CO\,(4$-$3) to CO\,(13$-12$)), using a temperature power-law  
distribution characterised by a power-lax index $b$, total number of CO molecules $N_{\rm CO}$, and minimum temperature $T_{\rm min}$. 
For IRAS05328$-$6827 and LMC\,053705$-$694741 the fits are very uncertain. Also listed are $\chi^2$ and number of degrees of freedom 
$N_{\rm f}$ for each fit.}
\label{tdist_pars}
\begin{tabular}{l@{\hspace{0mm}}|c@{\hspace{0mm}}|c@{\hspace{1mm}}|c@{\hspace{0mm}}|c@{\hspace{0mm}}|c}
\hline
Source ID            &$b$&$\log N_{\rm CO}$&$T_{\rm min}$&$\chi^2$&$N_{\rm f}$\\
                     &       & (mol)           &(K)          &        &           \\
		     \hline
\multicolumn{6}{c}{LMC YSOs}\\
\hline
       IRAS\,04514$-$6931&2.8$\pm$0.1&54.94$\pm$0.23&\al{1}1.1$\pm$	  3.9&\al{1}9.4&7\\
       IRAS\,05328$-$6827&3.2$\pm$1.9&54.41$\pm$0.43&\al{2}0.8$\pm$ 19.\ar{8}&      0.6&1\\
     LMC\,053705$-$694741&3.6$\pm$0.6&55.32$\pm$0.43&	   9.6$\pm$	  4.9&      2.0&1\\
                N113 YSO3&3.0$\pm$0.2&55.30$\pm$0.08&\al{1}9.4$\pm$	  2.8&      6.9&7\\
SAGE\,045400.9$-$691151.6&4.4$\pm$0.7&55.00$\pm$0.04&\al{3}8.8$\pm$	  4.3&      4.9&6\\
SAGE\,051351.5$-$672721.9&4.1$\pm$0.5&54.23$\pm$0.04&\al{4}1.6$\pm$	  4.1&      8.5&7\\
SAGE\,052202.7$-$674702.1&3.0$\pm$0.3&54.32$\pm$0.09&\al{2}6.9$\pm$	  5.3&      9.1&7\\
SAGE\,052212.6$-$675832.4&3.5$\pm$0.3&55.18$\pm$0.07&\al{2}1.8$\pm$	  2.8&      7.4&6\\
SAGE\,053054.2$-$683428.3&3.3$\pm$0.3&54.43$\pm$0.06&\al{3}1.0$\pm$	  4.4&      7.1&6\\
                   ST\,01&3.0$\pm$0.2&54.98$\pm$0.15&\al{1}3.9$\pm$	  3.2&\al{1}0.3&7\\
               N113 YSO-1&2.7$\pm$0.1&55.36$\pm$0.07&\al{2}1.4$\pm$	  3.1&      3.5&7\\
\hline
\multicolumn{6}{c}{SMC YSOs}\\
\hline
       IRAS\,00464$-$7322&3.1$\pm$0.3&54.15$\pm$0.07&\al{2}7.2$\pm$	  4.5&      7.0&6\\
       IRAS\,00430$-$7326&3.0$\pm$0.2&54.00$\pm$0.09&\al{2}5.6$\pm$	  4.3&      7.1&7\\
                    N\,81&2.7$\pm$0.2&55.26$\pm$0.13&	   4.0    	     &      8.8&6\\
     SMC\,012407$-$730904&2.6$\pm$0.1&54.74$\pm$0.16&\al{1}4.4$\pm$	  4.1&\al{1}2.5&7\\
\hline
\end{tabular}
\end{center}
\end{table}

\begin{figure*}
\begin{center}
\includegraphics[scale=1.04]{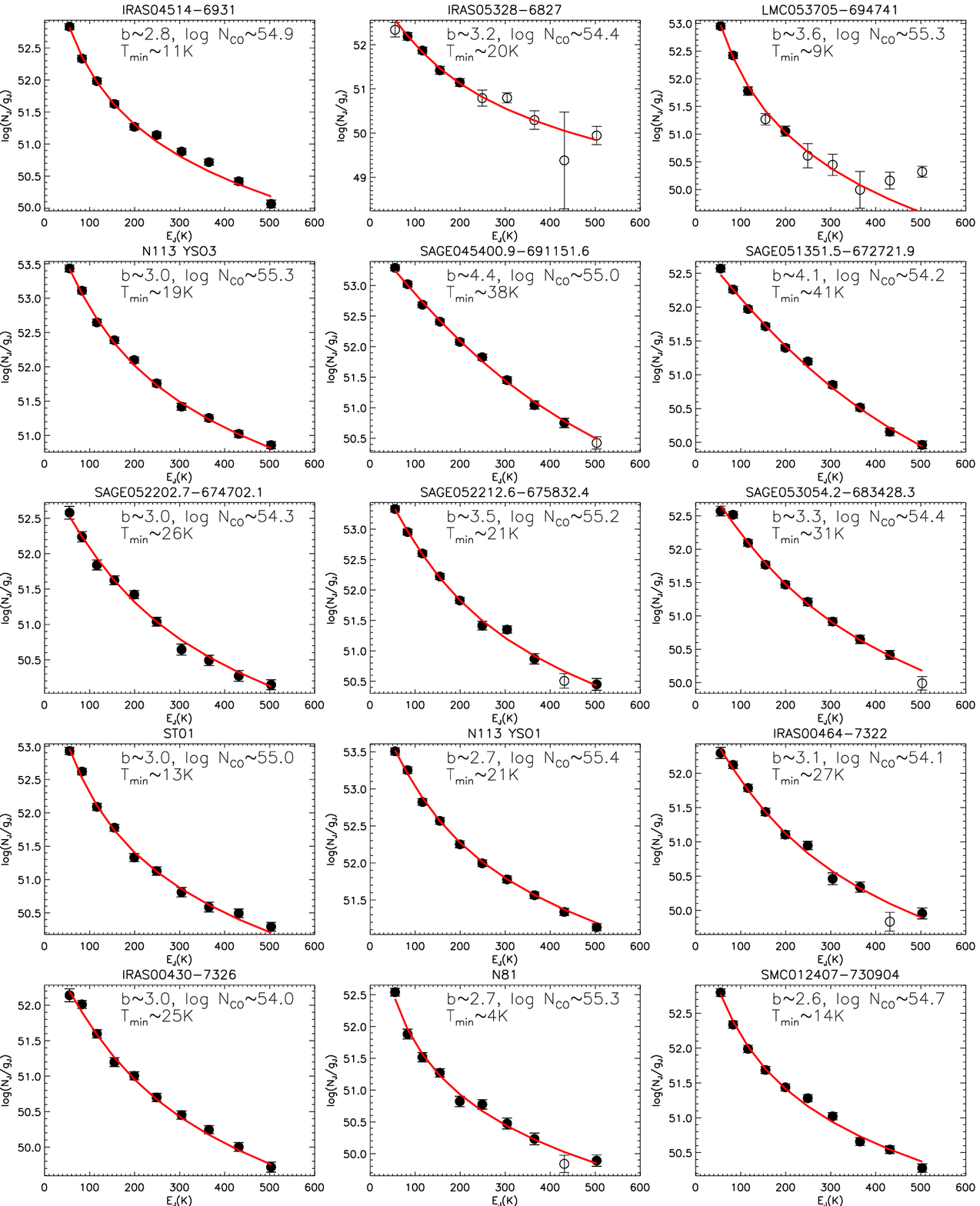}
\end{center}
\caption{CO rotational diagrams for the Magellanic YSO target sample (symbols as in
Fig.\,\ref{co_rotational1}). The data are fitted by a temperature power-law 
distribution characterised by a power-law index $b$, a total number of CO 
molecules $N_{\rm CO}$ and a minimum temperature $T_{\rm min}$ (solid red lines).
Approximate fit parameters are labelled in each panel; full fit results are listed in
Table\,\ref{tdist_pars}.}
\label{co_rotational2}
\end{figure*}

%\clearpage

%%%%%%%%%%%%%%%%%%%%%%%%%%%%%%%%%%%%%%%%%%%%%%%%%%
% Don't change these lines
\bsp	% typesetting comment
\label{lastpage}

\end{document}